\documentclass[]{spie}  

\usepackage{amsmath,amsfonts,amssymb}
\usepackage{graphicx}
\usepackage[colorlinks=true, allcolors=blue]{hyperref}

\graphicspath{{images/}}

\title{A Comparison of Trapped Particle Models in Low Earth Orbit}

\author[a,b,c]{Jakub \v{R}\'{i}pa}
\author[d]{Giuseppe Dilillo}
\author[e,f]{Riccardo Campana}
\author[a]{G\'abor Galg\'oczi}
\affil[a]{Institute of Physics, E\"otv\"os Lor\'and University, Budapest, Hungary}
\affil[b]{Department of Theoretical Physics and Astrophysics, Faculty of Science, Masaryk University, Brno, Czech Republic}
\affil[c]{Astronomical Institute of Charles University, Prague, Czech Republic}
\affil[d]{Universita di Udine, Dipartimento di Matematica, Informatica e Fisica, Udine, Italy}
\affil[e]{INAF/OAS, Via Gobetti 101, I-40129, Bologna, Italy}
\affil[f]{INFN-Sezione di Bologna, Viale Berti Pichat 6/2, I-40127, Bologna, Italy}

\authorinfo{Further author information: (Send correspondence to J.R.)\\
J.R.: E-mail: \url{jripa@caesar.elte.hu}\\
G.D.: E-mail: \url{dilillo.giuseppe@spes.uniud.it}\\
G.G.: E-mail: \url{galgoczi@caesar.elte.hu}\\
R.C.: E-mail: \url{riccardo.campana@inaf.it}}

\begin{document}
\maketitle

\begin{abstract}
Space radiation is well-known to pose serious issues to solid-state high-energy sensors. Therefore, radiation models play a key role in the preventive assessment of the radiation damage, duty cycles, performance and lifetimes of detectors. In the context of HERMES-SP mission we present our investigation of AE8/AP8 and AE9/AP9 specifications of near-Earth trapped radiation environment. We consider different circular Low-Earth orbits. Trapped particles fluxes are obtained, from which maps of the radiation regions are computed, estimating duty cycles at different flux thresholds. Outcomes are also compared with published results on in-situ measurements.

This study is done on behalf of the HERMES-SP collaboration.
\end{abstract}


\section{INTRODUCTION}
\label{sec:intro}

Thanks to features such as compactness, light-weight, small power consumption and low cost, silicon-based high-energy sensors are enabling miniaturised spacecraft to pursue ambitious scientific objectives, once reachable only by larger missions.
However, space is an hazardous environment: the effect of space radiation lead to damage of solid-state sensors and to a consequent worsening of its performance \cite{Segneri2009,DelMonte2014}. Space radiation models play a key role in the preventive assessment of the radiation damage and henceforth detector duty cycle, performance and lifetime.

In the context of \emph{High Energy Rapid Modular Ensemble of Satellites} - Scientific Pathfinder (HERMES-SP) mission \cite{Fuschino2019} we present an investigation of the AE8/AP8 \cite{Sawyer1976,Vette1991a,Vette1991b} and IRENE (International Radiation Environment Near Earth) AE9/AP9 \cite{Johnston2014,Johnston2015,Ginet2013,OBrien2018} models specifications of near-Earth trapped radiation environment. We consider 36 different circular Low-Earth Orbits (LEO) at three different altitudes (500, 550 and 600\,km) and twelve inclinations ($0^\circ$, $5^\circ$, $10^\circ$, $15^\circ$, $20^\circ$, $30^\circ$, $40^\circ$, $50^\circ$, $60^\circ$, $70^\circ$, $80^\circ$ and $90^\circ$).

For each simulated orbit we obtain differential and integral trapped particles fluxes, from which we compute maps of the trapped particles region and estimate duty cycles at different flux thresholds.

Differences between AE8/AP8 and AE9/AP9 models are quantified. Trapped particles fluxes computed by AP9/AE9 models are found to be generally higher than its AP8/AE8 counterparts, especially at low inclinations. The models are also compared with published results on in-situ measurements performed by the Particle Monitor instrument onboard BeppoSAX\cite{Campana2014}.

The outcomes of this study are used in an investigation of GAGG:Ce scintillator afterglow emission (see Ref.~\citenum{Dilillo2020}) and in a software toolkit to simulate activation background for high energy detectors onboard satellites (see Ref.~\citenum{Galgoczi2020}).

\section{METHODS}
\label{sec:methods}

\subsection{AE8/AP8 and AE9/AP9 models}

The fluxes of geomagnetically trapped electrons and protons inside the inner van Allen radiation belt \cite{vanAllen1958} contribute to the overall detected instrumental background and they are especially important when a satellite at LEO passes through the South Atlantic Anomaly (SAA) or polar regions.

Several models describing the fluxes of the trapped particles around the Earth, based on measurements from several space missions, have been developed over last decades, for example the NASA’s AE8 and AP8 models of trapped electrons and protons, respectively. These models are based on data from more than 20 satellites from the early sixties to the mid-seventies and are available on the ESA's SPace ENVironment Information System (SPENVIS)\footnote{\url{www.spenvis.oma.be}}. The SPENVIS tool is an Internet interface to models of the space environment and its effects, developed by a consortium led by the Royal Belgian Institute for Space Aeronomy (BIRA-IASB). The more recent AE9 and AP9 models of trapped electrons and protons are based on 33 satellite datasets from 1976 to 2011, and currently are available on SPENVIS only for evaluation purposes. At variance with respect to the AE8/AP8 models, the AE9/AP9 models include the flux uncertainties due to the statistical variations, instrument errors as well as the variations due to the changing space weather. The AE9/AP9 models also provide a more detailed spatial resolution.

For the AE8 model we choose the MAX condition which means the solar cycle maximum. The flux of
trapped electrons is on average highest near the maximum of the solar activity. For the AP8 model we choose the MIN condition which means the solar cycle minimum. The flux of protons is on average highest near the minimum of the solar activity \cite{Huston1996}. Therefore, this is a conservative estimation of fluxes. For the AE8/AP8 models the MAX/MIN of the solar cycle are the only two available versions.

For the AE9/AP9 models the 50\,\% confidence level (CL 50) and 90\,\% confidence level (CL 90) of fluxes were calculated using the Monte Carlo (MC) mode with 100 runs. The MC mode accounts for the uncertainty due to the random perturbations as well as the flux variations due to the space weather \cite{Ginet2013}.

\subsection{Calculation of trapped particle fluxes}

We employed the software provided by the U.S. Air Force Research Laboratory\footnote{\url{https://www.vdl.afrl.af.mil/programs/ae9ap9}} v1.50 to calculate the differential and integral fluxes (flux of particles with energy higher than a given energy) of both the AE8/AP8 and the AE9/AP9 models.

For the geomagnetic field, the International Geophysical Reference Field (IGRF) model \cite{Thebault2015} for the ``main'' magnetic field was used, in conjunction with the Olson-Pfitzer Quiet (OPQ77) model \cite{Olson1977} for the ``external'' magnetic field, fixed at 2020 January 01\cite{Johnston2017}.

The basic ``Kepler'' orbit propagator with ``J2'' perturbation effects was used to generate the ephemeris of a satellite with a given altitude and inclination. The ``J2'' perturbation accounts for long-term variations in the orbit due to oblateness of the Earth \cite{Johnston2017}.

The time sampling of the trapped particle fluxes was every 10 s with the total orbit duration of 60 days for the time period from 2020-01-01 to 2020-03-01. In order to have a more uniform sampling of the trapped particle regions, i.e. to avoid repeat ground-track orbits, the right ascension of ascending node (RAAN) was increased by $5^\circ$ in a stepwise manner every 10 days starting from RAAN = $0^\circ$ and ending with RAAN = $25^\circ$.

\section{Trapped Particle Maps}
\label{sec:maps}

\begin{figure}[p]
\begin{center}
\begingroup
\setlength{\tabcolsep}{0pt} 
\begin{tabular}{ccc}
\includegraphics[width=0.335\linewidth]{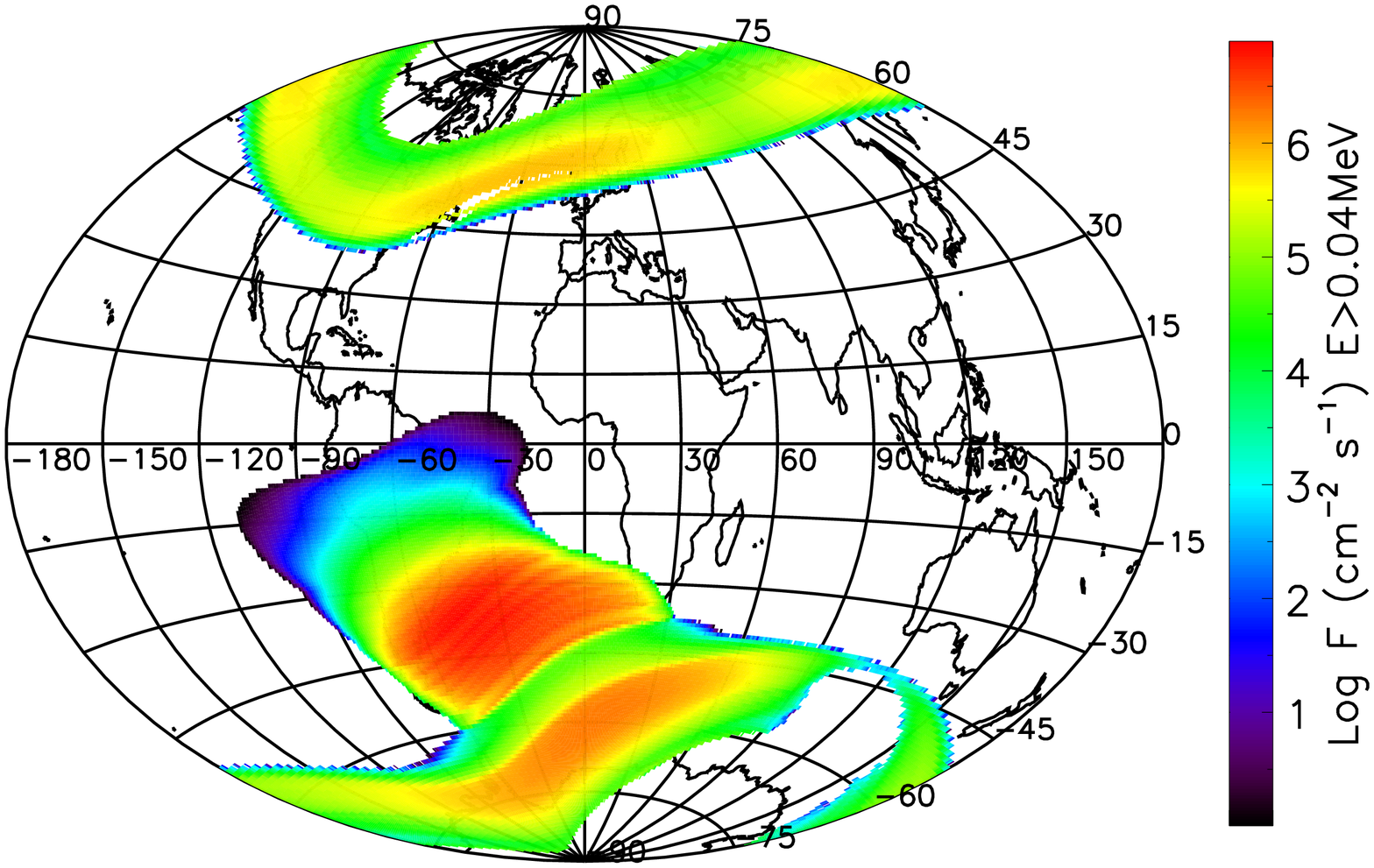}&
\includegraphics[width=0.335\linewidth]{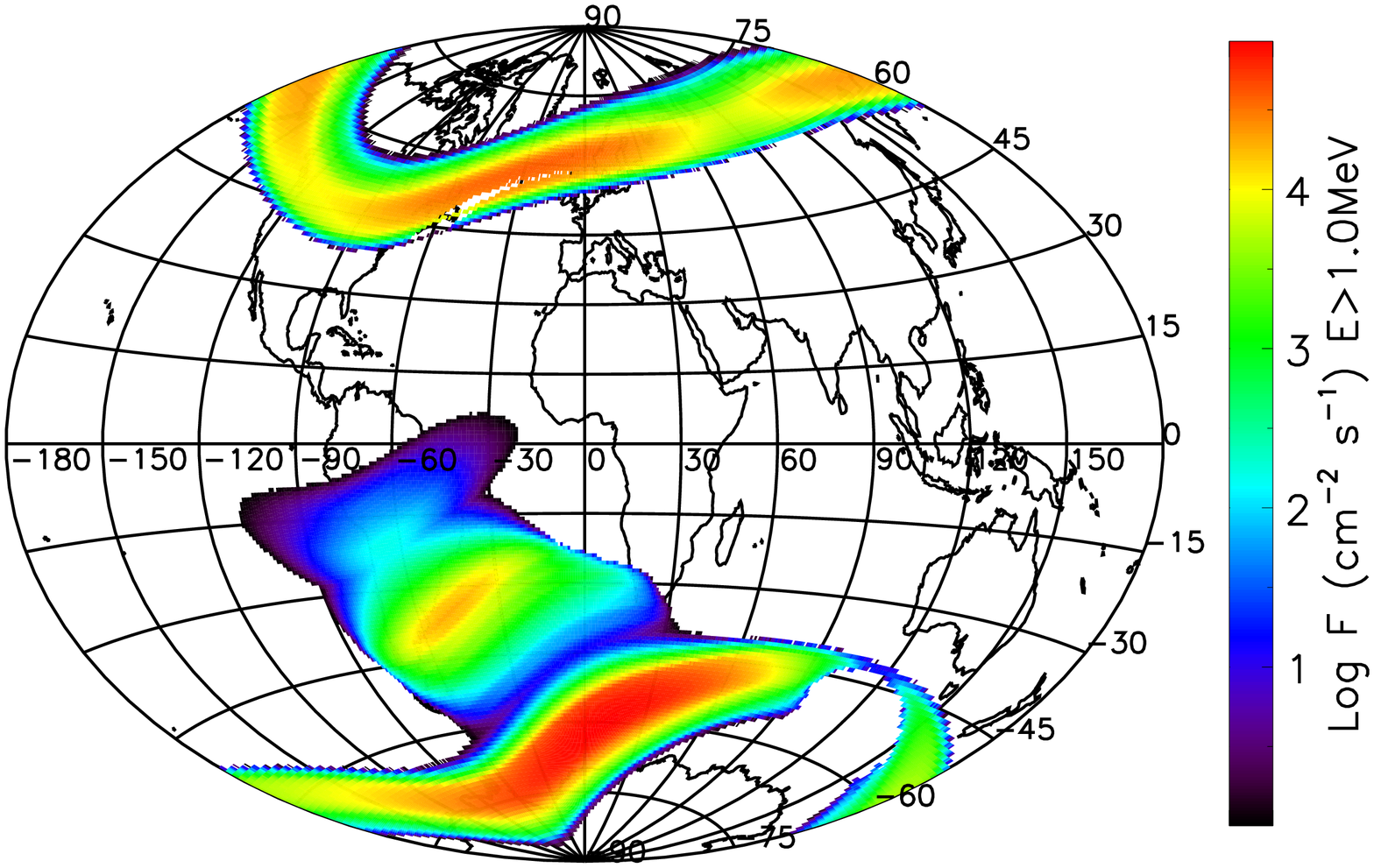}&
\includegraphics[width=0.335\linewidth]{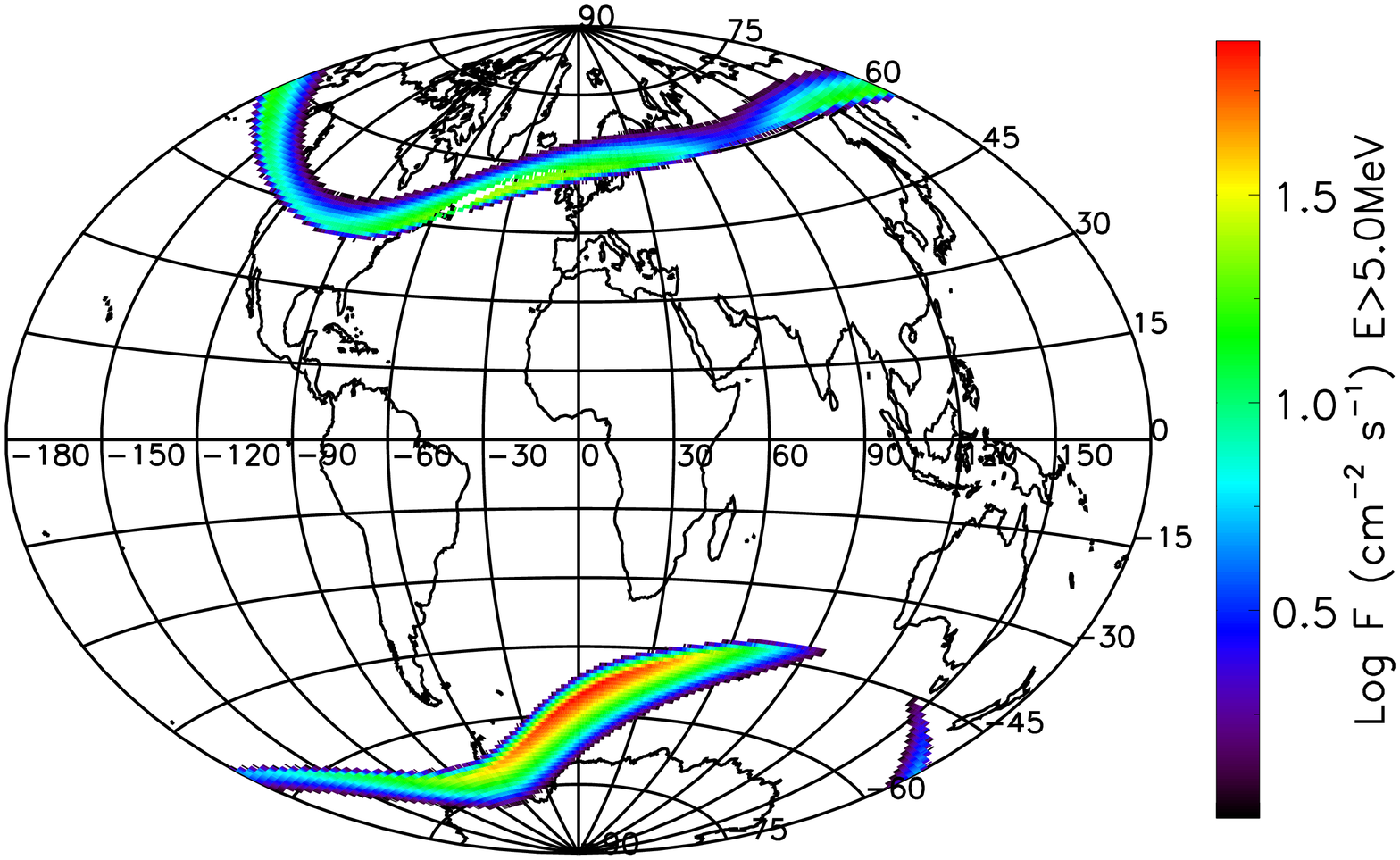}
\\
\includegraphics[width=0.335\linewidth]{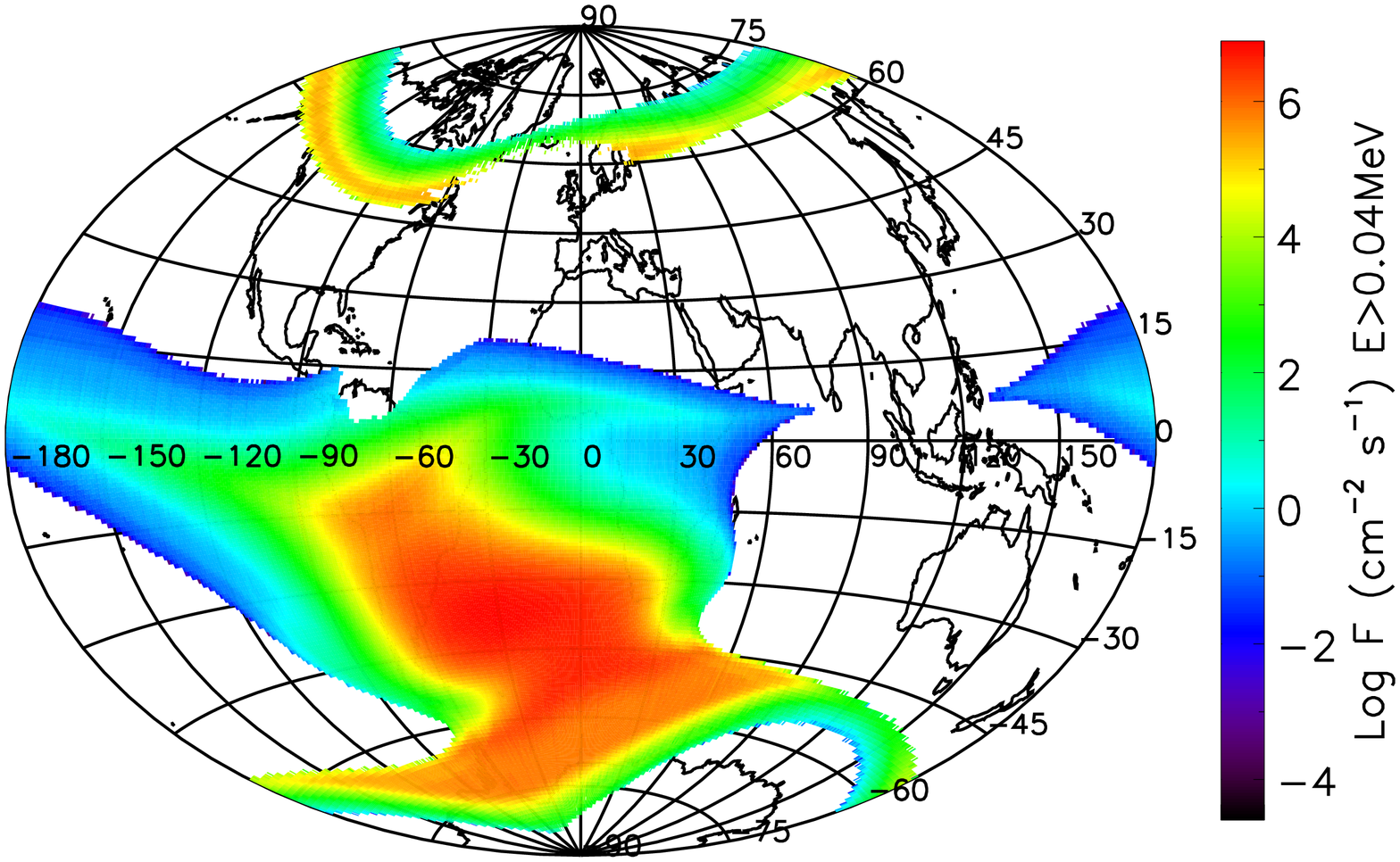}&
\includegraphics[width=0.335\linewidth]{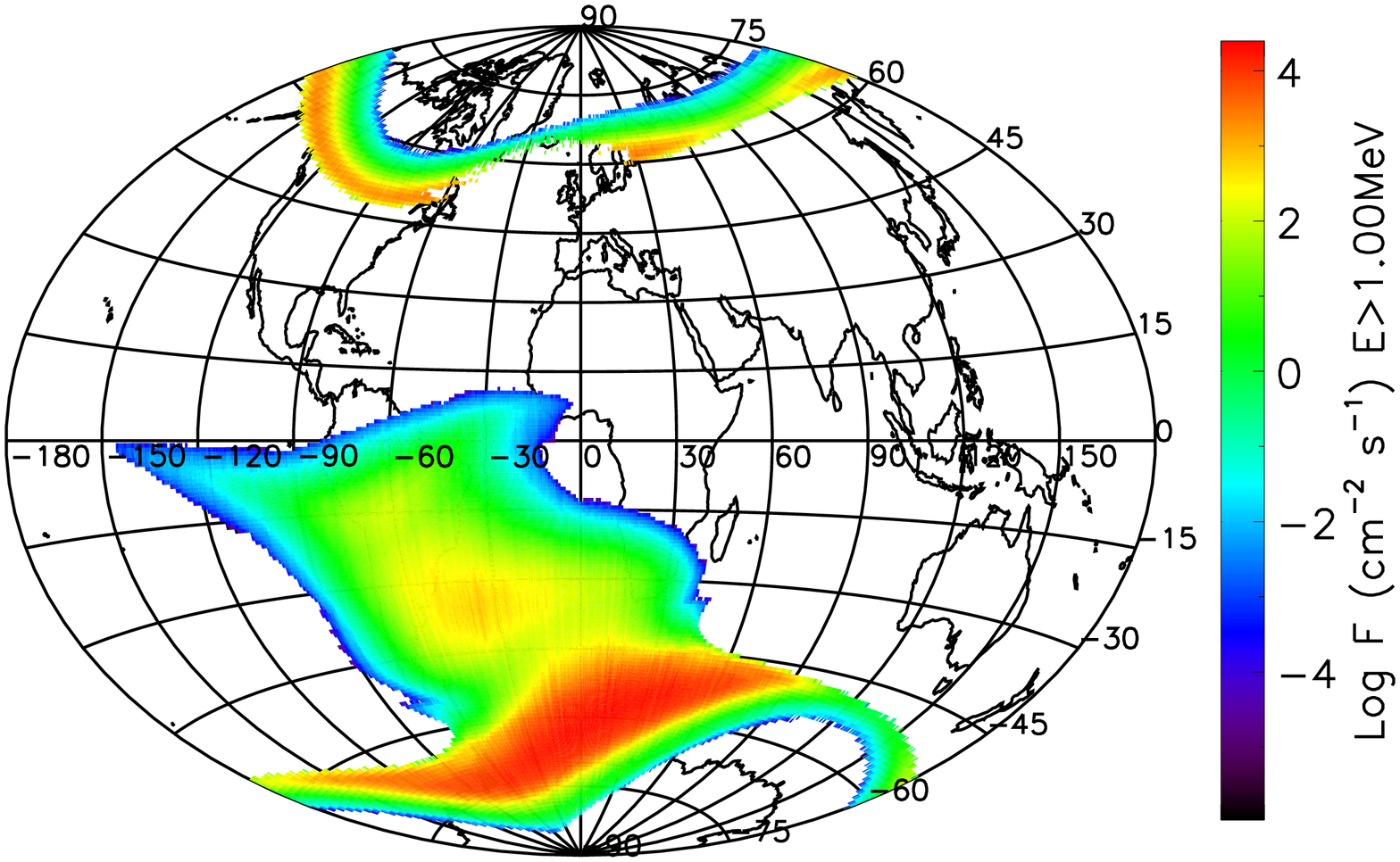}&
\includegraphics[width=0.335\linewidth]{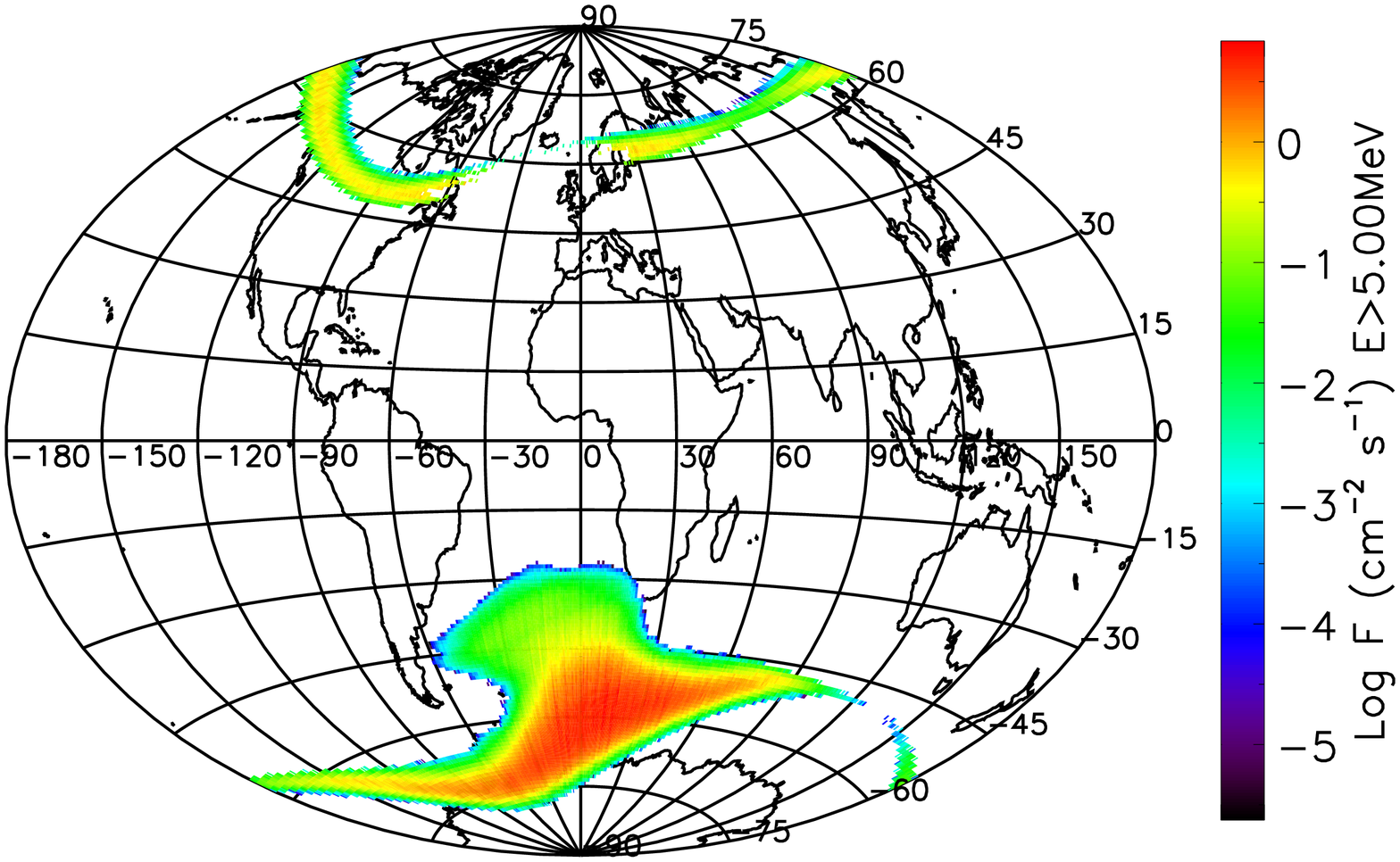}
\end{tabular}
\endgroup
\end{center}
\caption{\label{fig:maps_e}
Integral flux maps of trapped electrons for AE8 MAX model (top panels) compared with the flux maps for AE9 50\,\% CL model (bottom panels) at 550\,km altitude with low energy thresholds of 40\,keV, 1\,MeV and 5\,MeV, respectively from left to right.}
\end{figure}

\begin{figure}[p]
\begin{center}
\begingroup
\setlength{\tabcolsep}{0pt} 
\begin{tabular}{ccc}
\includegraphics[width=0.335\linewidth]{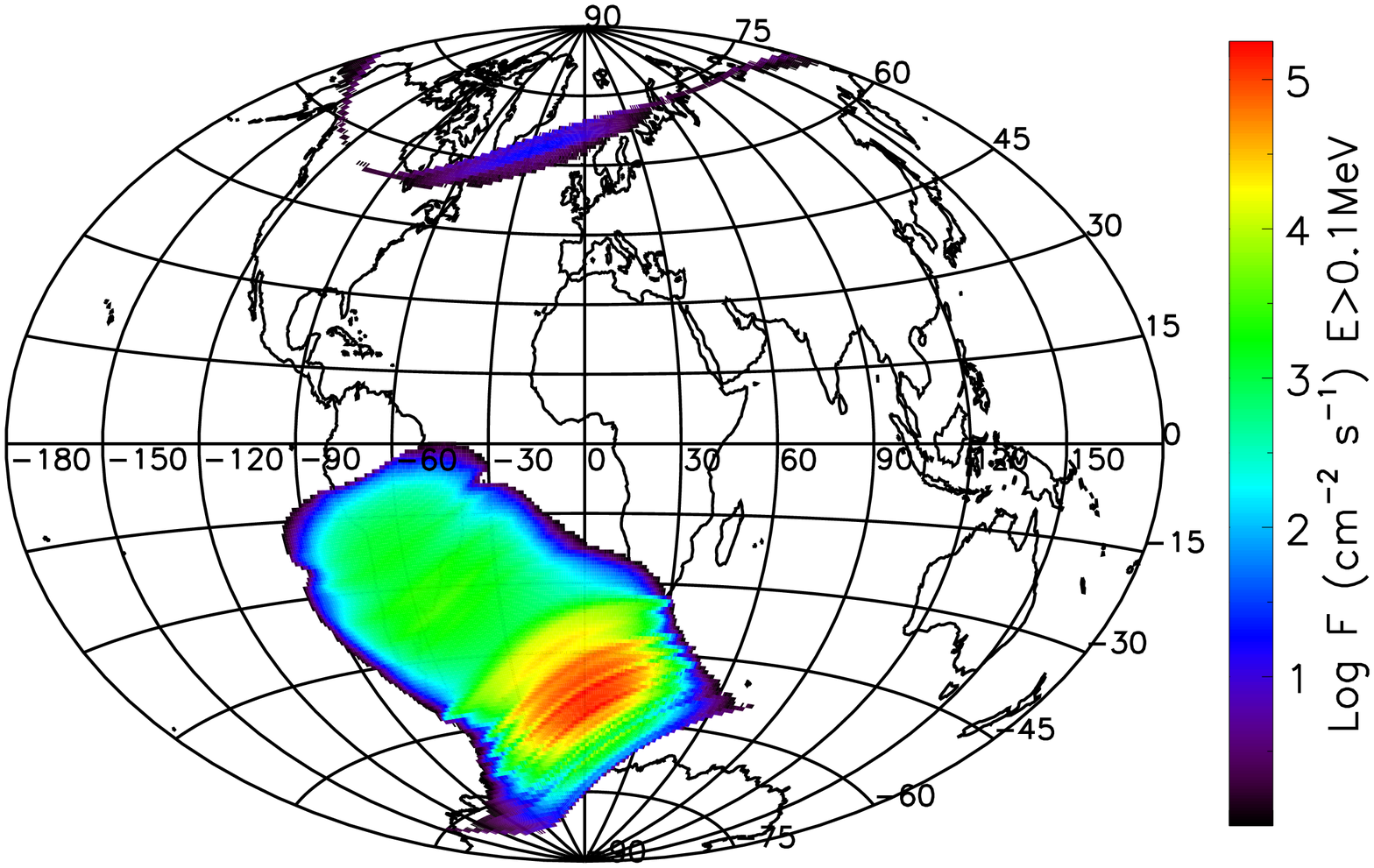}&
\includegraphics[width=0.335\linewidth]{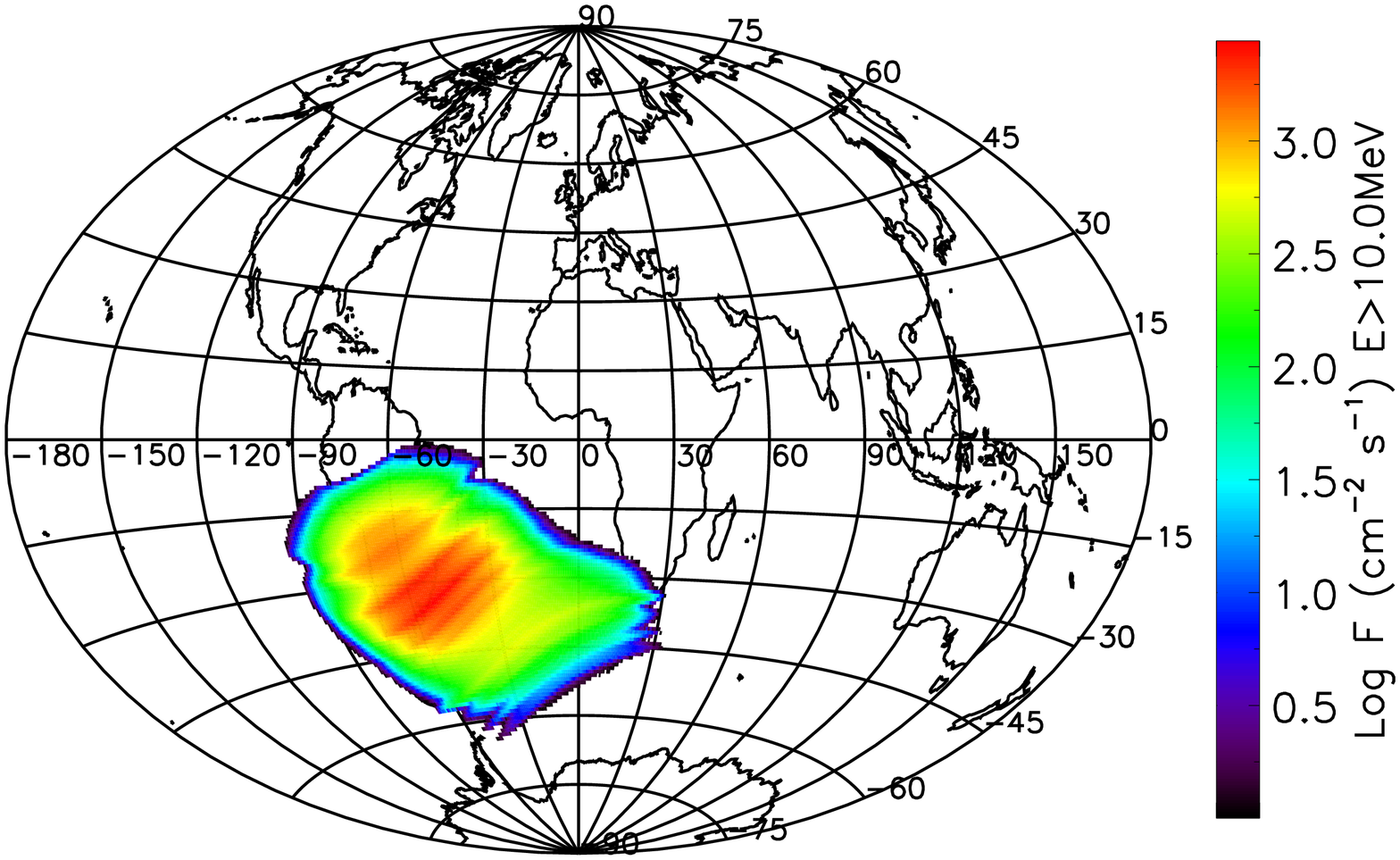}&
\includegraphics[width=0.335\linewidth]{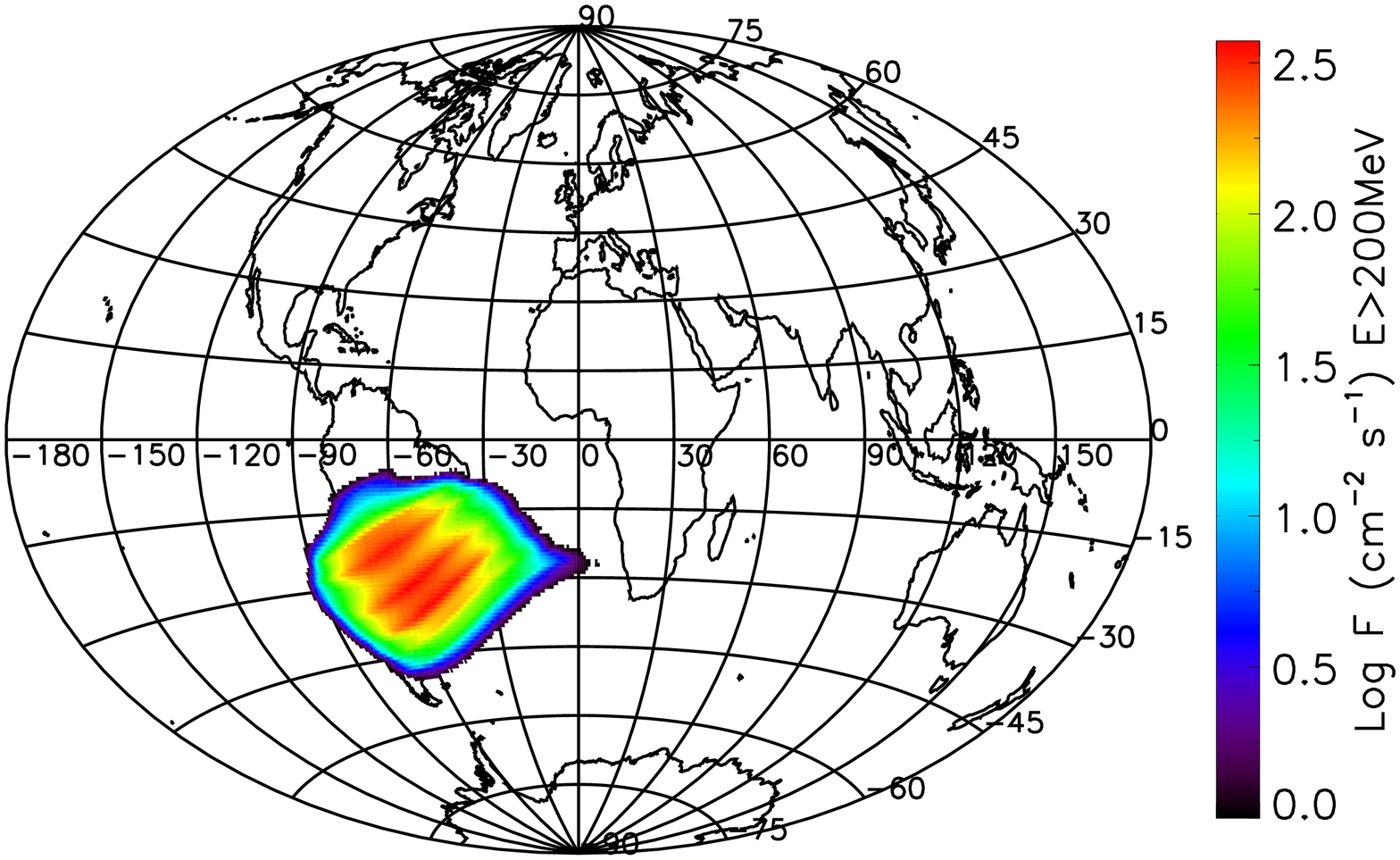}
\\
\includegraphics[width=0.335\linewidth]{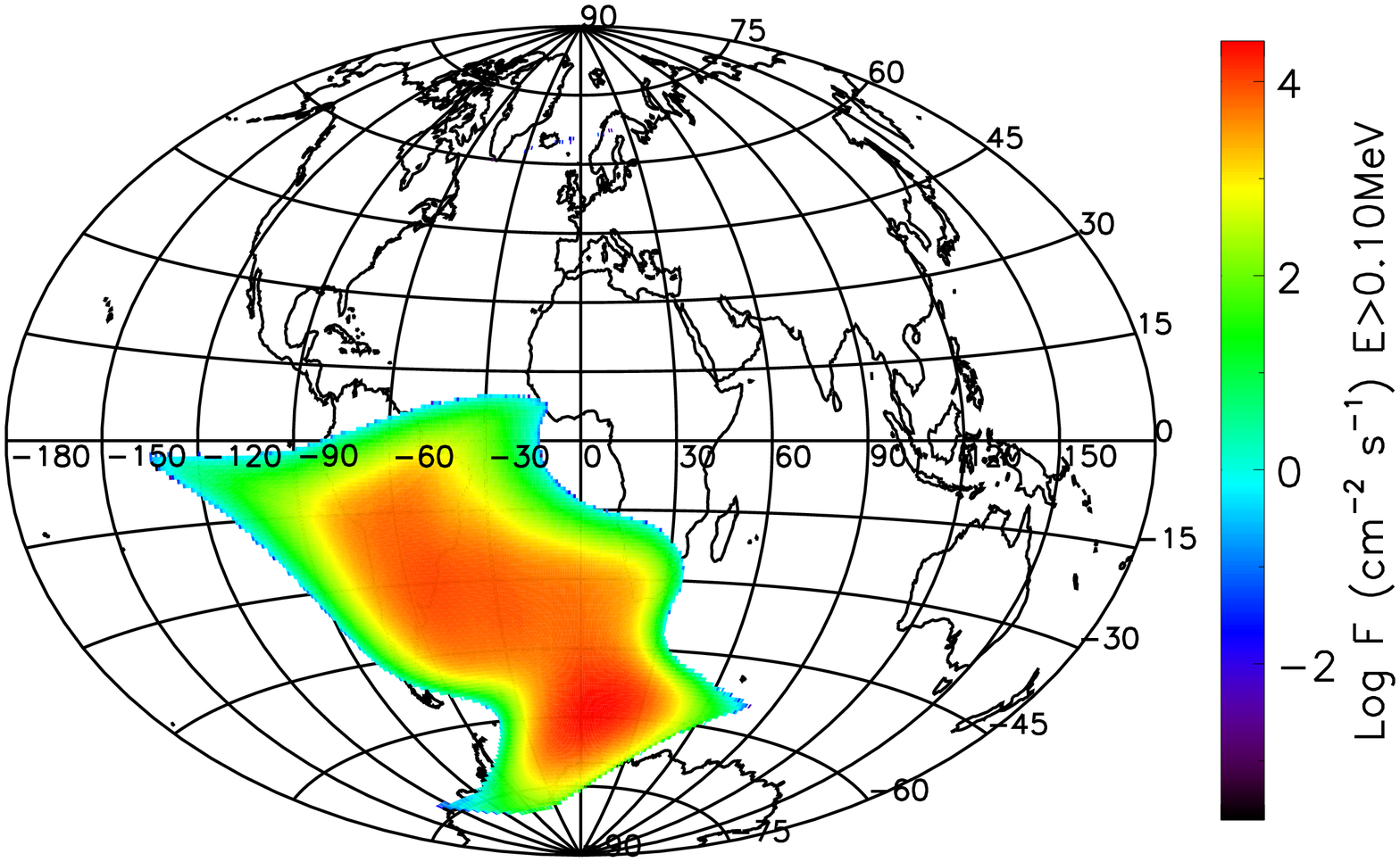}&
\includegraphics[width=0.335\linewidth]{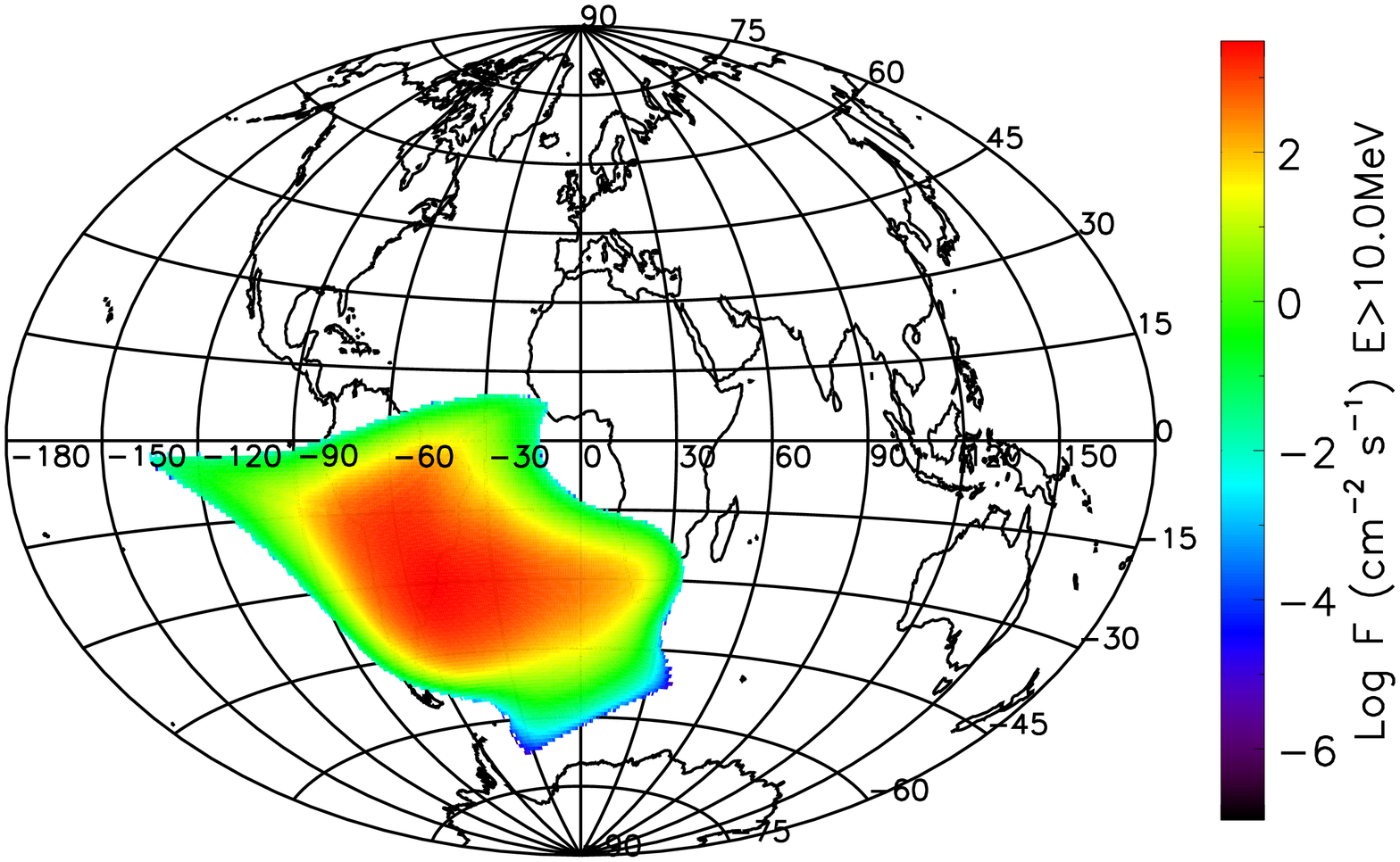}&
\includegraphics[width=0.335\linewidth]{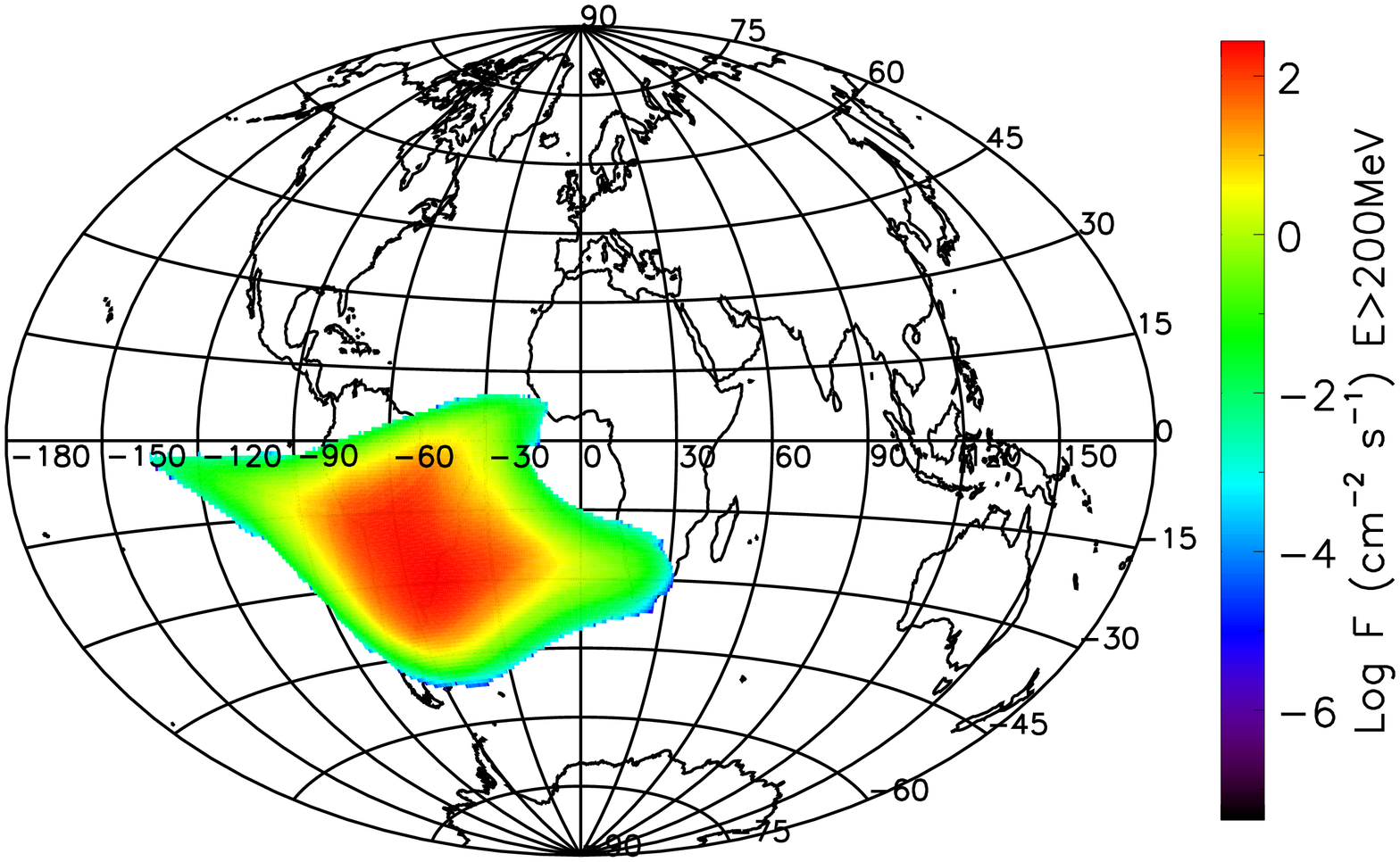}
\end{tabular}
\endgroup
\end{center}
\caption{\label{fig:maps_p}
Integral flux maps of trapped protons for AP8 MIN model (top panels) compared with the flux maps for AP9 50\,\% CL model (bottom panels) at 550\,km altitude with low energy thresholds of 0.1\,MeV, 10\,MeV and 200\,MeV, respectively from left to right.}
\end{figure}

Maps of the radiation regions were computed using the trapped particle models. Figure~\ref{fig:maps_e} shows maps of the integral fluxes of trapped electrons for AE8 MAX model compared with the flux maps for AE9 50\,\% CL model for various low energy thresholds at 550\,km altitude. Figure~\ref{fig:maps_p} shows maps of the integral flux maps of trapped protons for AP8 MIN model compared with the flux maps for AP9 50\,\% CL model at the same altitude.

\begin{figure}[p]
\begin{center}
\begingroup
\setlength{\tabcolsep}{0pt} 
\begin{tabular}{ccc}
\includegraphics[width=0.333\linewidth]{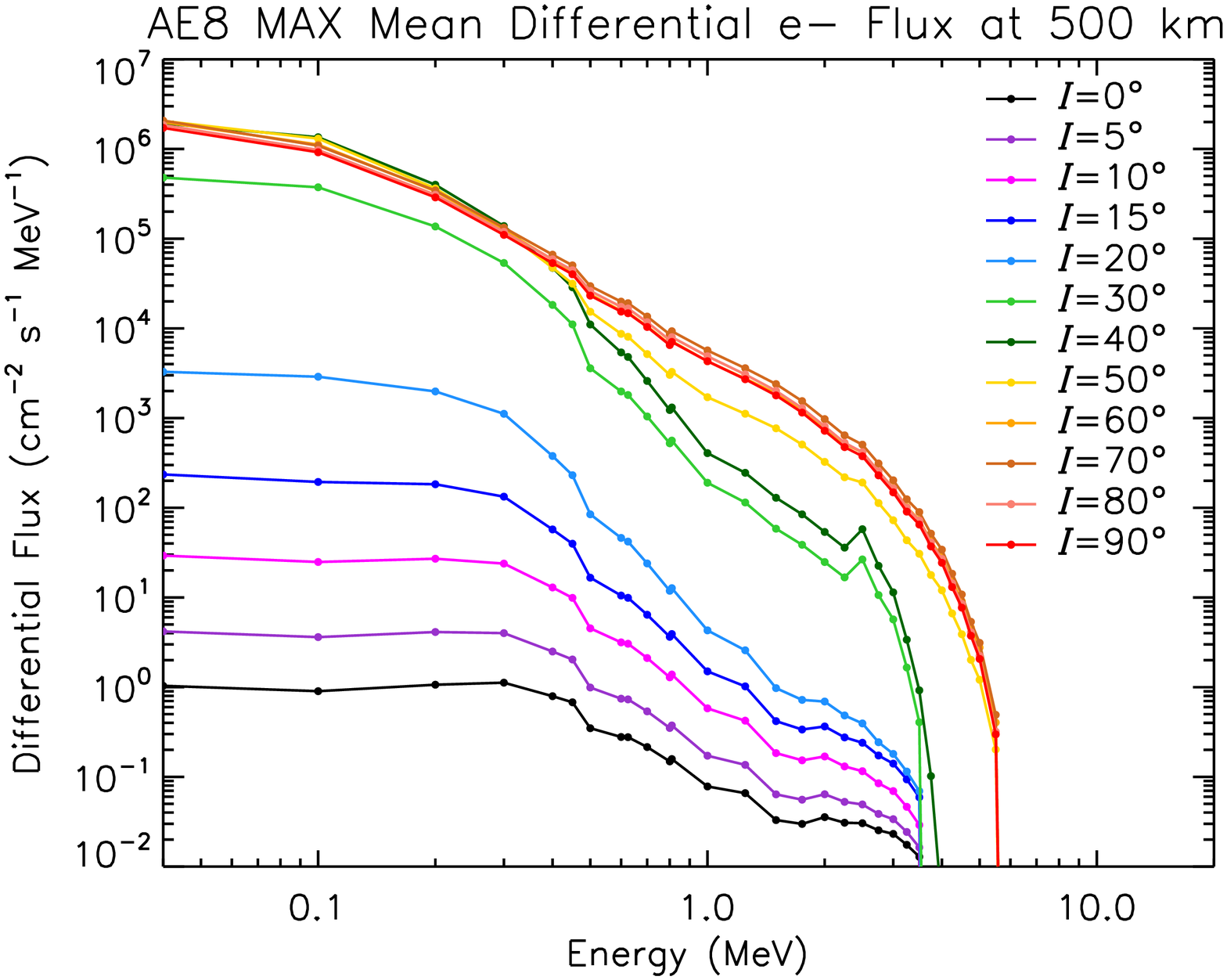}&
\includegraphics[width=0.333\linewidth]{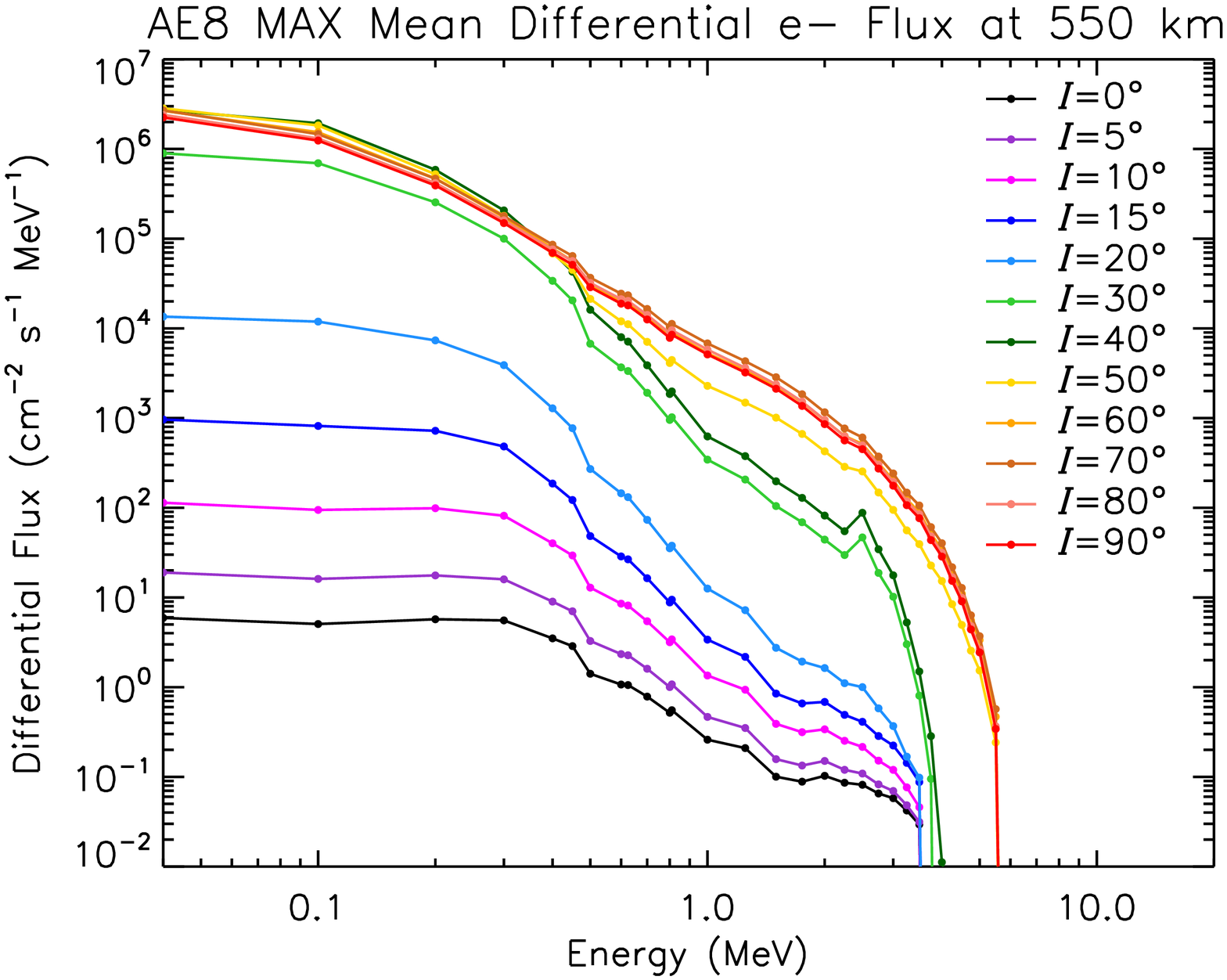}&
\includegraphics[width=0.333\linewidth]{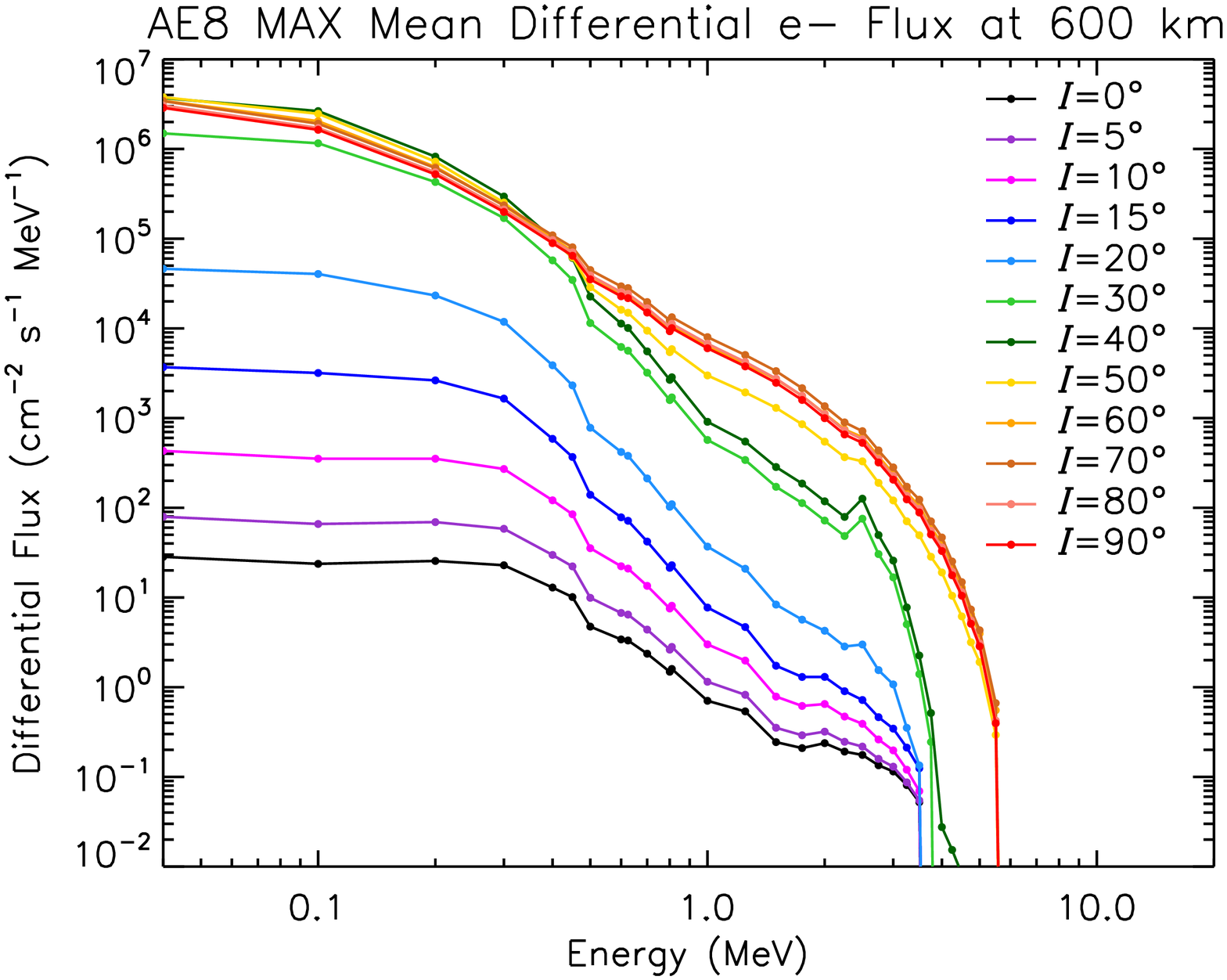}
\\
\includegraphics[width=0.333\linewidth]{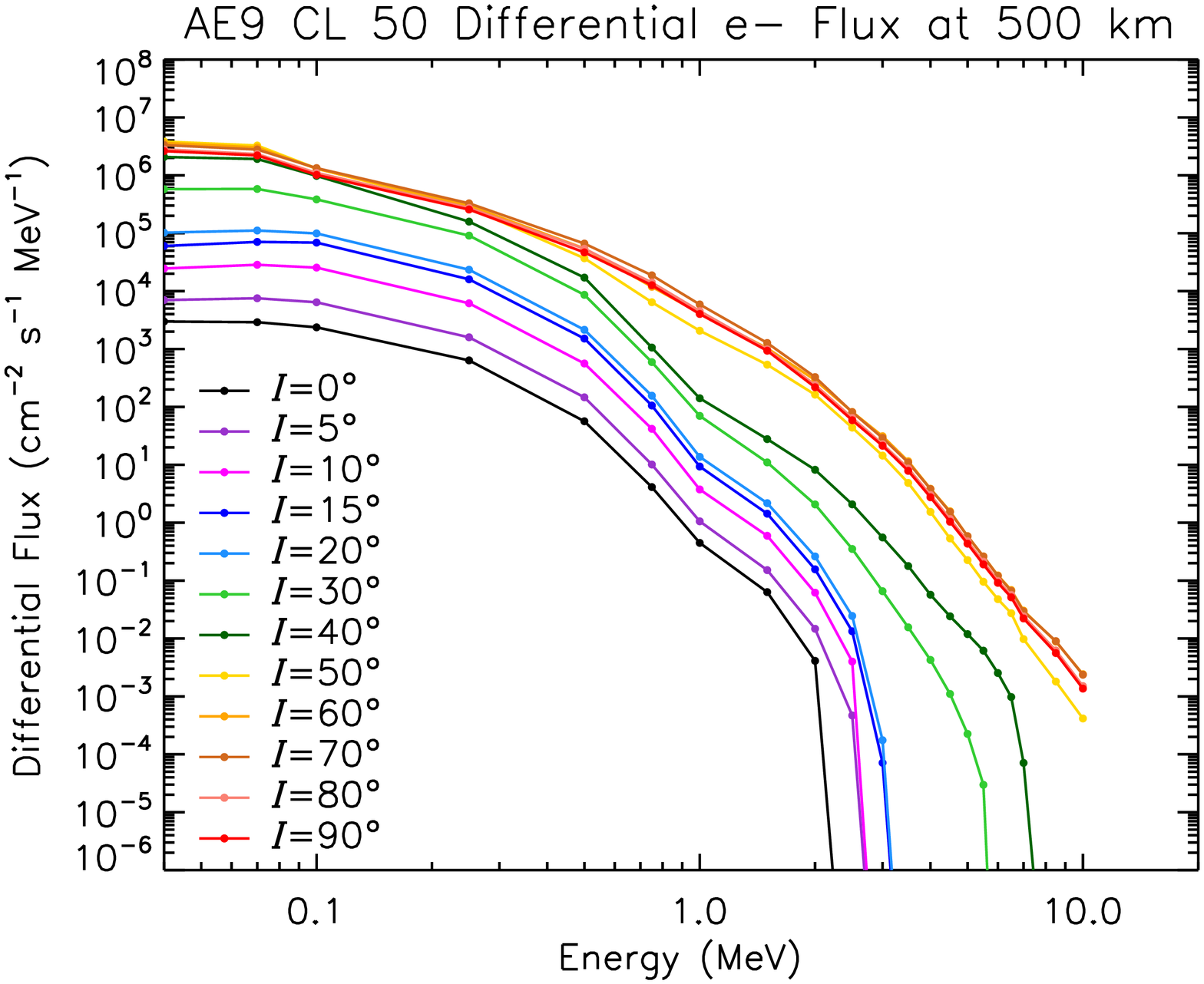}&
\includegraphics[width=0.333\linewidth]{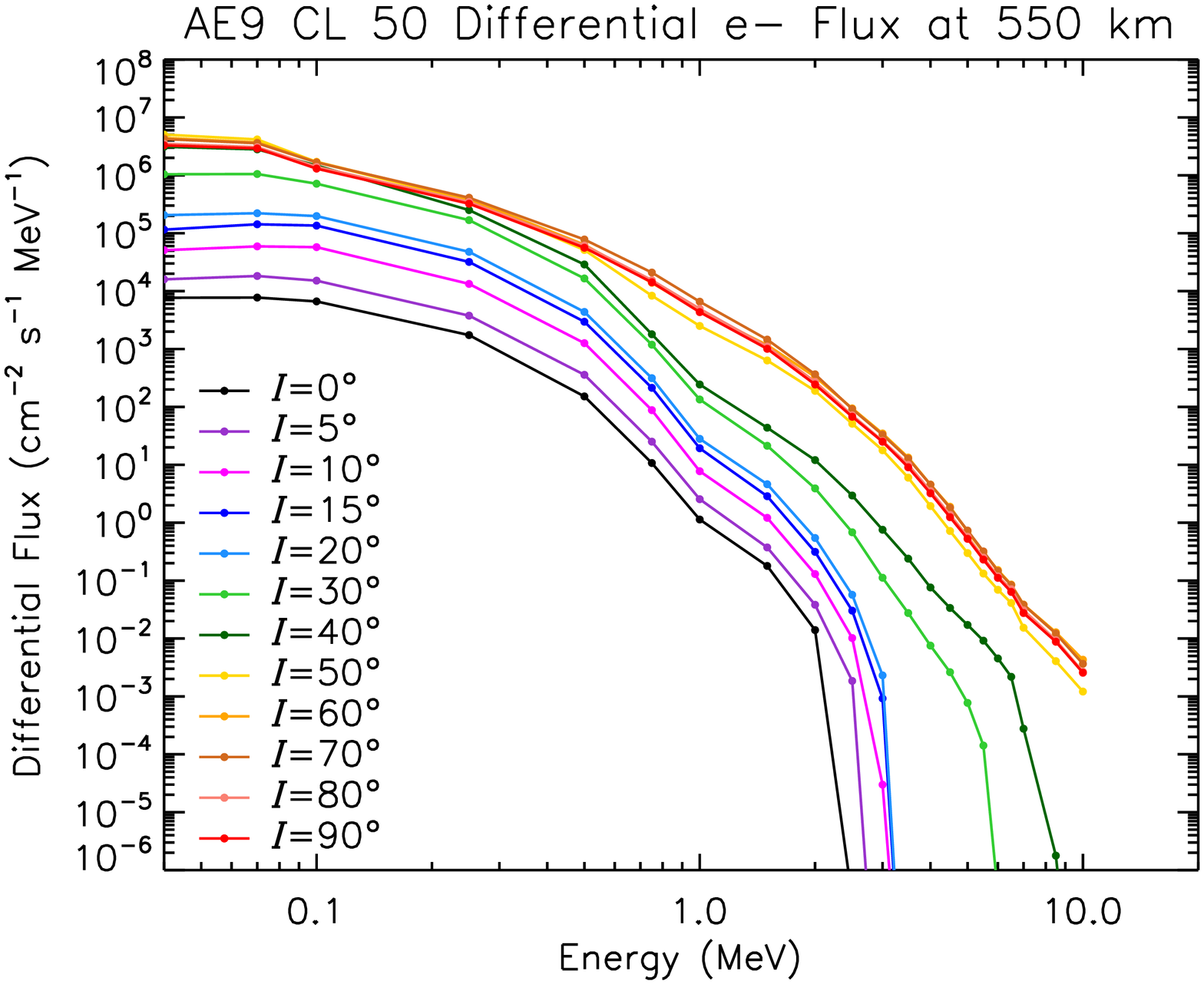}&
\includegraphics[width=0.333\linewidth]{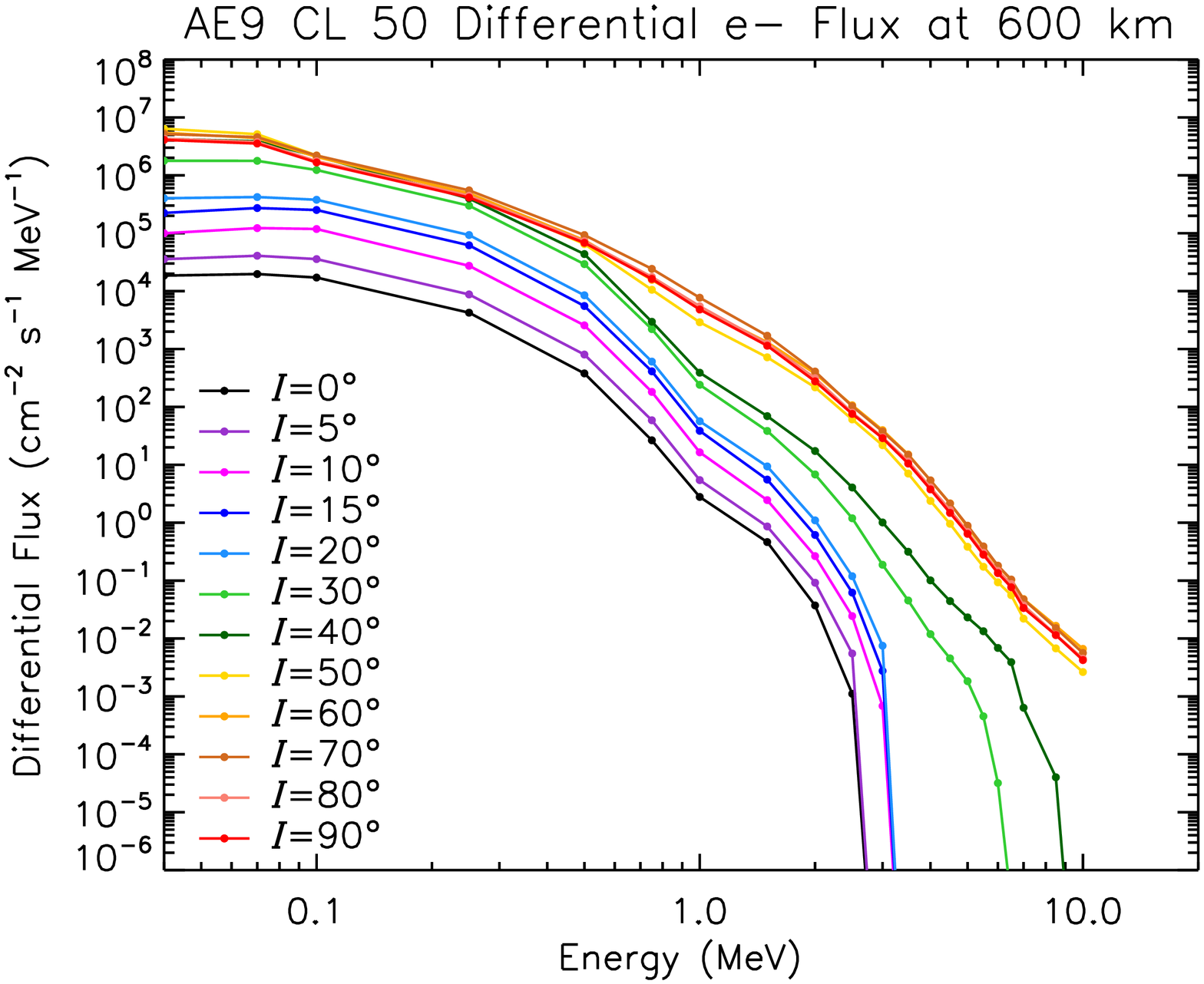}
\end{tabular}
\endgroup
\end{center}
\caption{\label{fig:diff_spectra_e}
Orbit averaged differential fluxes of trapped electrons for mean AE8 MAX (top panels) and AE9 50\,\% CL (bottom panels) models computed for altitudes 500, 550 and 600\,km altitudes, respectively, from left to right and for different orbital inclinations $I$.}
\end{figure}

\begin{figure}[p]
\begin{center}
\begingroup
\setlength{\tabcolsep}{0pt} 
\begin{tabular}{ccc}
\includegraphics[width=0.333\linewidth]{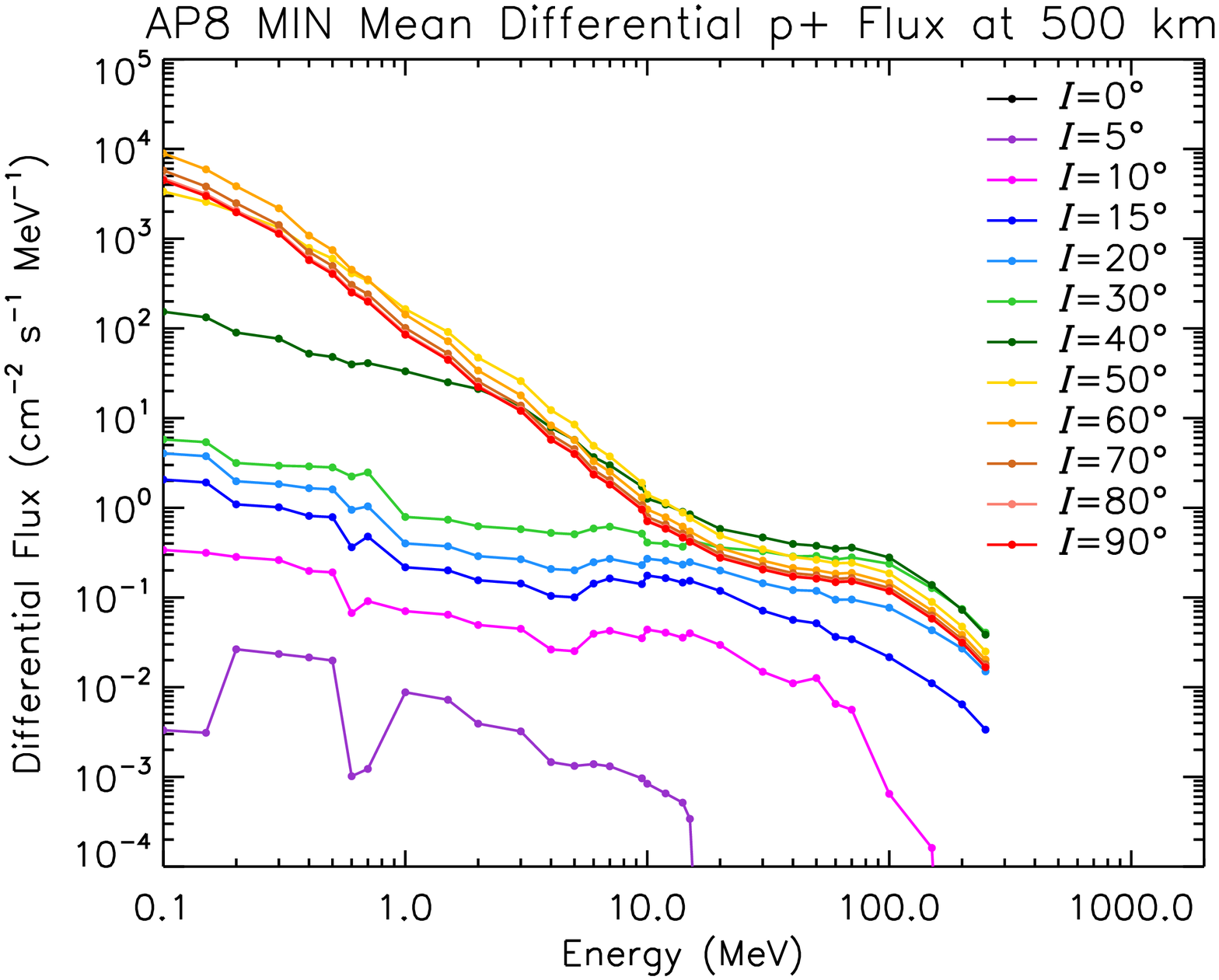}&
\includegraphics[width=0.333\linewidth]{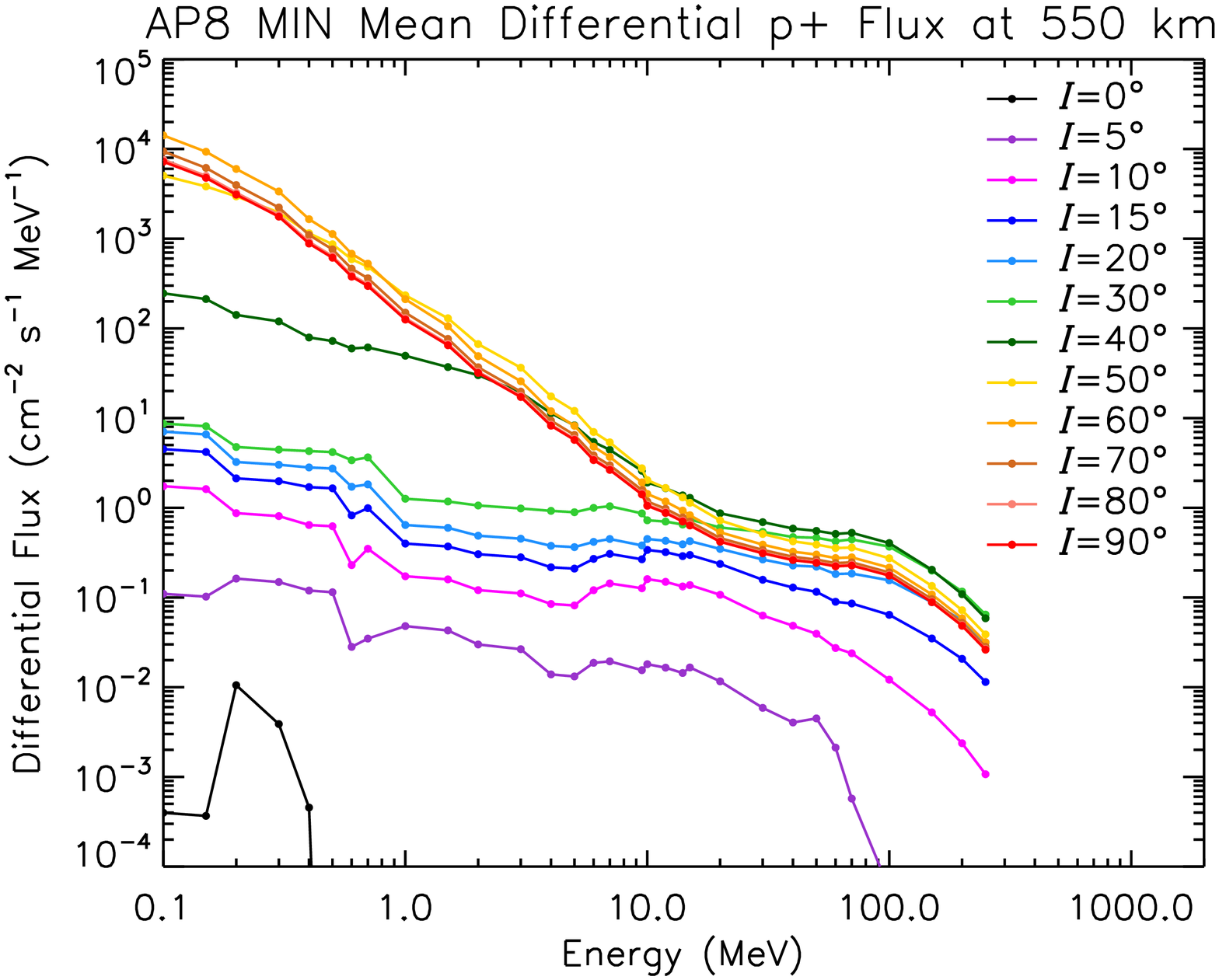}&
\includegraphics[width=0.333\linewidth]{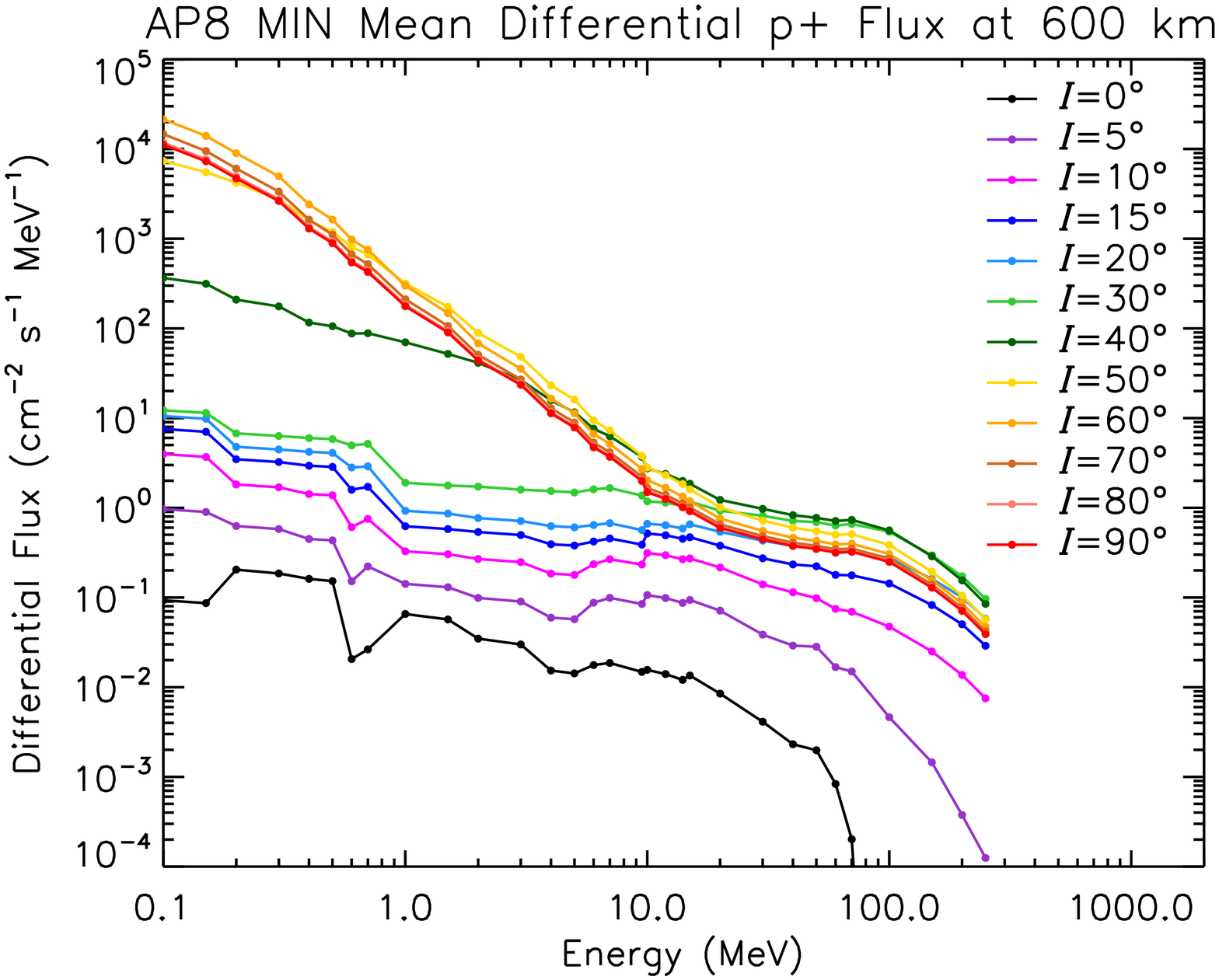}
\\
\includegraphics[width=0.333\linewidth]{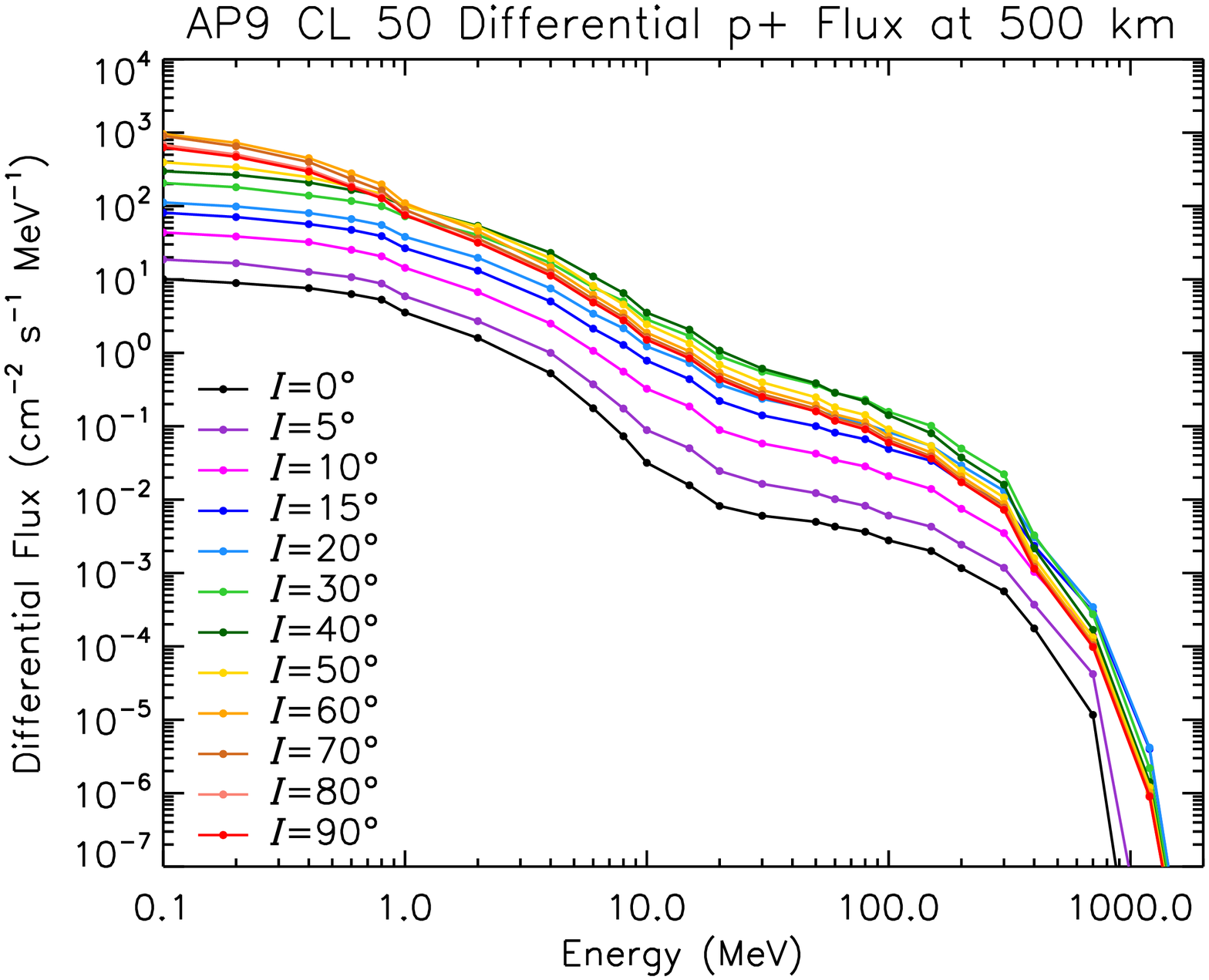}&
\includegraphics[width=0.333\linewidth]{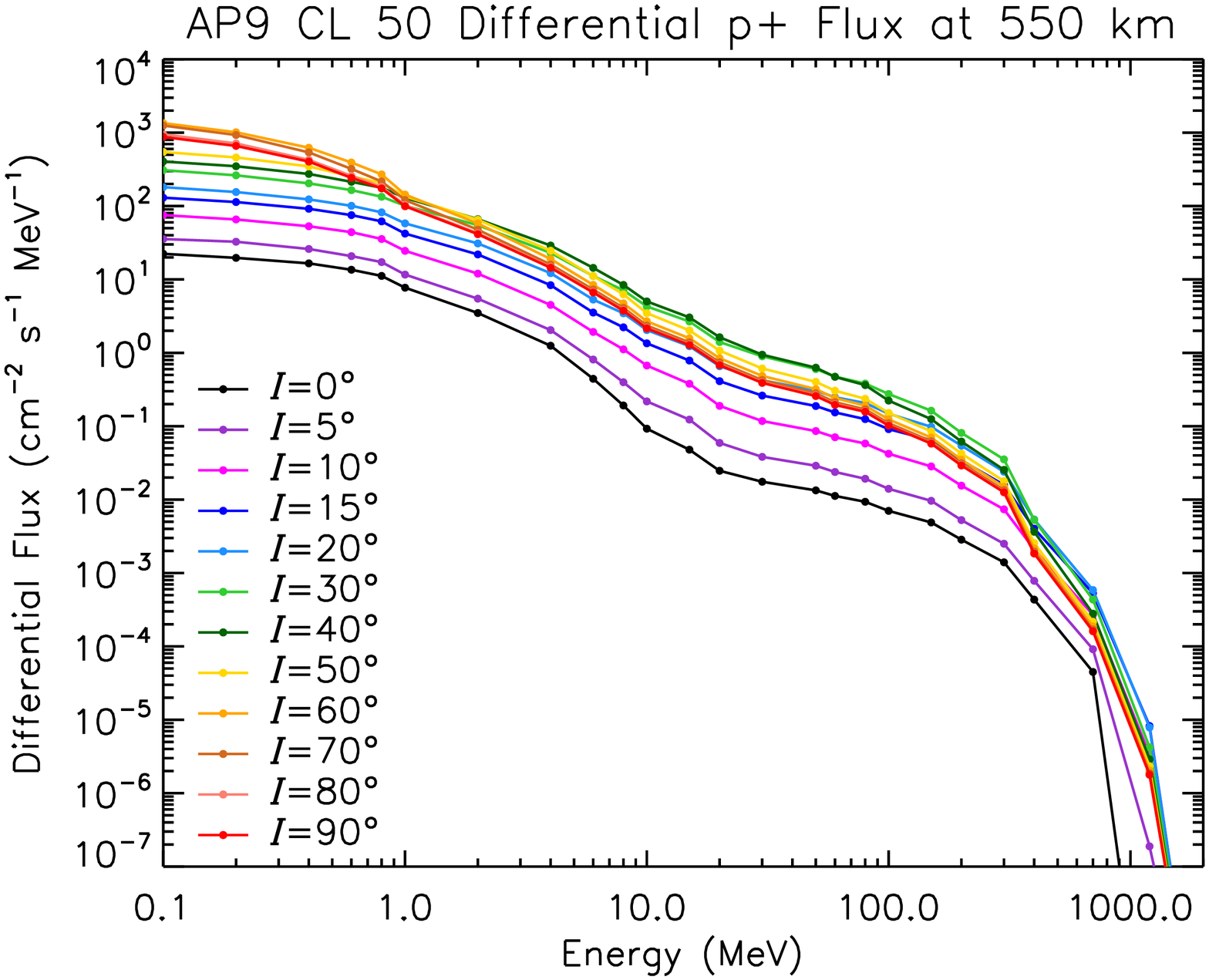}&
\includegraphics[width=0.333\linewidth]{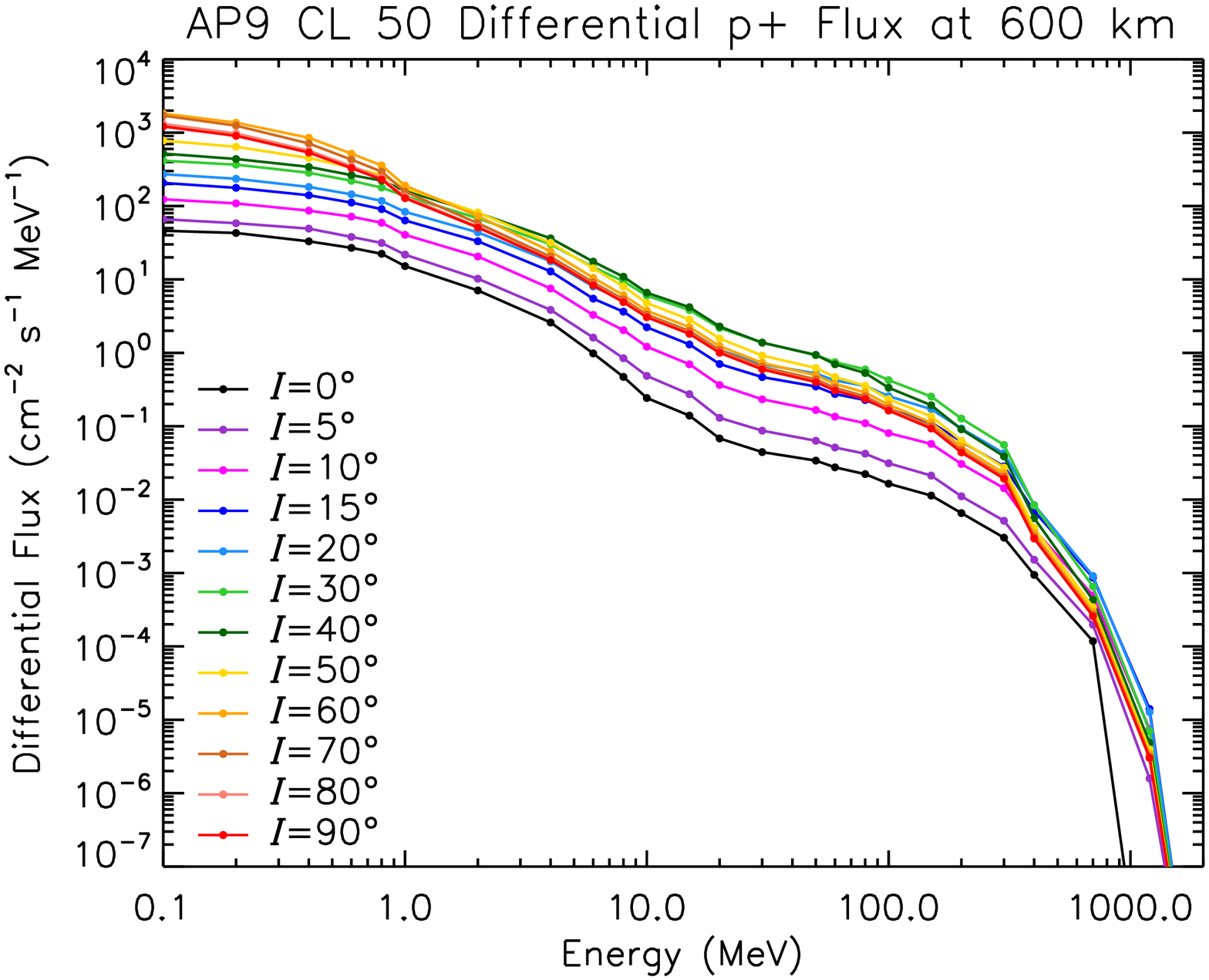}
\end{tabular}
\endgroup
\end{center}
\caption{\label{fig:diff_spectra_p} 
Orbit averaged differential fluxes of trapped protons for mean AP8 MIN (top panels) and AP9 50\,\% CL (bottom panels) models computed for altitudes 500, 550 and 600\,km altitudes, respectively, from left to right and for different orbital inclinations $I$.}
\end{figure}

\section{ORBIT AVERAGED SPECTRA}
\label{sec:spectra}

The mean fluxes averaged along the whole 60-days long trajectory have been calculated for three different altitudes (500, 550 and 600\,km). Figure~\ref{fig:diff_spectra_e} shows total-trajectory-averaged differential fluxes of trapped electrons for AE8 MAX and AE9 50\,\% CL models for different inclinations and comparing different altitudes. Similarly, Figure~\ref{fig:diff_spectra_p} shows averaged differential fluxes of trapped protons for AP8 MIN and AP9 50\,\% CL models.

\subsection{Comparison of Models}
\label{sec:compar_models}

\begin{figure}[p]
\begin{center}
\begingroup
\setlength{\tabcolsep}{0pt} 
\begin{tabular}{ccc}
\includegraphics[width=0.333\linewidth]{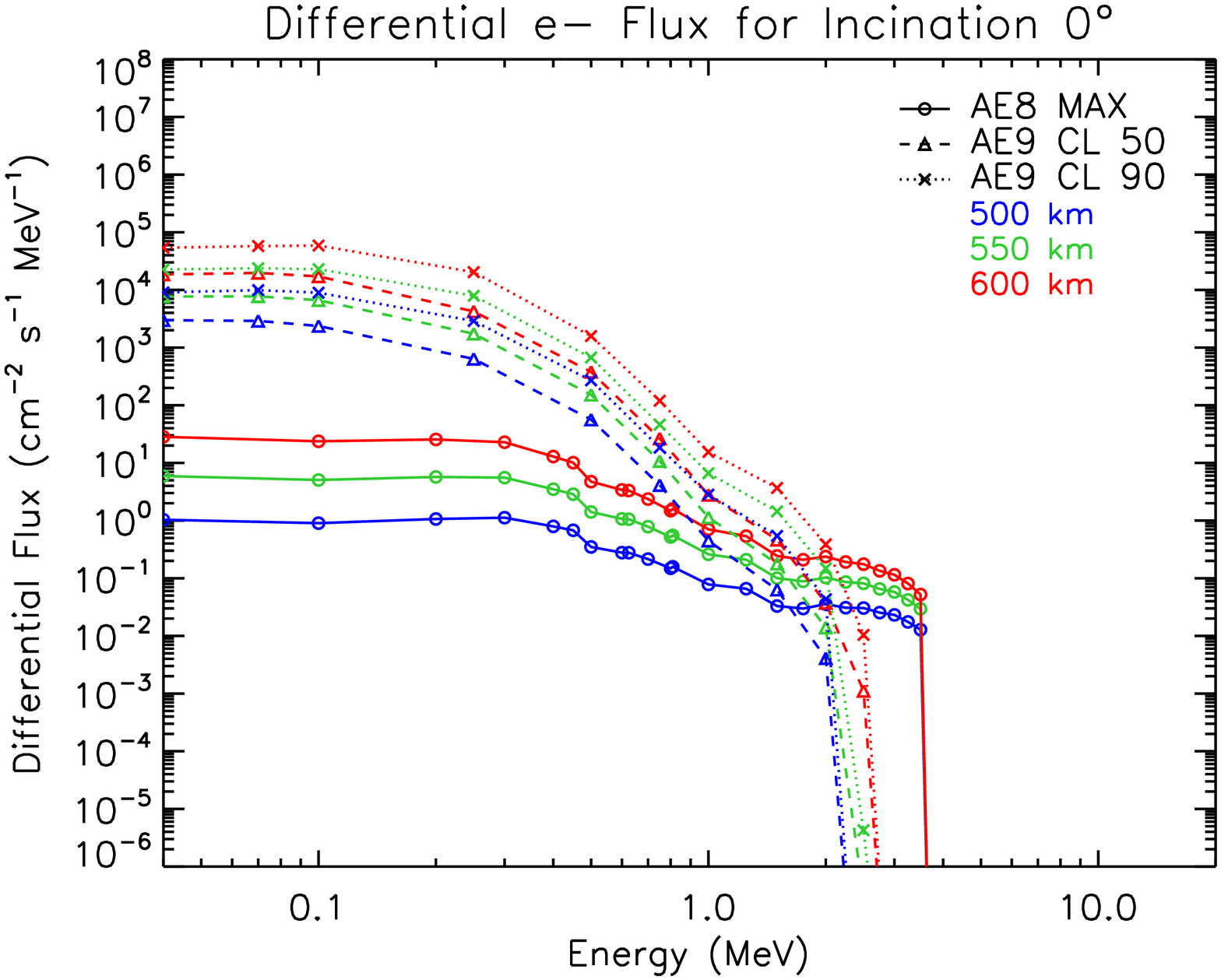}&
\includegraphics[width=0.333\linewidth]{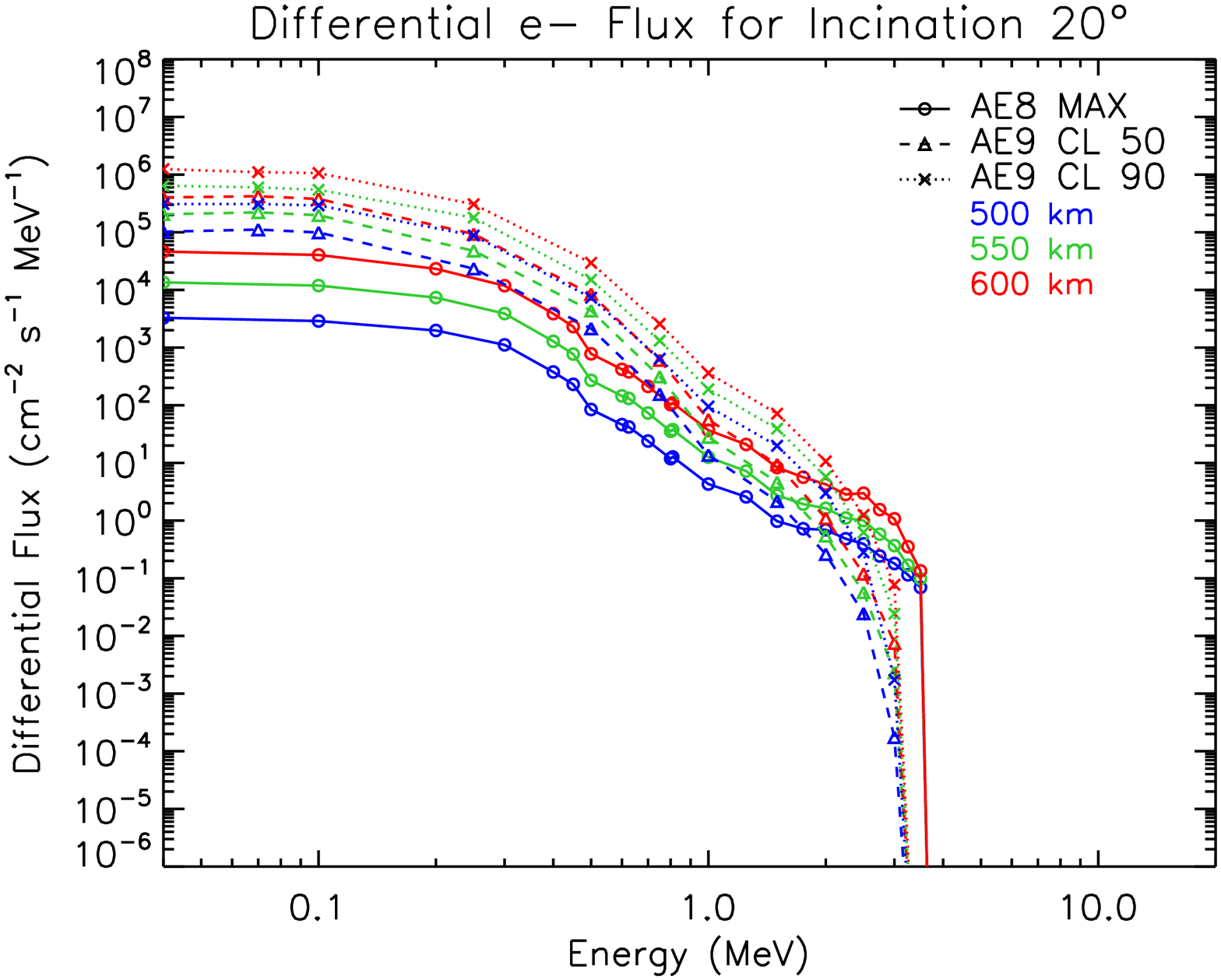}&
\includegraphics[width=0.333\linewidth]{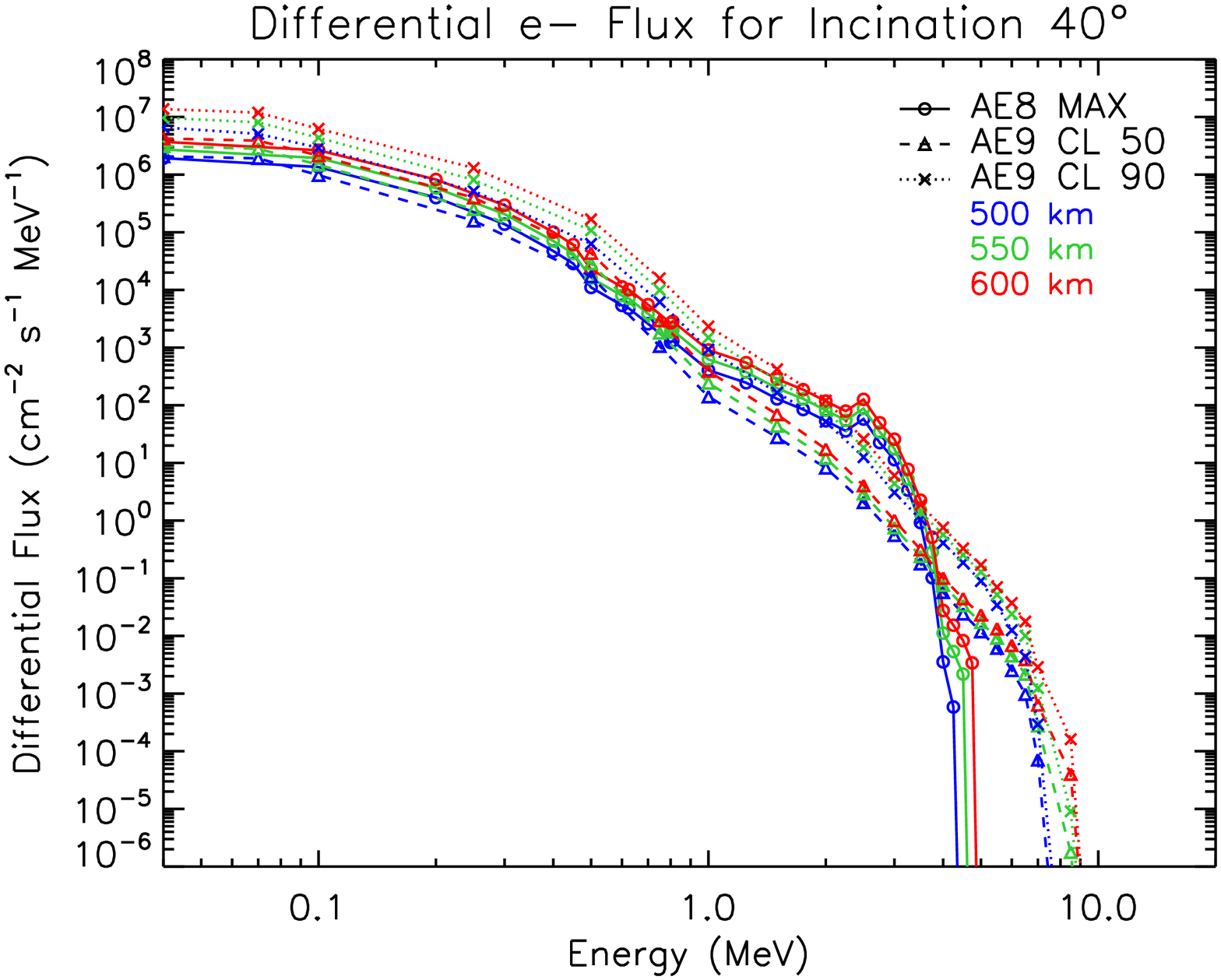}
\\
\includegraphics[width=0.333\linewidth]{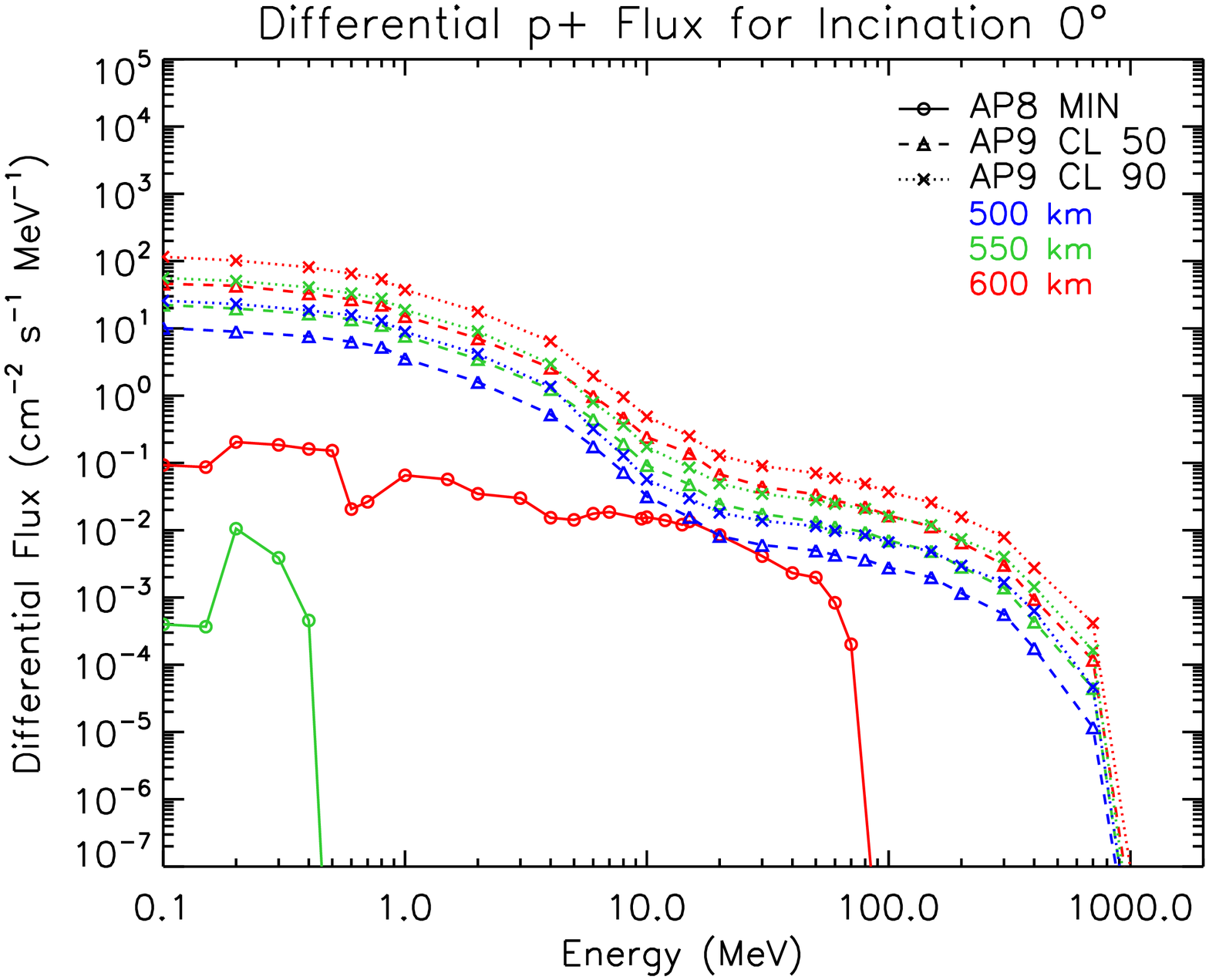}&
\includegraphics[width=0.333\linewidth]{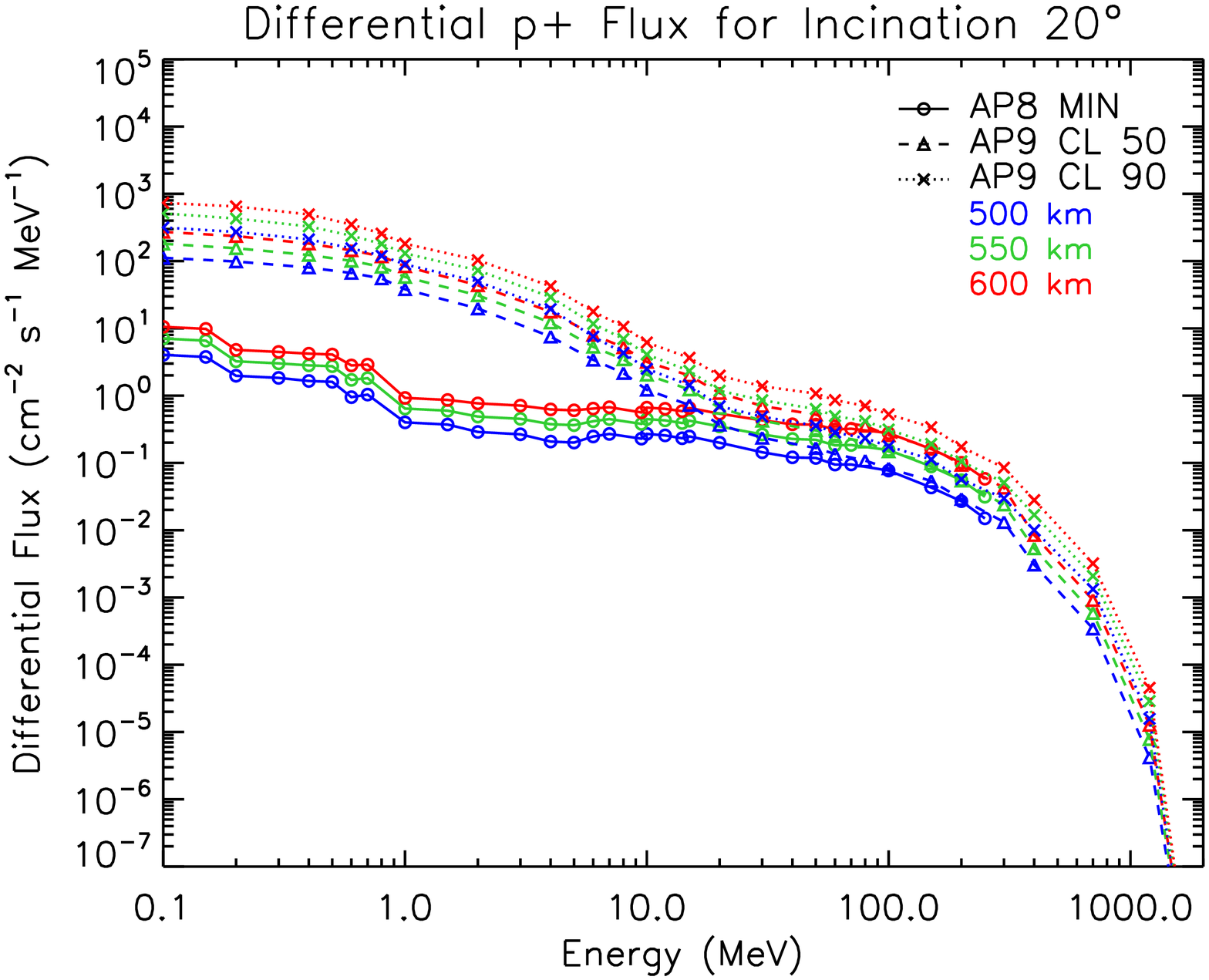}&
\includegraphics[width=0.333\linewidth]{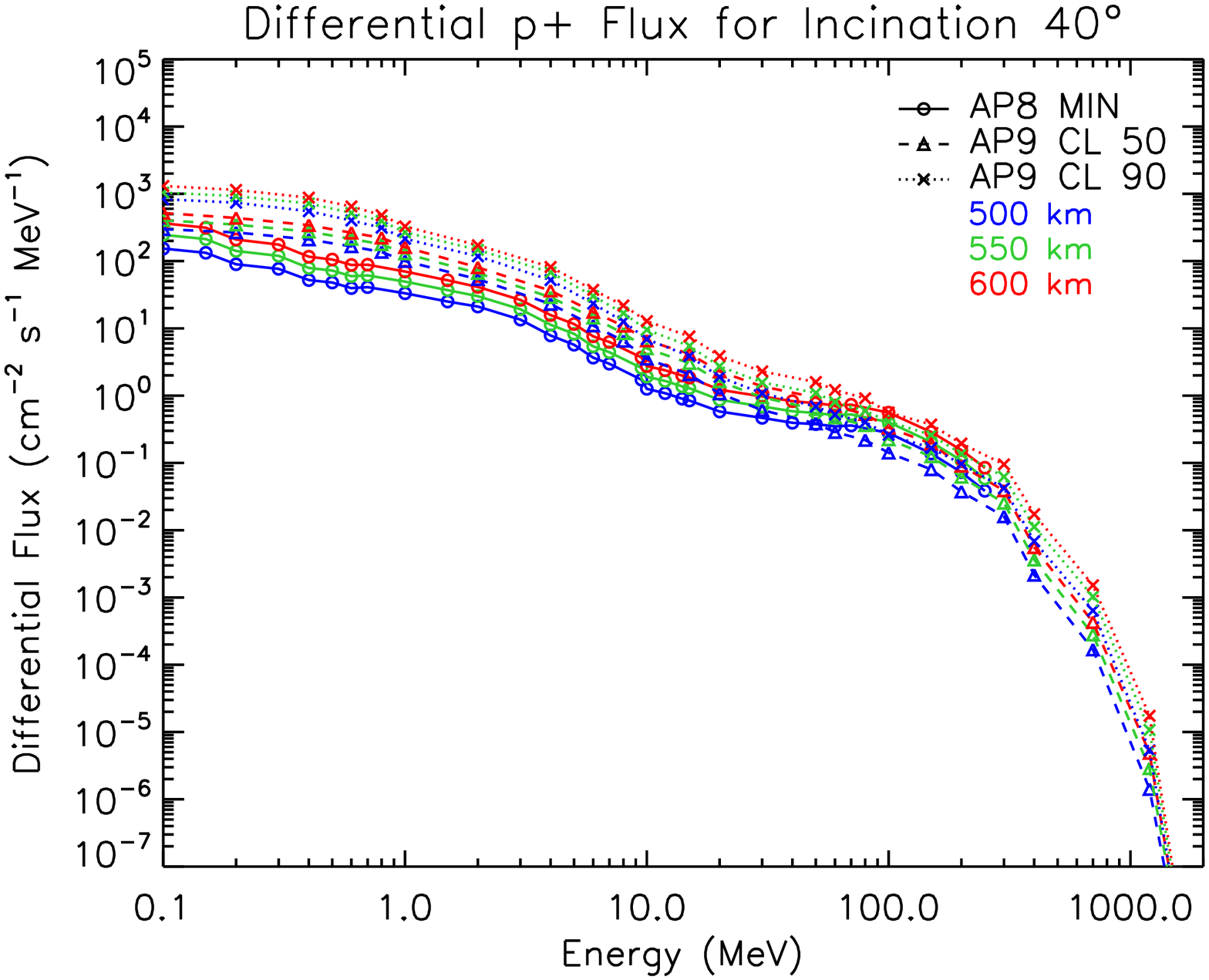}
\end{tabular}
\endgroup
\end{center}
\caption{\label{fig:compar_diff_spectra} 
Comparison of mean differential fluxes of trapped electrons (top panels) and protons (bottom panels) for AE8 MAX and AP8 MIN and 50\,\% CL fluxes for AE9 and AP9 models.}
\end{figure}

\begin{figure}[p]
\begin{center}
\begingroup
\setlength{\tabcolsep}{0pt} 
\begin{tabular}{ccc}
\includegraphics[width=0.333\linewidth]{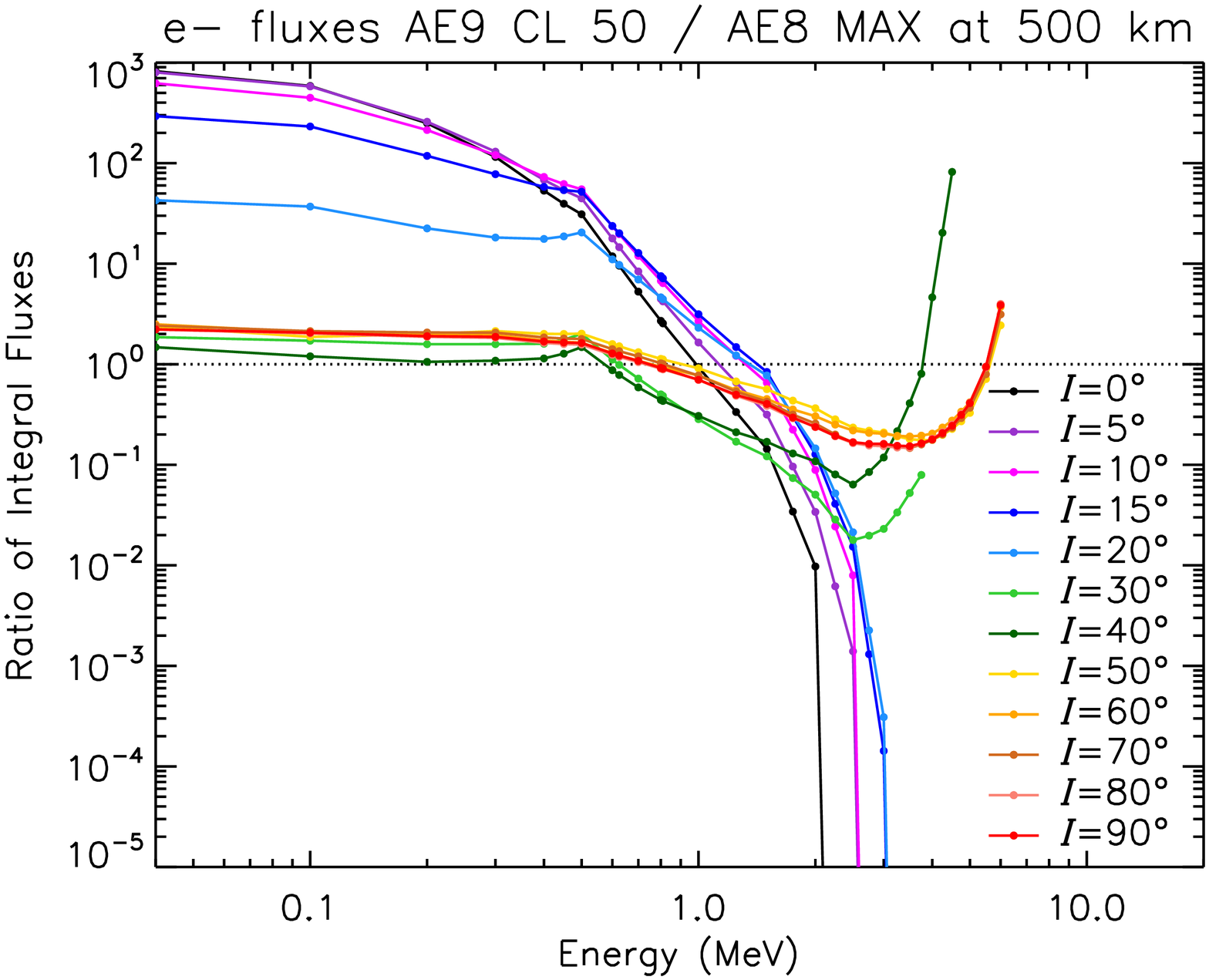}&
\includegraphics[width=0.333\linewidth]{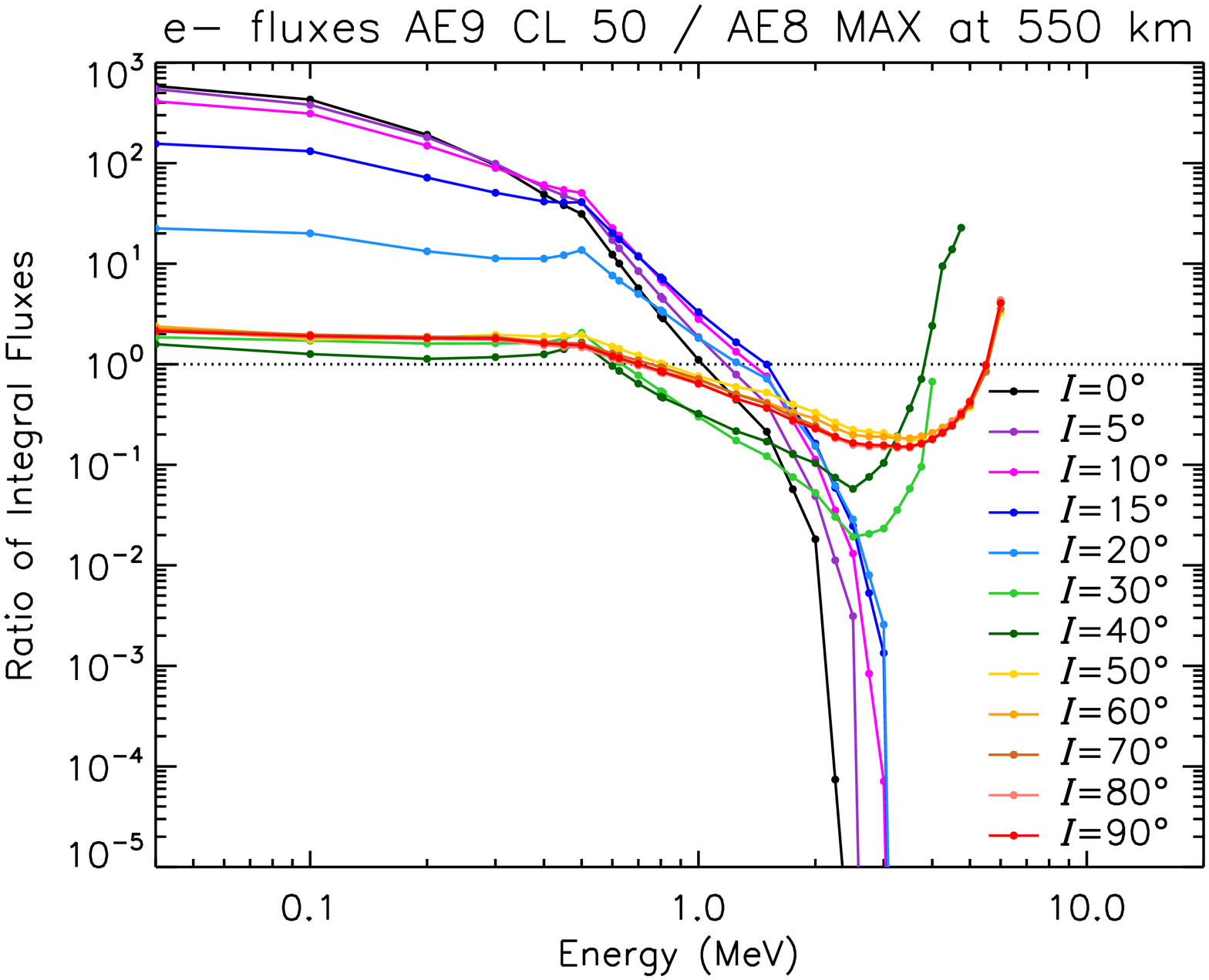}&
\includegraphics[width=0.333\linewidth]{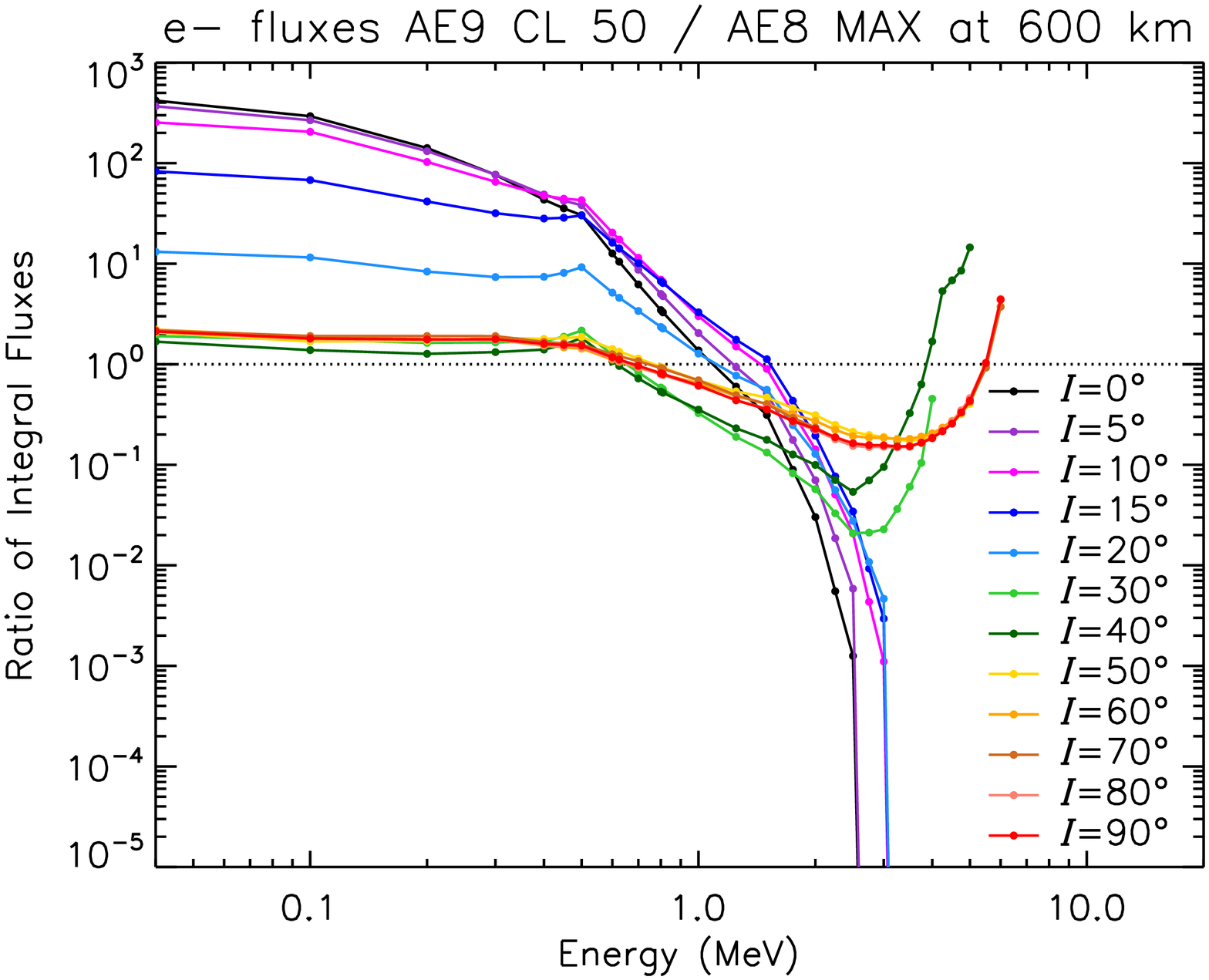}
\\
\includegraphics[width=0.333\linewidth]{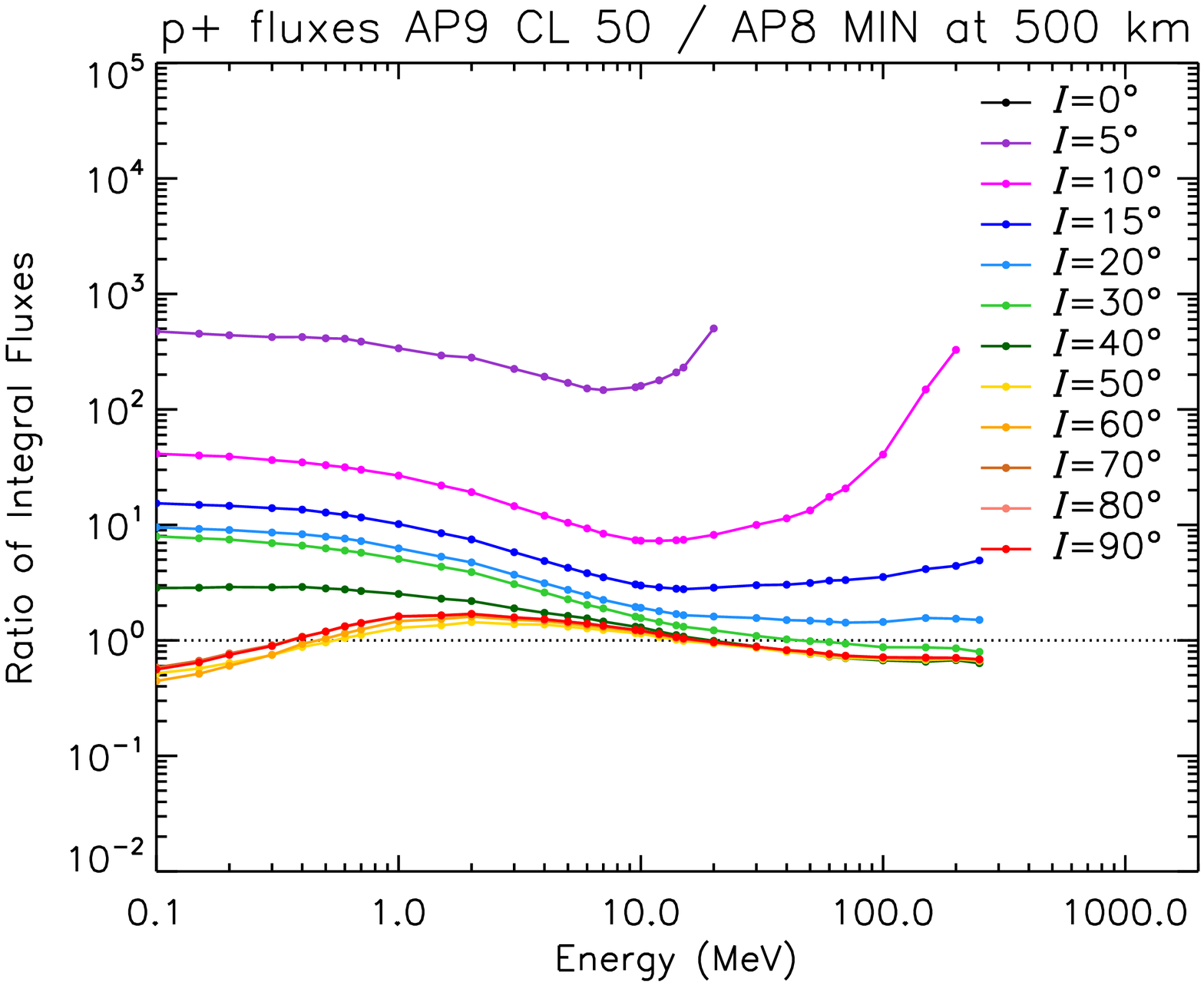}&
\includegraphics[width=0.333\linewidth]{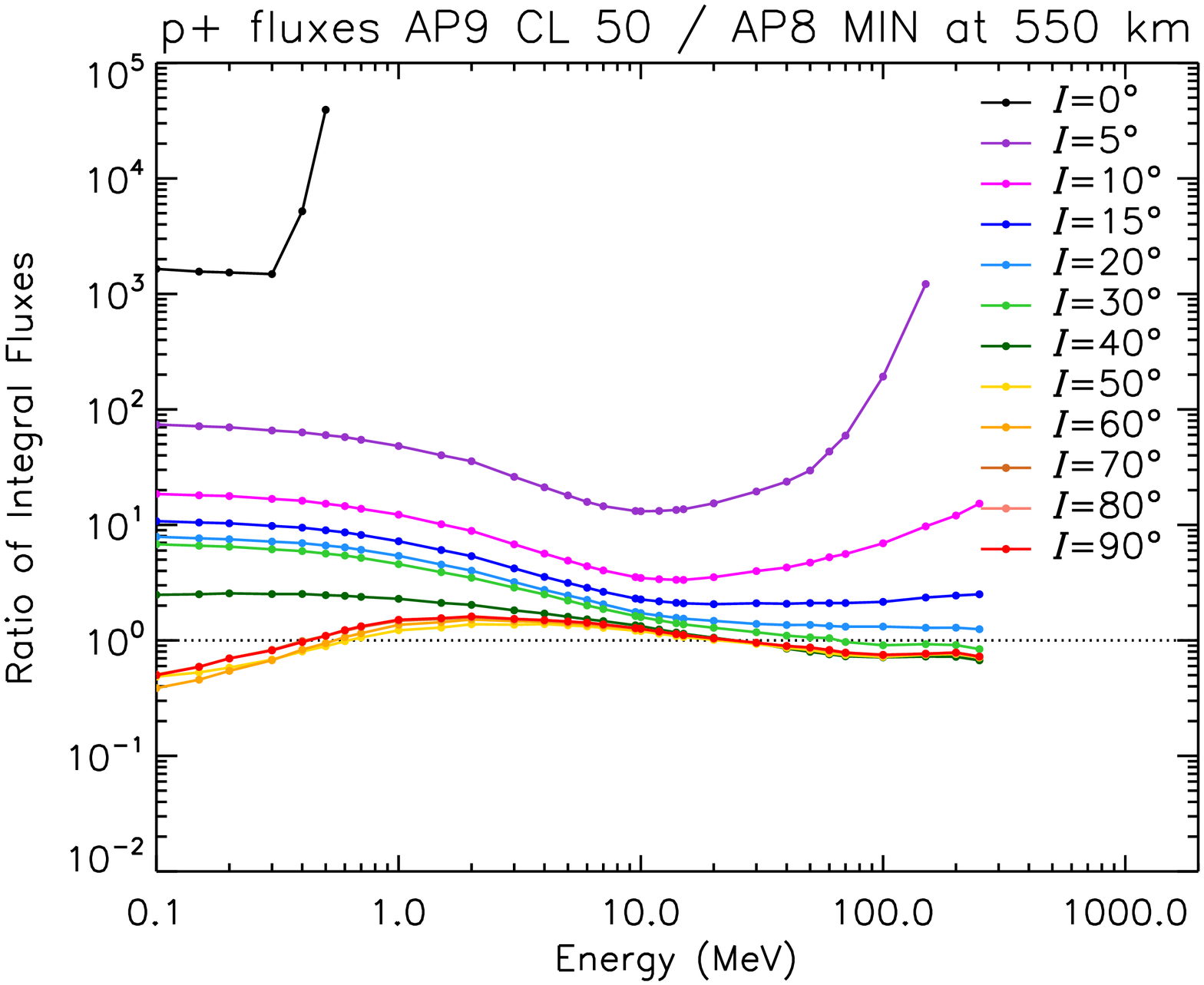}&
\includegraphics[width=0.333\linewidth]{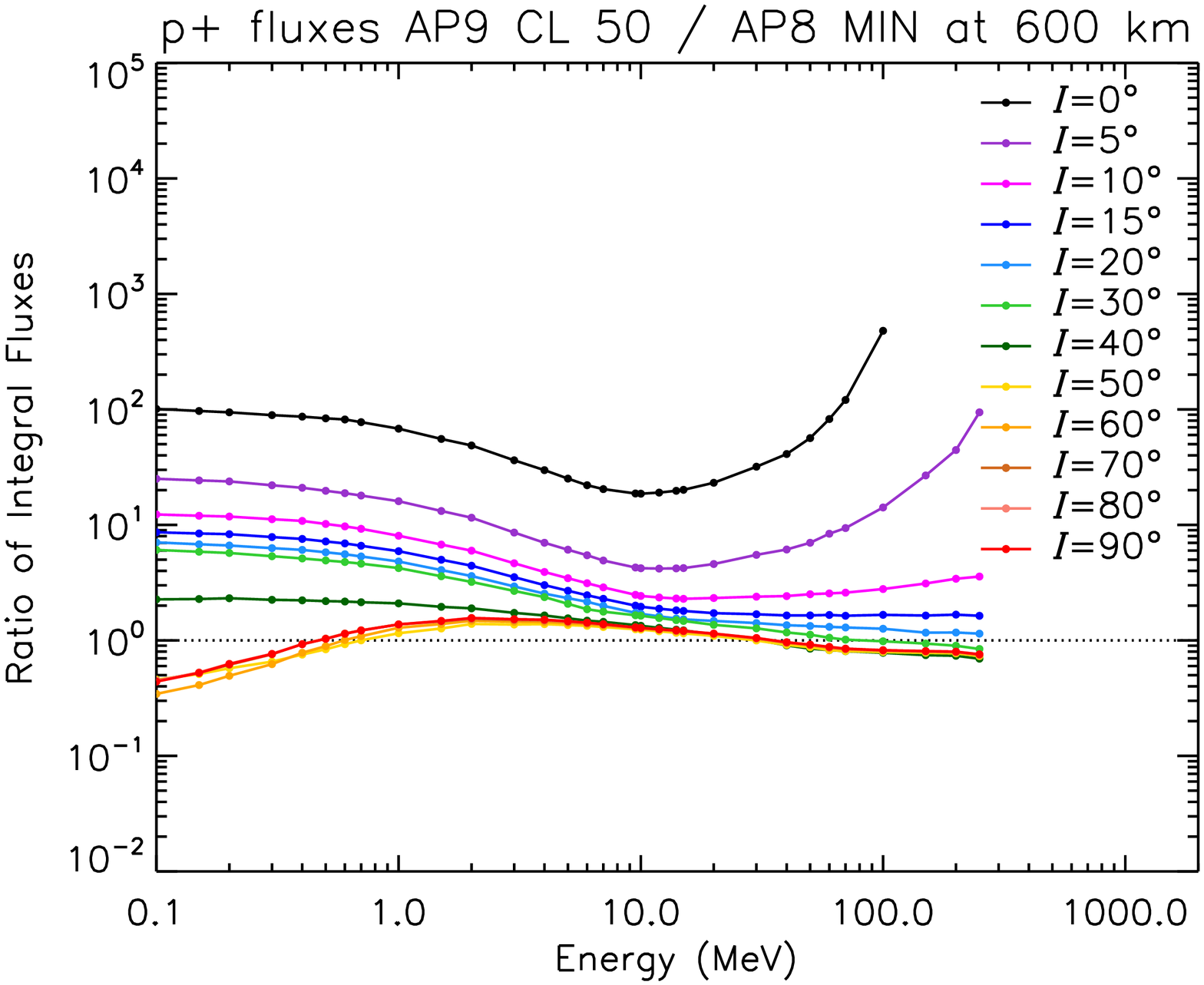}
\end{tabular}
\endgroup
\end{center}
\caption{\label{fig:compar_diff_spectra_ratio} 
Ratio of integral fluxes of trapped electrons (AE9 50\,\% CL / AE8MAX) in the top panels and protons (AP9 50\,\% CL / AP8MIN) in the bottom panels for altitudes 500, 550 and 600\,km, respectively from left to right, and for different inclinations $I$.}
\end{figure}

Figure~\ref{fig:compar_diff_spectra} compares the differential fluxes of trapped electrons for AE8 MAX, AE9 50\,\% CL and 90\,\% CL models and fluxes of trapped protons for AP8 MIN, AP9 50\,\% CL and 90\,\% CL models. Figure~\ref{fig:compar_diff_spectra_ratio} shows the ratio of the integral fluxes given by models AE9 50\,\% CL / AE8 MAX of trapped electrons and the ratio of the integral fluxes given by models AP9 50\,\% CL / AP8 MIN of trapped protons for different altitudes and inclinations.

The standard Ax8 models and the recent IRENE Ax9 models give dramatically different trapped particle fluxes, especially for low inclinations and low energies.

\section{DUTY CYCLE}
\label{sec:duty_cycle}

We define the \emph{duty cycle} as the fraction of time a satellite spends in a region with particle flux lower than a given flux threshold. From simulations of different circular orbits with different altitudes and inclinations and from the integral fluxes given by the trapped particle models, we computed the duty cycles for various particle flux thresholds and low-energy thresholds.

\begin{figure}[p]
\begin{center}
\begin{tabular}{cc}
\includegraphics[width=0.333\linewidth]{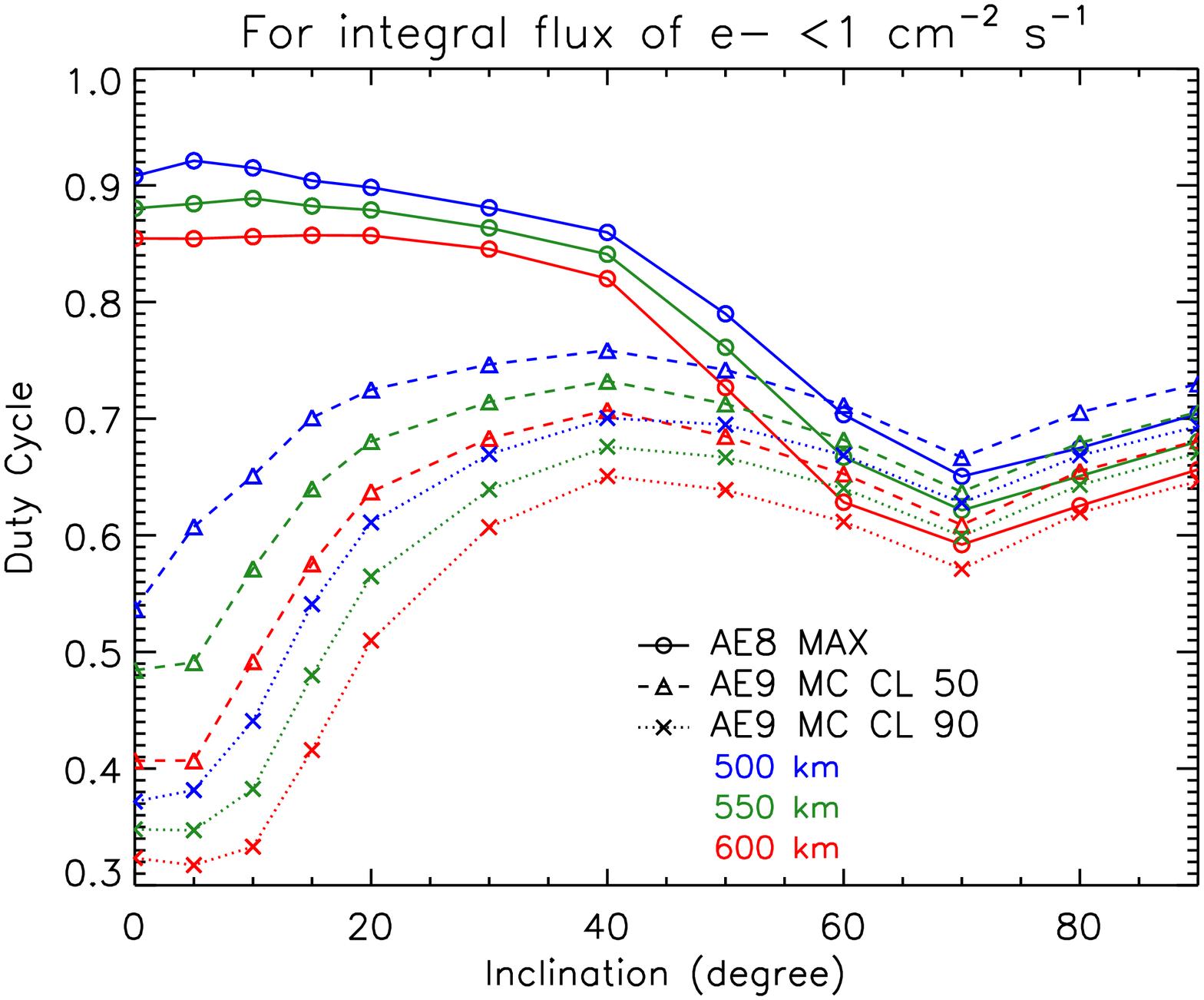}&
\includegraphics[width=0.333\linewidth]{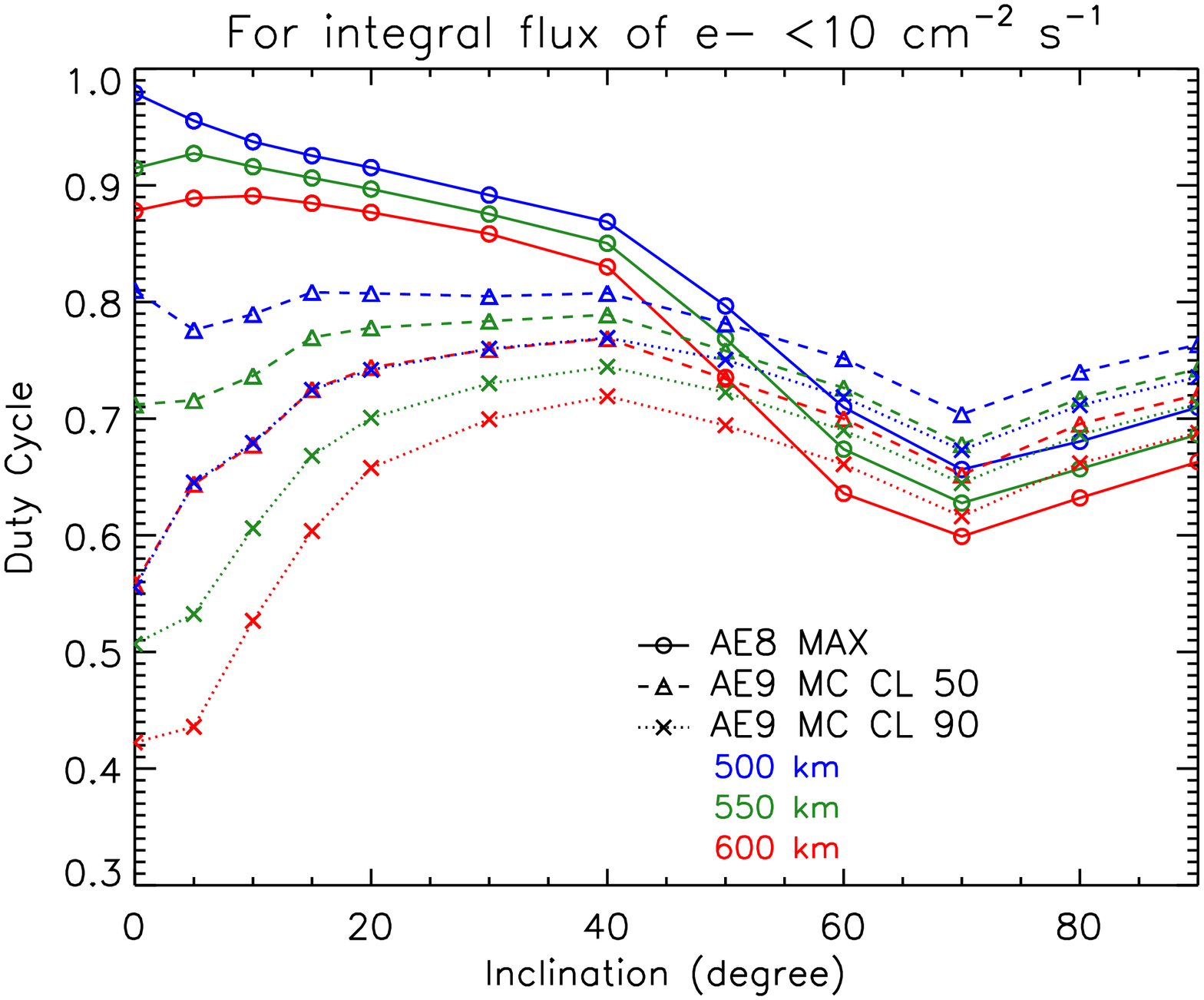}
\\
\includegraphics[width=0.333\linewidth]{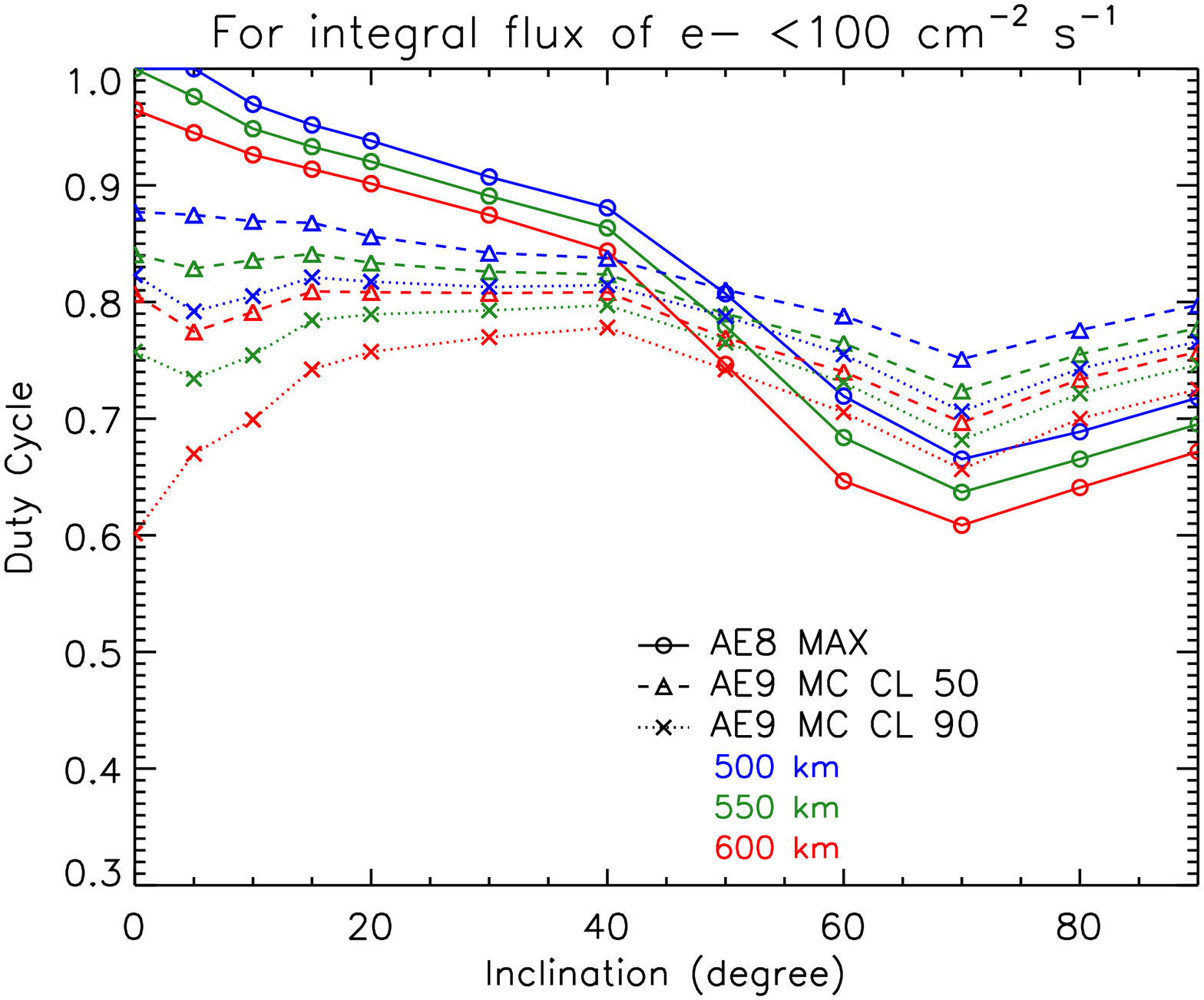}&
\includegraphics[width=0.333\linewidth]{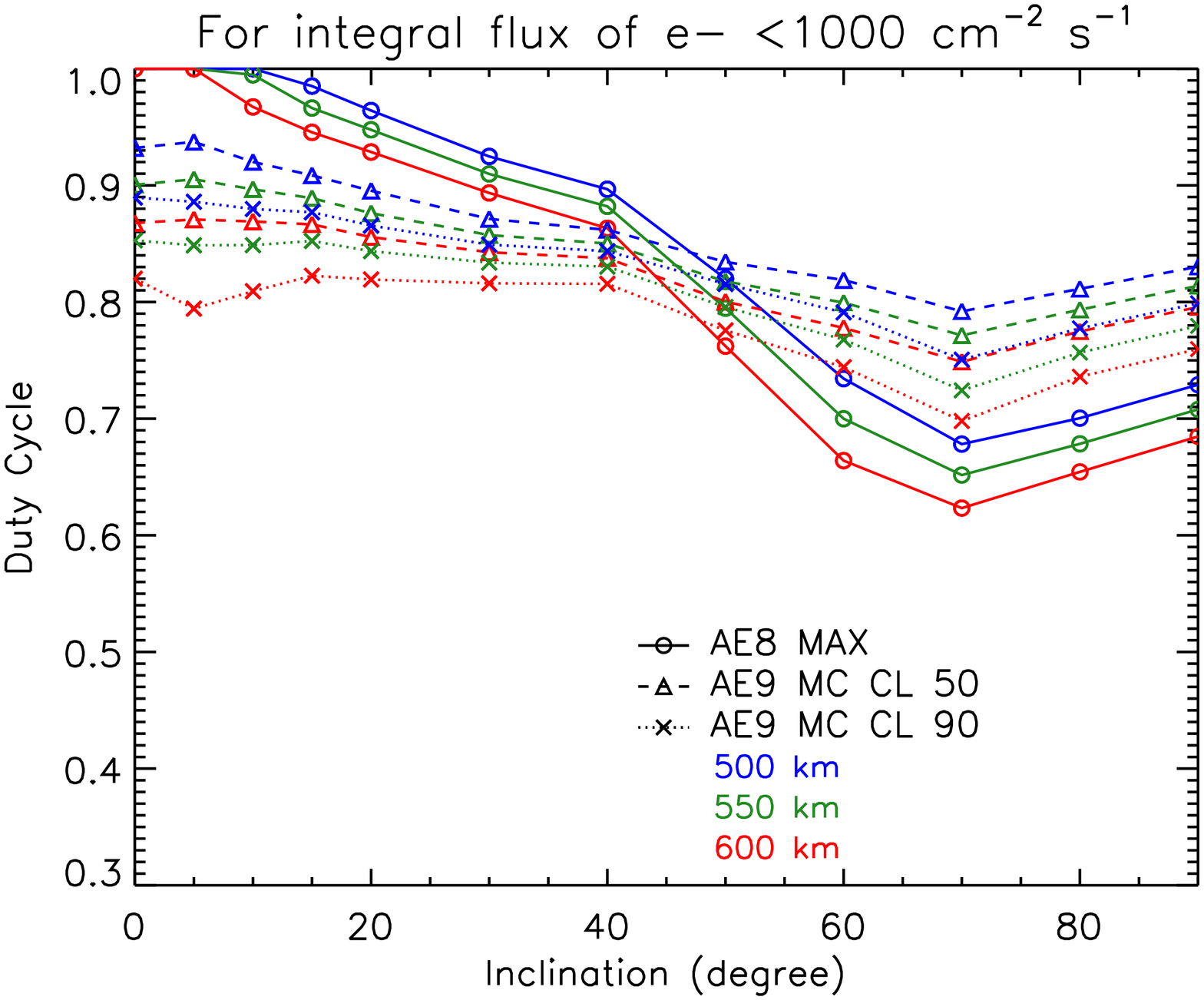}
\end{tabular}
\end{center}
\caption{\label{fig:duty_inc_e} 
Comparison of duty cycle as a function of orbital inclination for different models of trapped electrons for low-energy threshold of 0.04\,MeV and for different flux thresholds and altitudes.}
\end{figure}

\begin{figure}[p]
\begin{center}
\begin{tabular}{cc}
\includegraphics[width=0.333\linewidth]{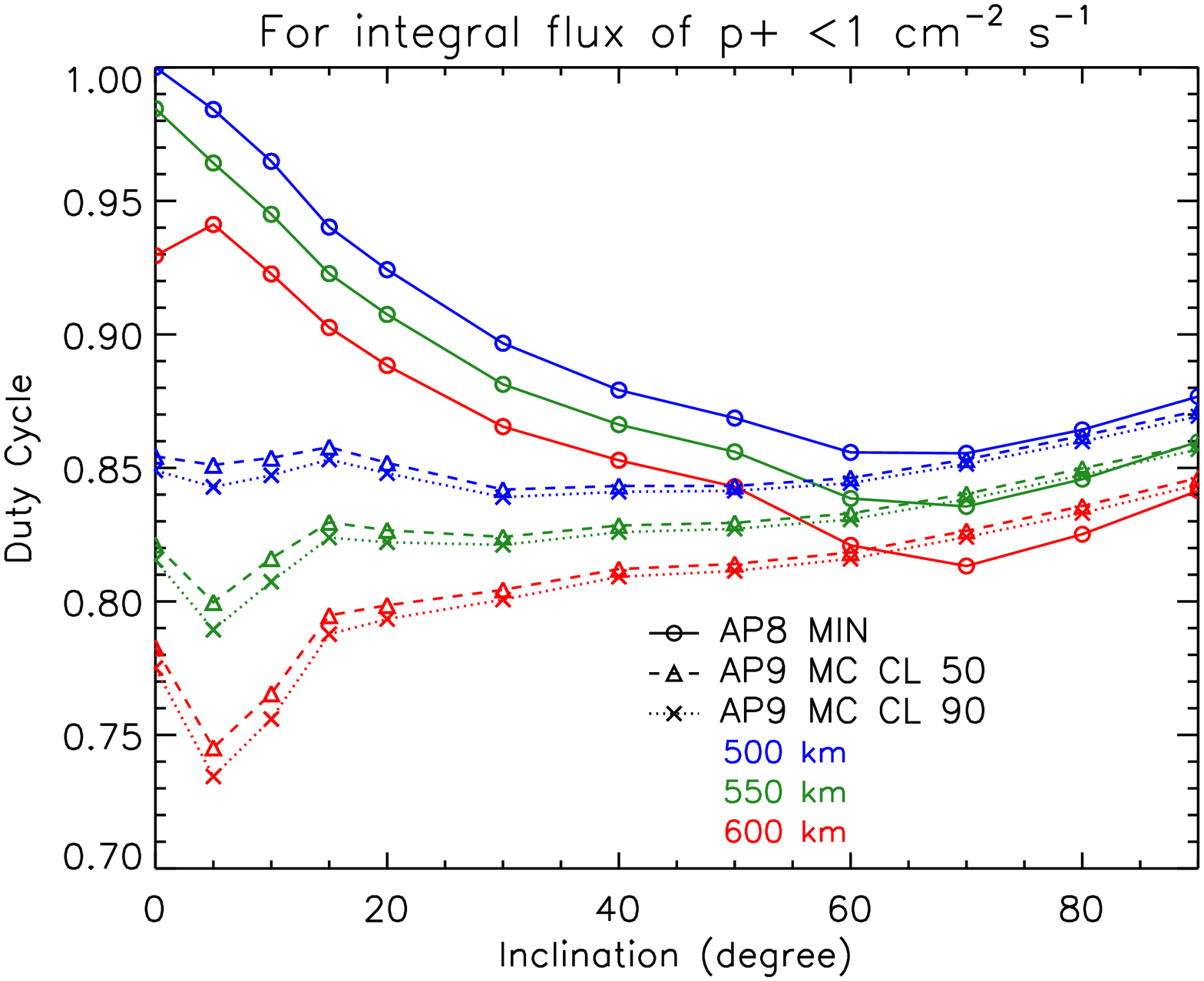}&
\includegraphics[width=0.333\linewidth]{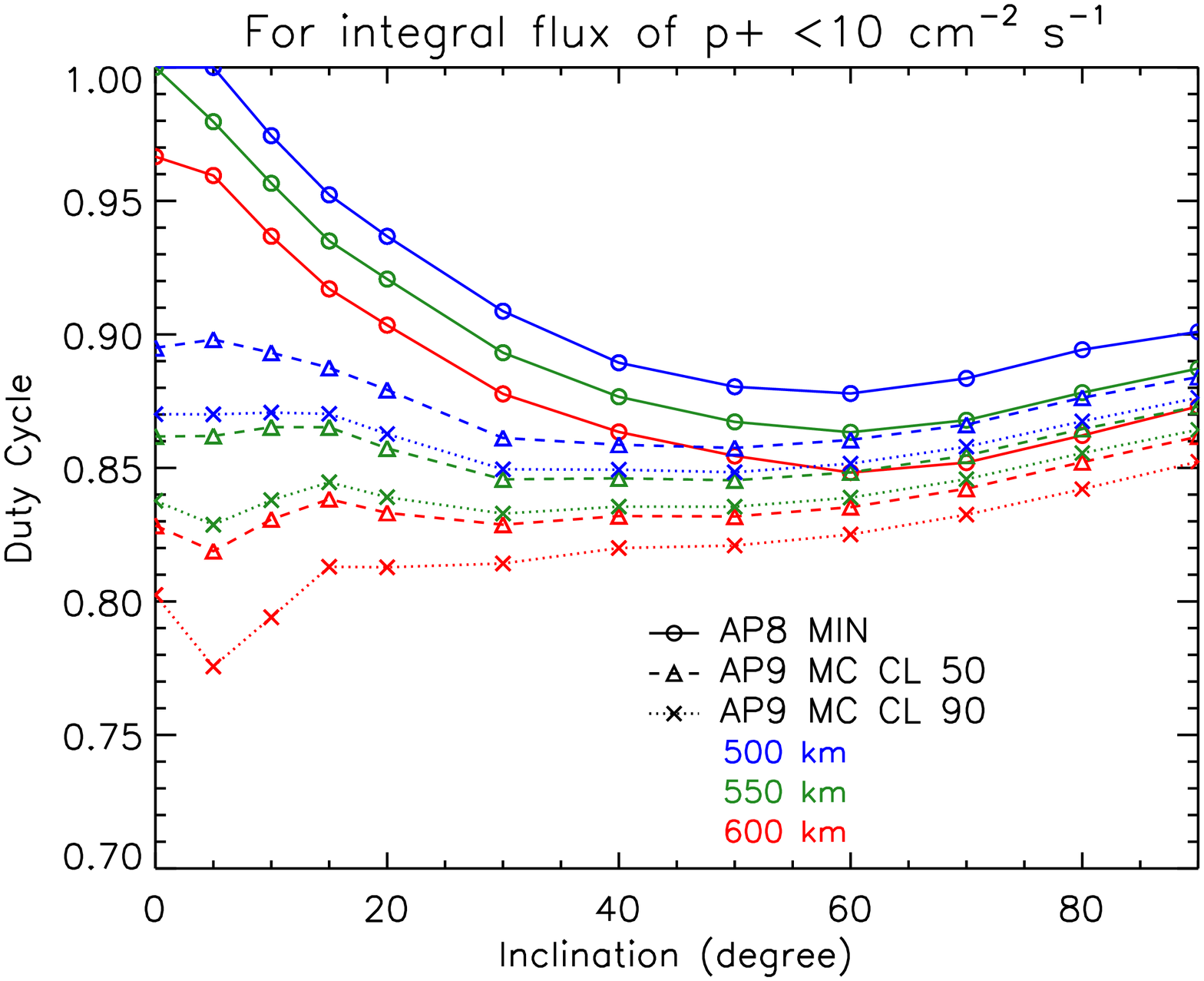}
\\
\includegraphics[width=0.333\linewidth]{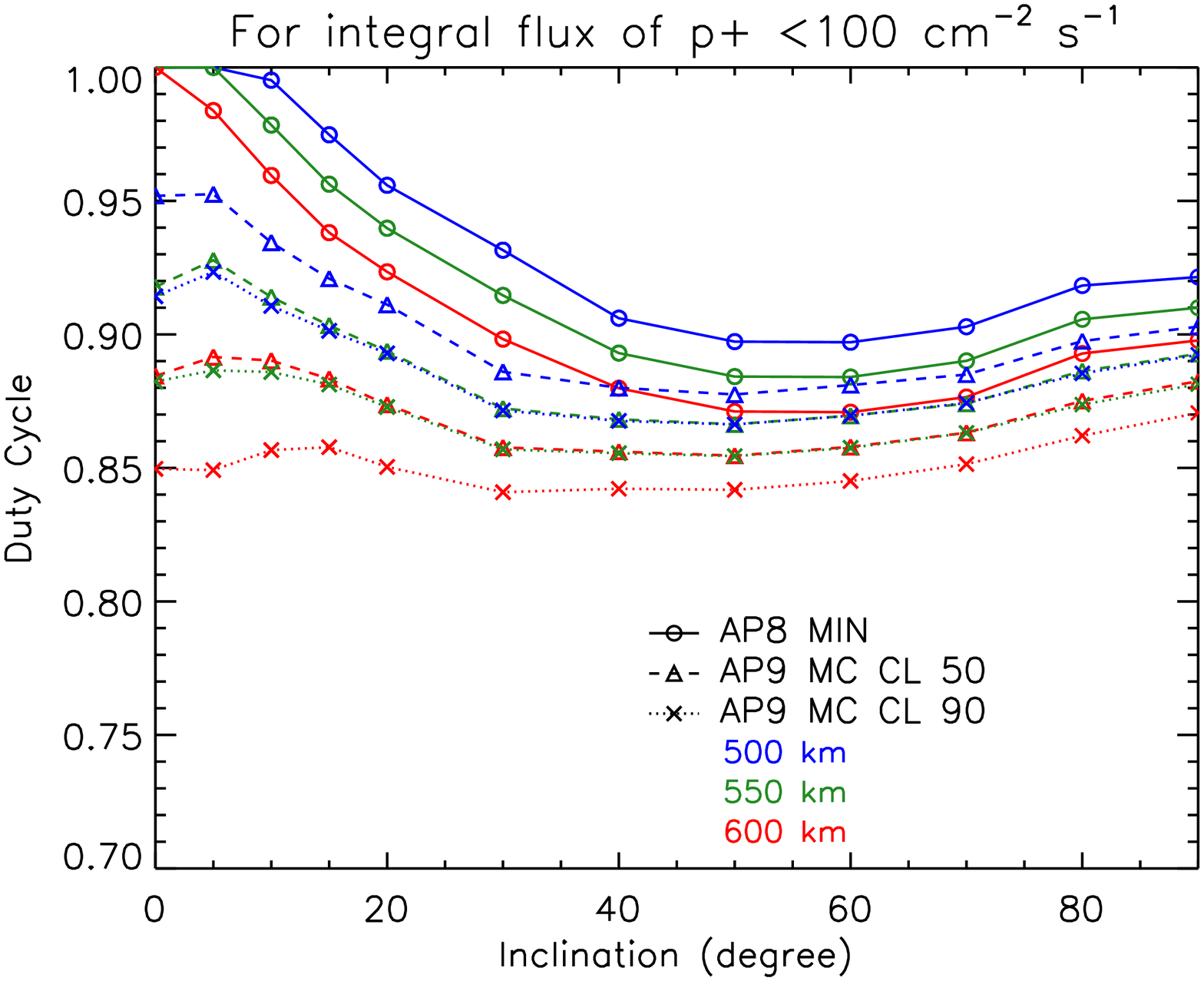}&
\includegraphics[width=0.333\linewidth]{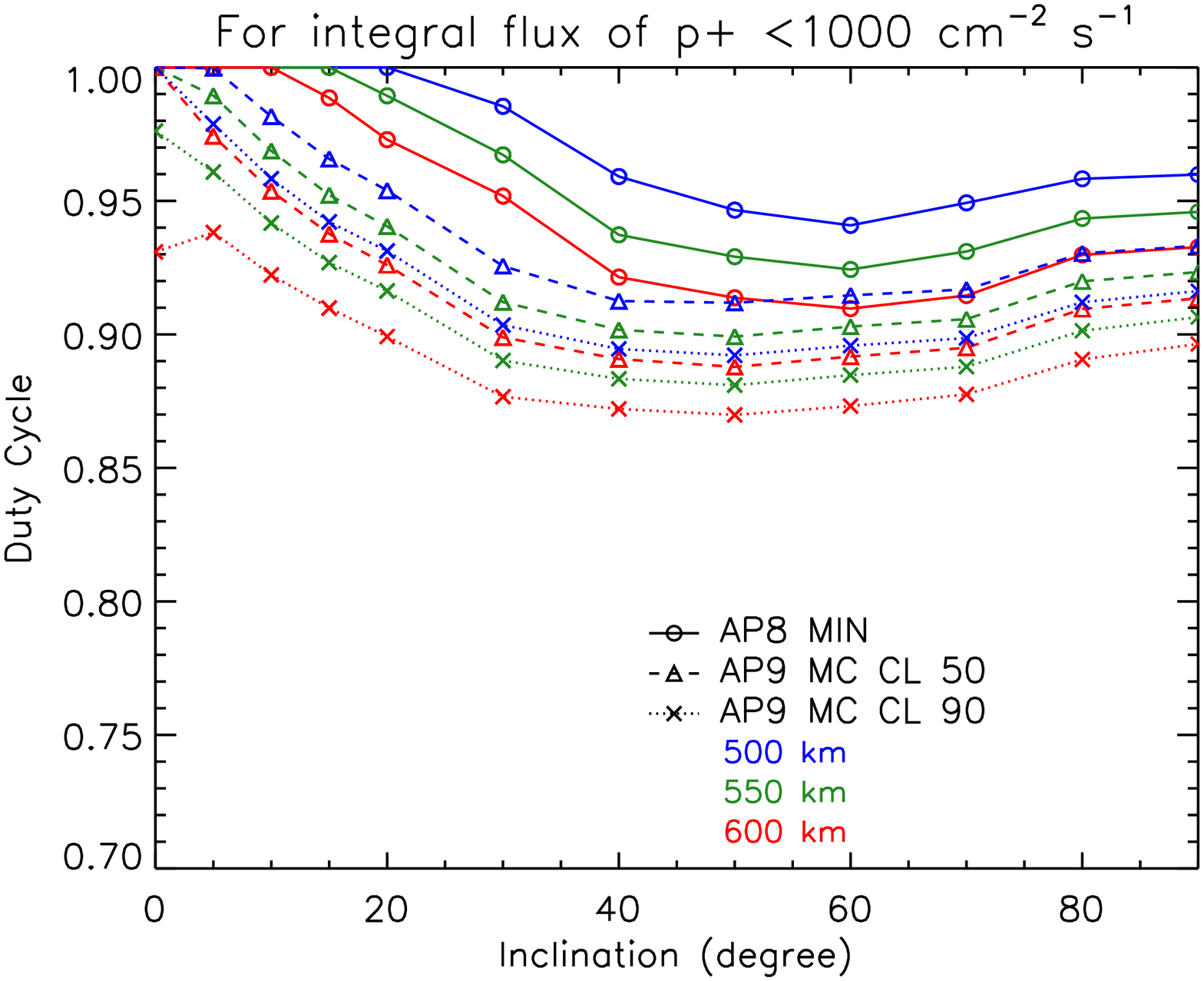}
\end{tabular}
\end{center}
\caption{\label{fig:duty_inc_p} 
Comparison of duty cycle as a function of orbital inclination for different models of trapped protons for low-energy threshold of 0.1\,MeV and for different flux thresholds and altitudes.}
\end{figure}

Figure~\ref{fig:duty_inc_e} shows a comparison of duty cycles as a function of orbital inclination for different models of trapped electrons for low-energy threshold of 0.04\,MeV and for different flux thresholds and altitudes. Similarly, Figure~\ref{fig:duty_inc_p} shows a comparison of duty cycles for trapped protons for low-energy threshold of 0.1\,MeV.

\begin{figure}[p]
\begin{center}
\begin{tabular}{cc}
\includegraphics[width=0.333\linewidth]{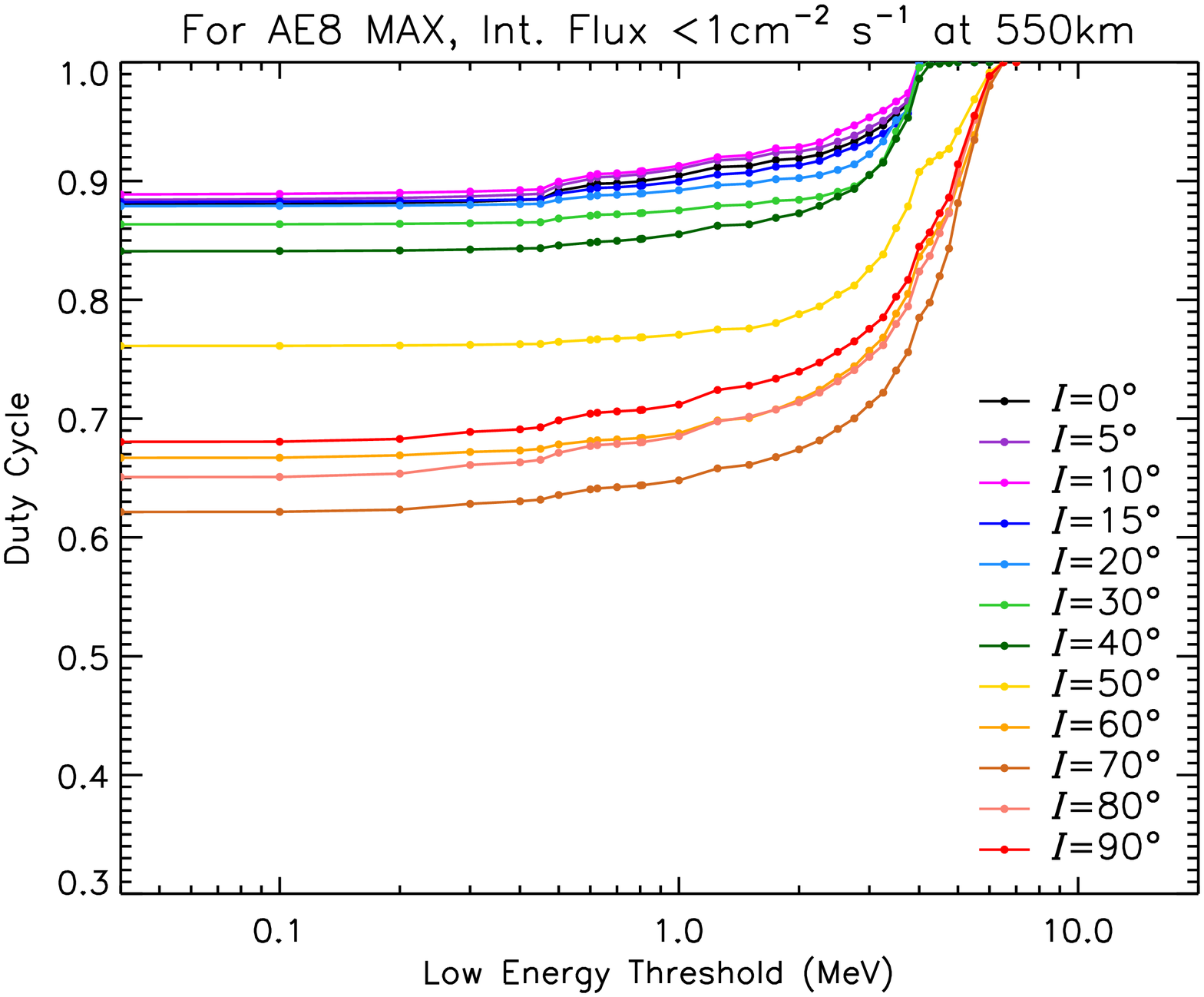}&
\includegraphics[width=0.333\linewidth]{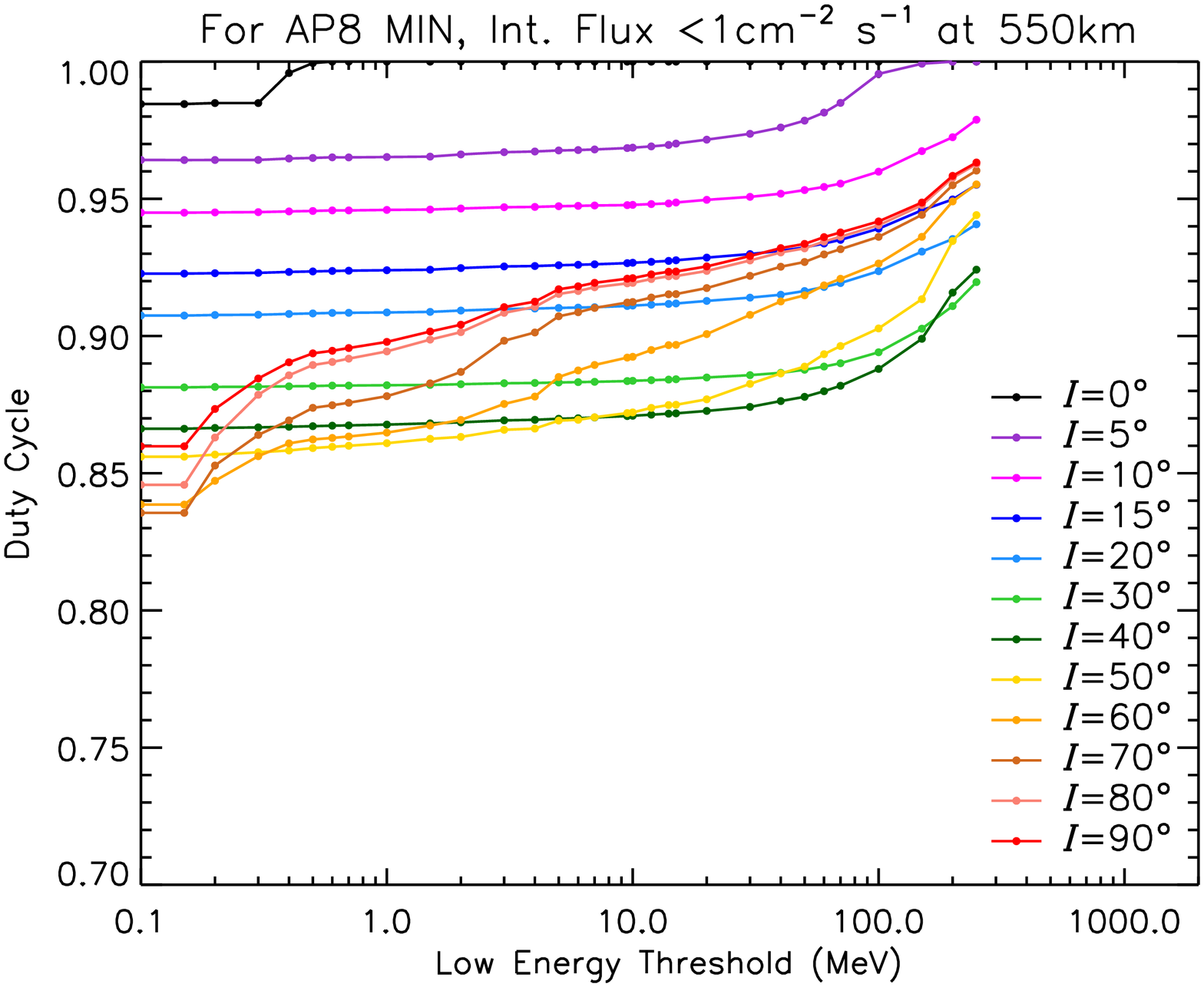}
\\
\includegraphics[width=0.333\linewidth]{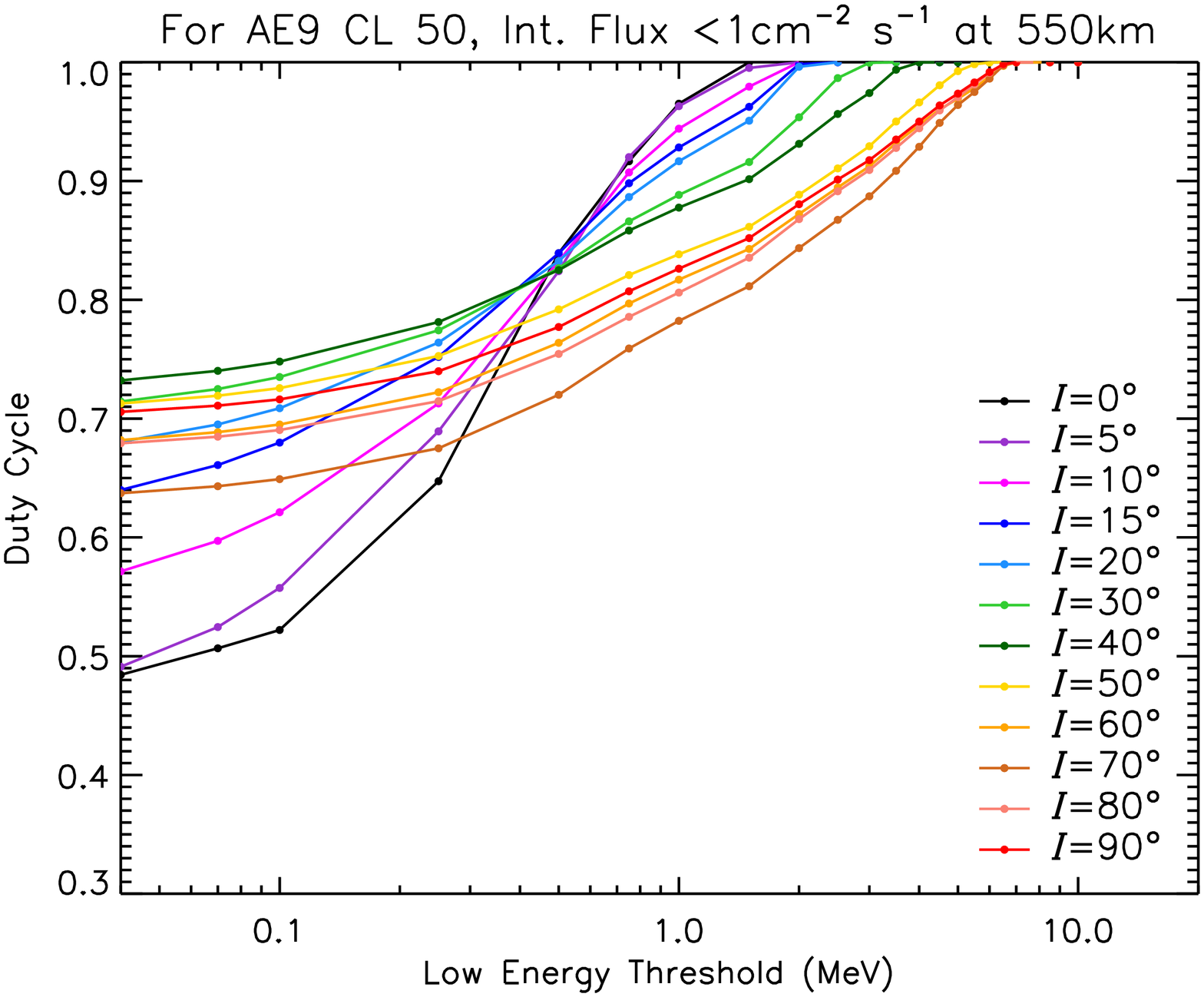}&
\includegraphics[width=0.333\linewidth]{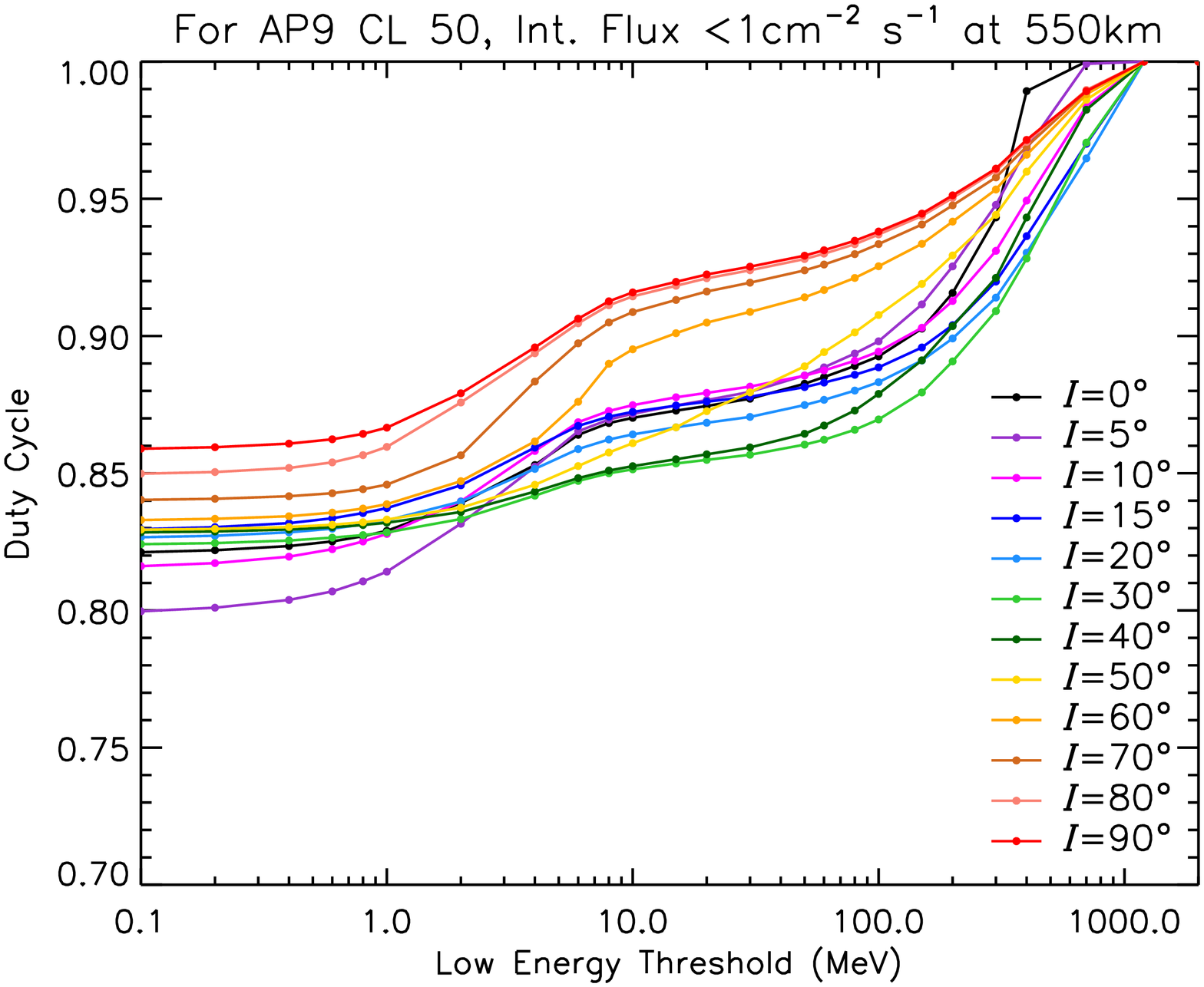}
\end{tabular}
\end{center}
\caption{\label{fig:duty_energy} 
Duty cycle for particle flux $<1$\,cm$^{-2}$s$^{-1}$ for different orbital inclination as a function of the low-energy threshold for AE8MAX (top left), AE9 50\,\% CL (bottom left) models of trapped e$^-$ and AP8MIN (top right), AP9 50\,\% CL (bottom right) models of trapped p$^+$.}
\end{figure}

Figure~\ref{fig:duty_energy} shows duty cycle for particle flux $<1$\,cm$^{-2}$s$^{-1}$ for different orbital inclination as a function of the low-energy threshold for models of trapped electrons and protons at 550\,km altitude.

\section{IN-SITU MEASUREMENTS}
\label{sec:in_situ}

\begin{figure}[p]
\begin{center}
\begingroup
\setlength{\tabcolsep}{1pt} 
\begin{tabular}{ccc}
\includegraphics[width=0.328\linewidth,trim=7 0 7 0,clip]{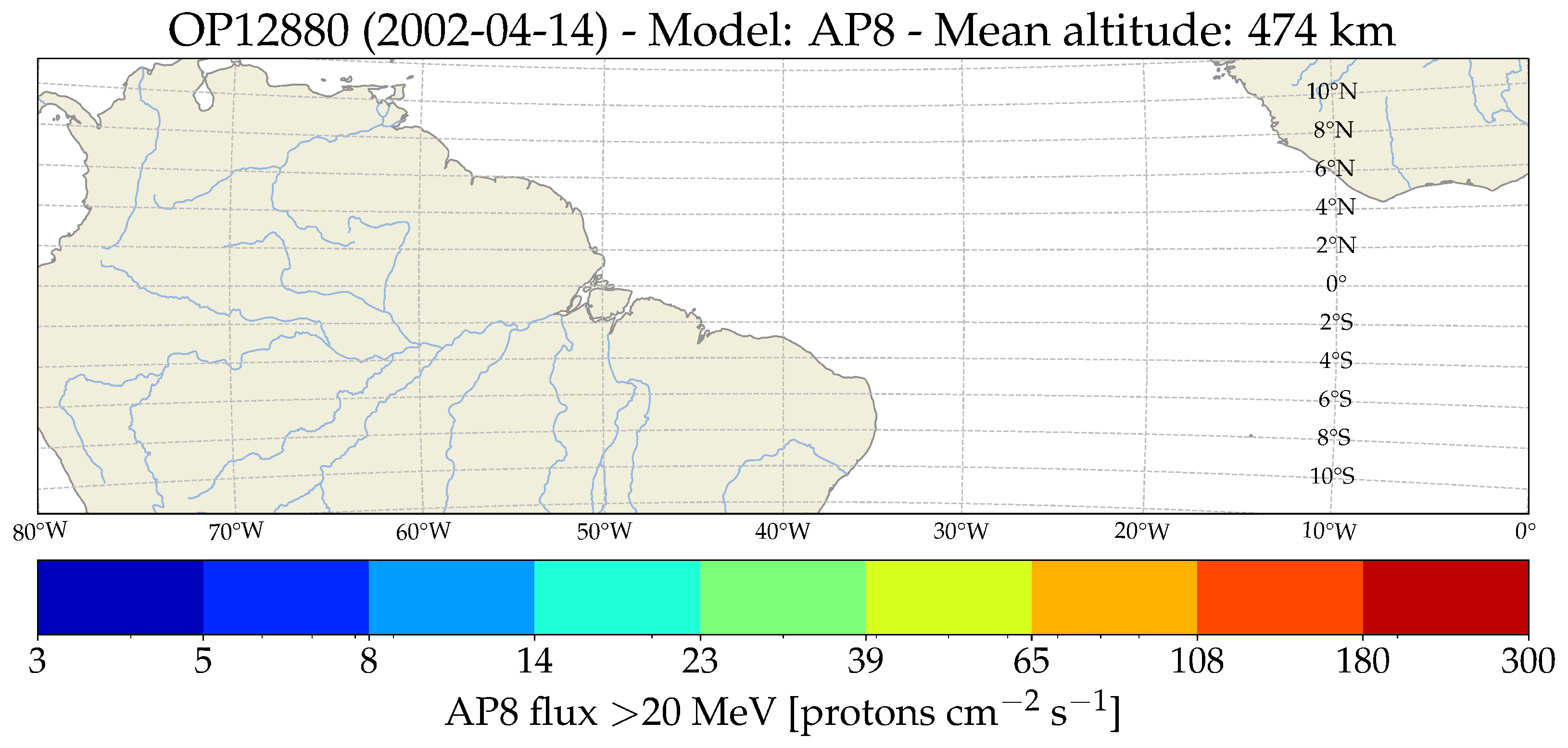}&
\includegraphics[width=0.328\linewidth,trim=7 0 7 0,clip]{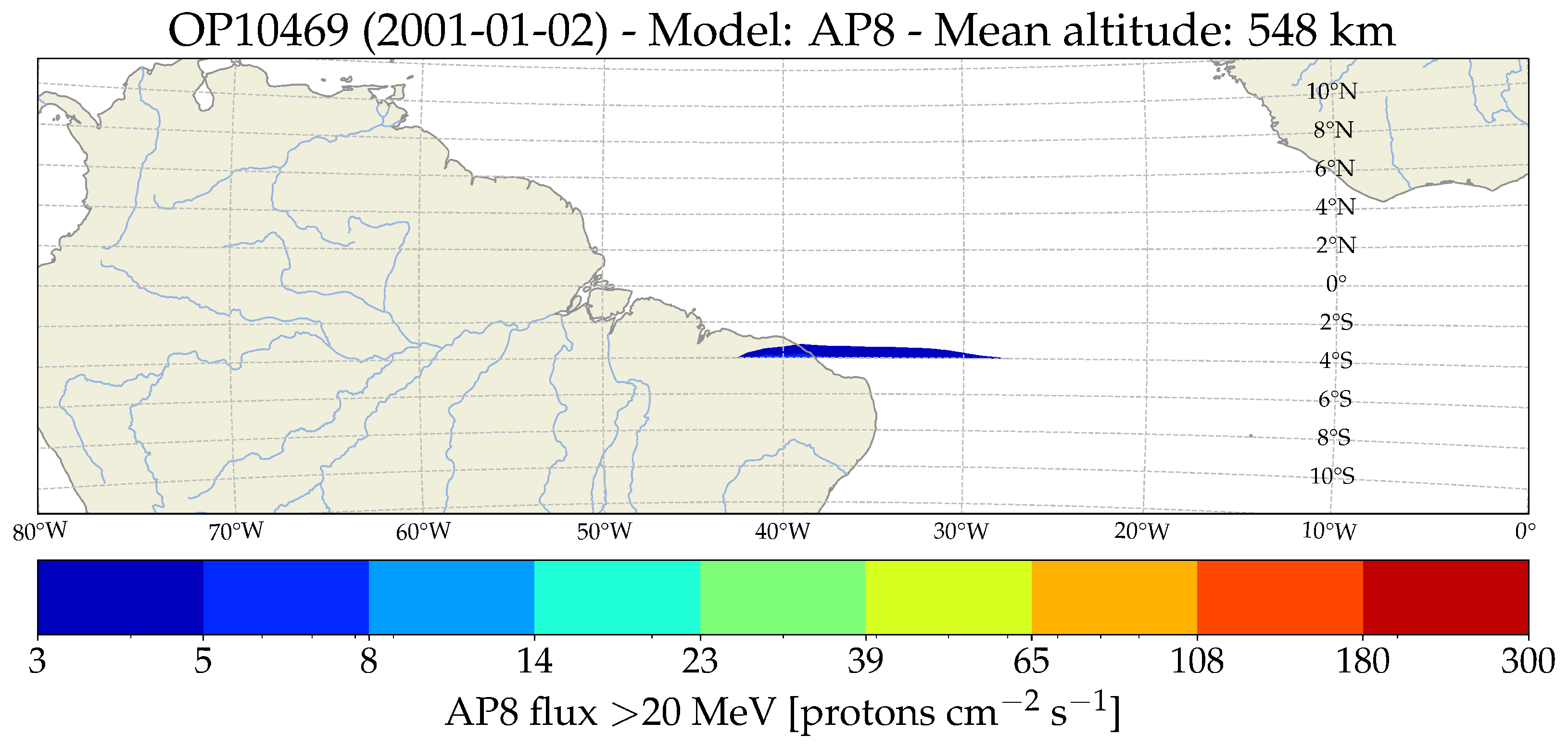}&
\includegraphics[width=0.328\linewidth,trim=7 0 7 0,clip]{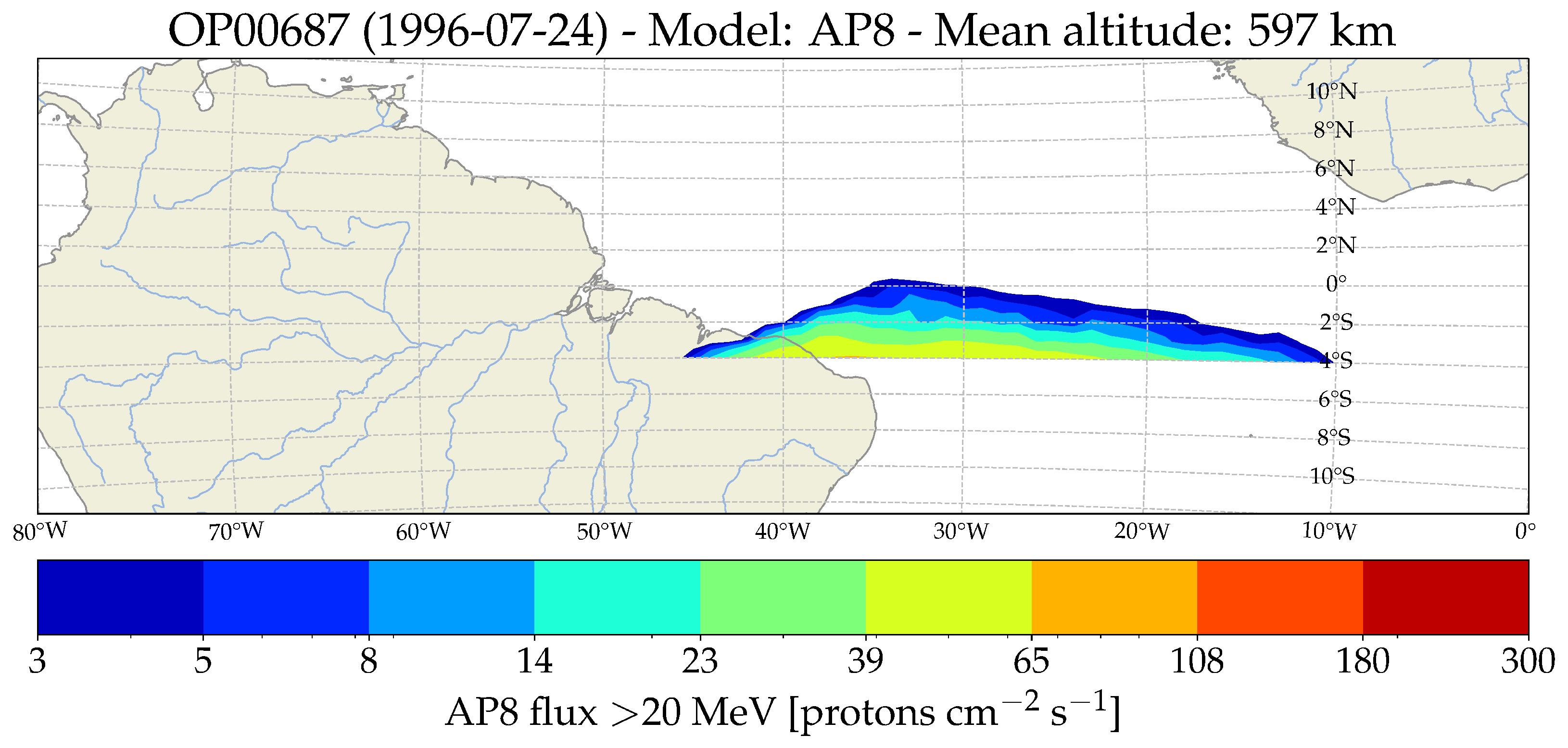}
\\
\includegraphics[width=0.328\linewidth,trim=7 0 7 0,clip]{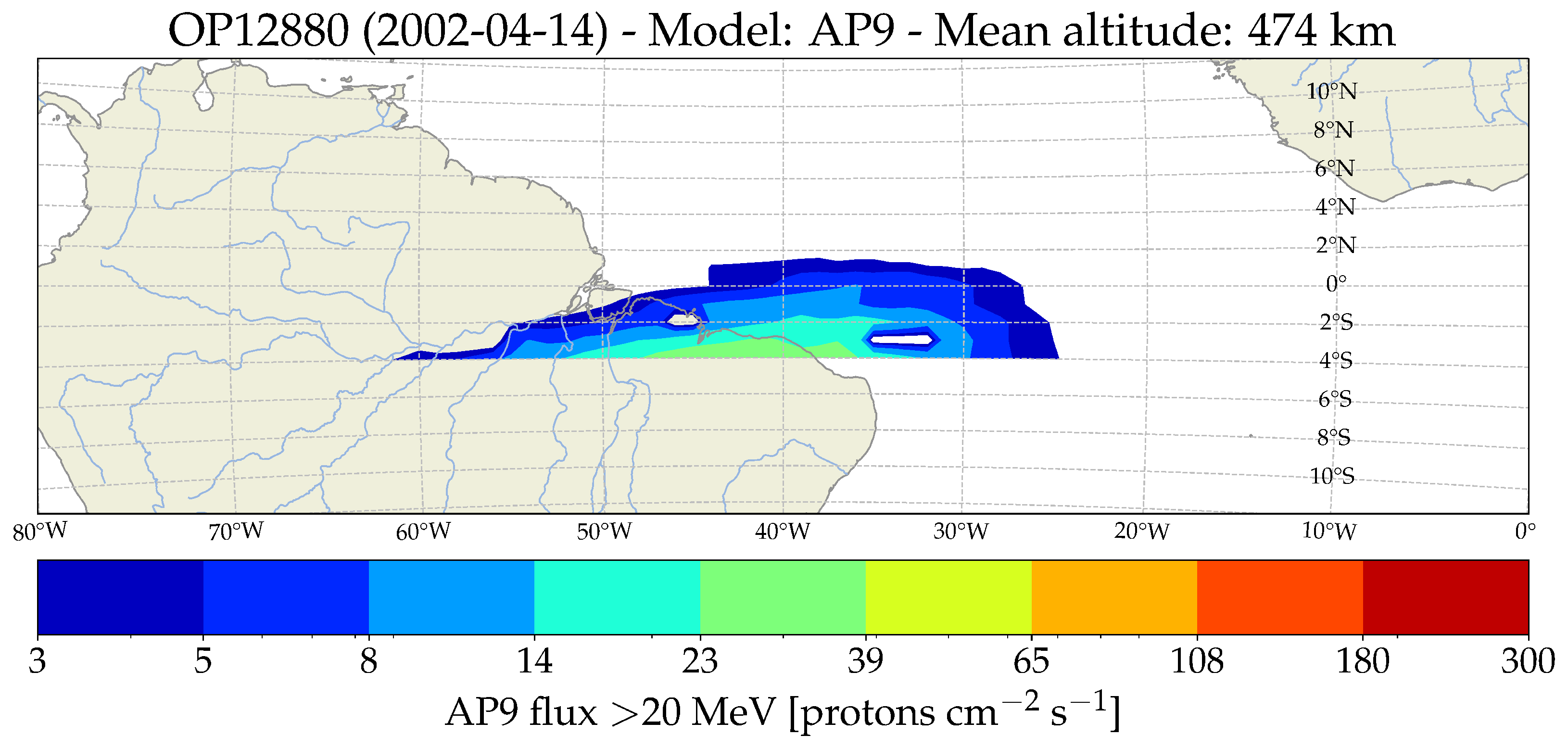}&
\includegraphics[width=0.328\linewidth,trim=7 0 7 0,clip]{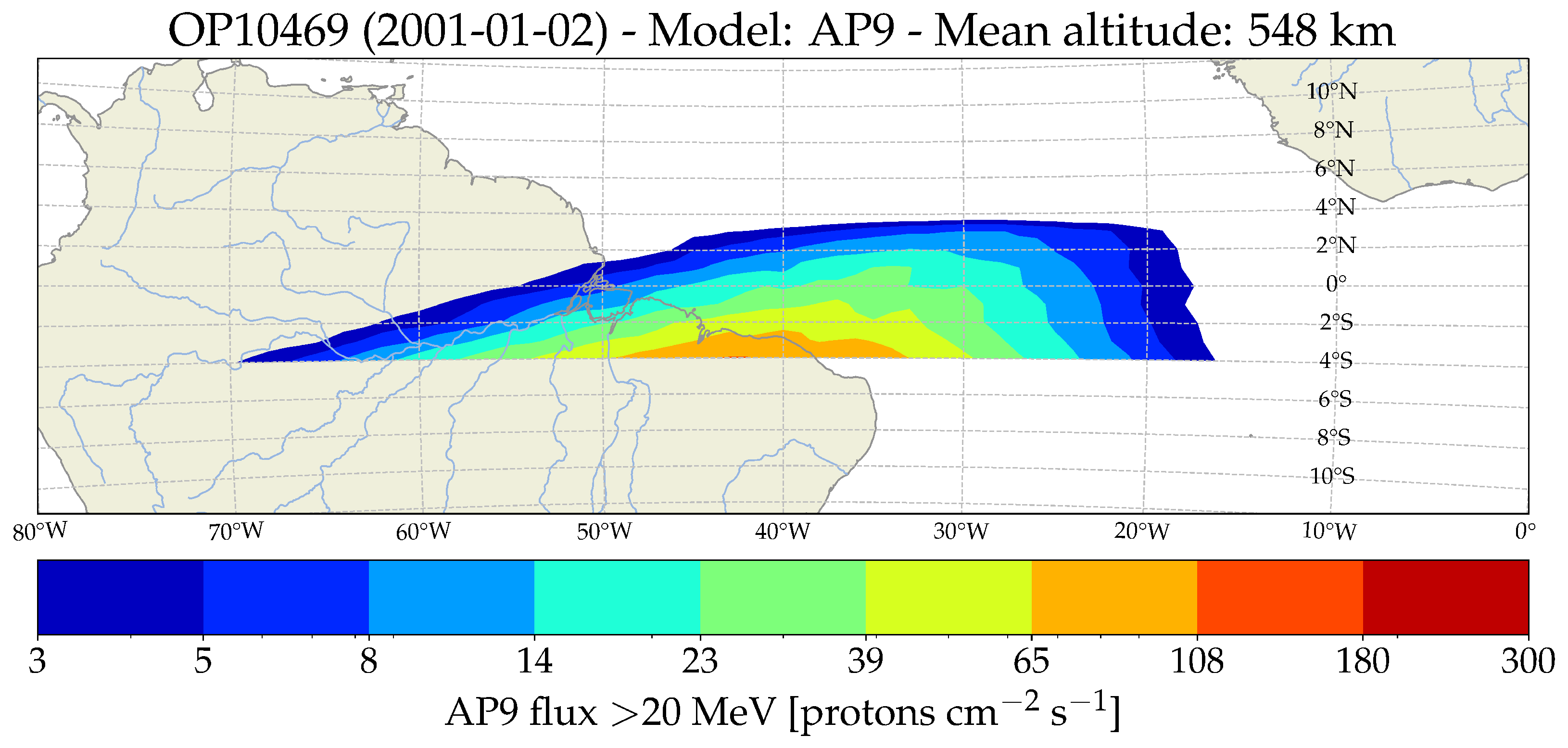}&
\includegraphics[width=0.328\linewidth,trim=7 0 7 0,clip]{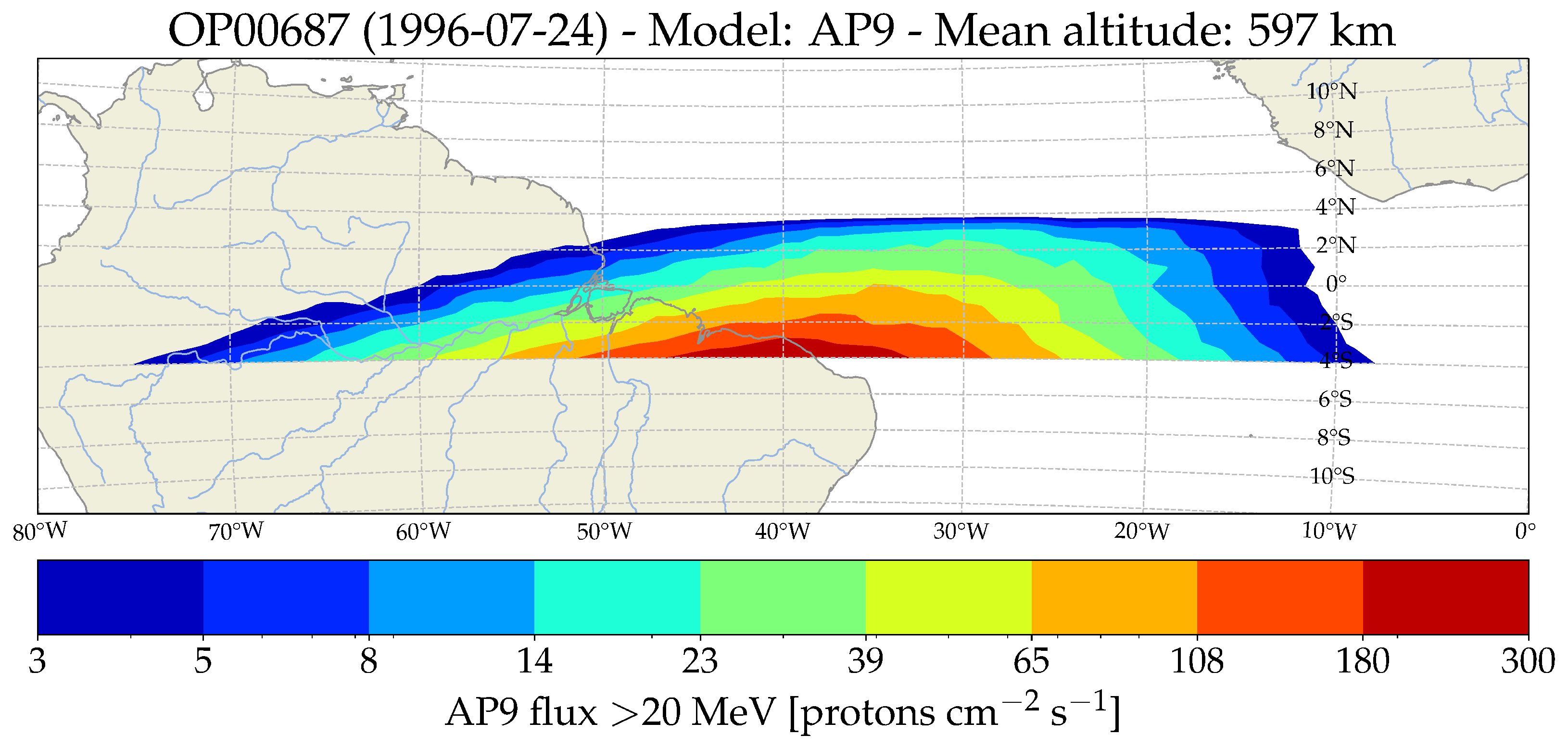}
\\
\includegraphics[width=0.328\linewidth,trim=7 0 7 0,clip]{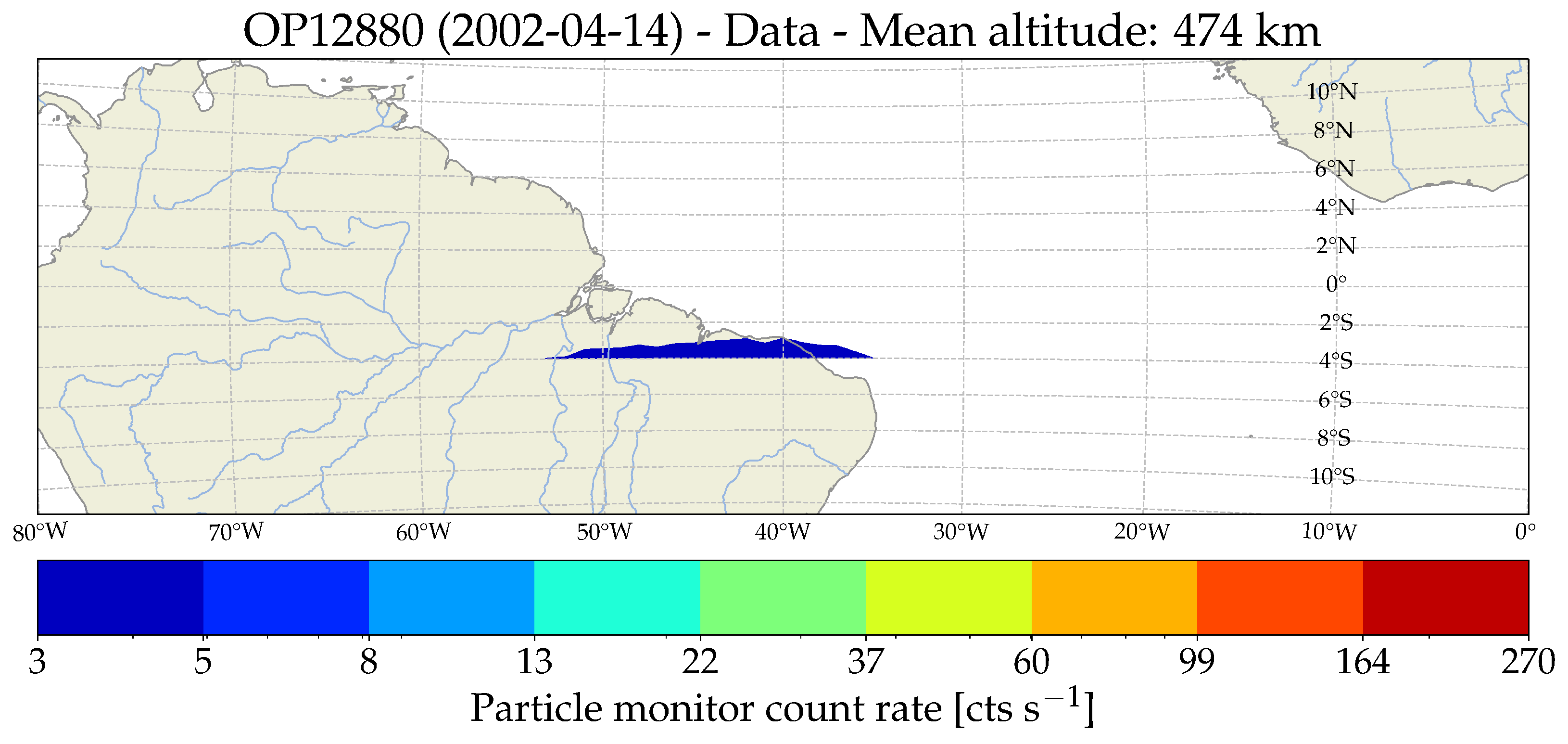}&
\includegraphics[width=0.328\linewidth,trim=7 0 7 0,clip]{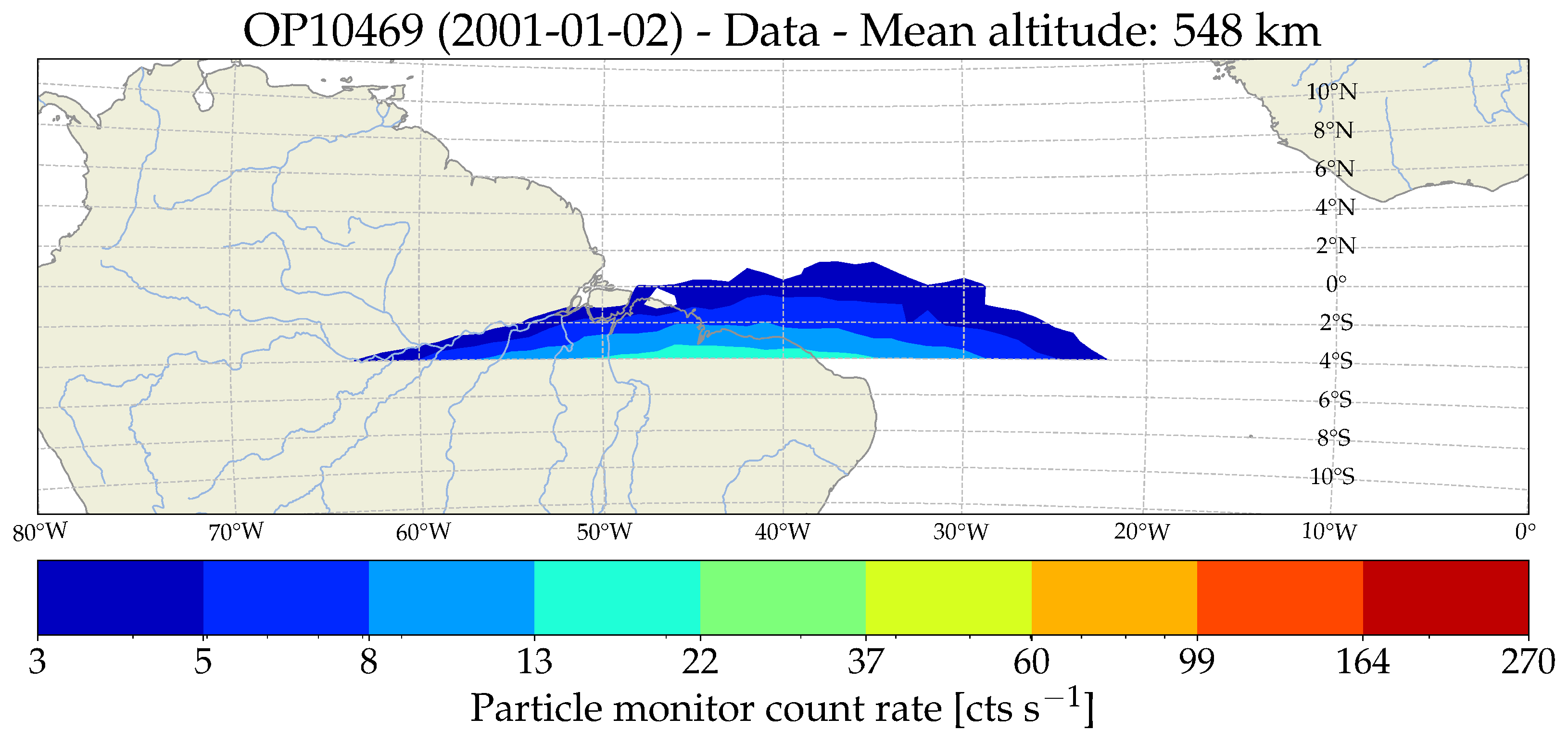}&
\includegraphics[width=0.328\linewidth,trim=7 0 7 0,clip]{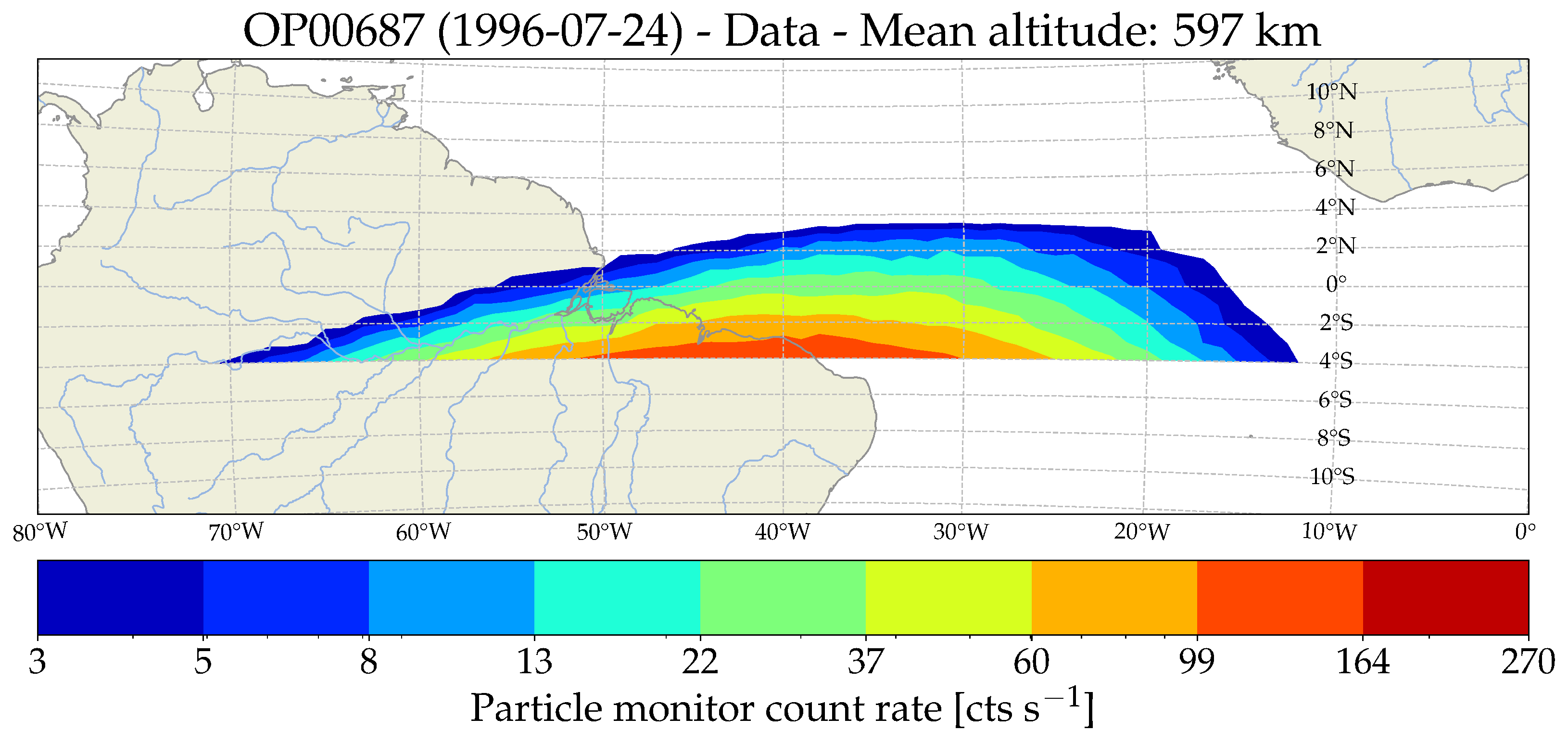}
\end{tabular}
\endgroup
\end{center}
\caption{\label{fig:beppo_sax} 
Comparison of trapped proton fluxes in the South Atlantic Anomaly by the AP8 MIN (top panels) and AP9 mean (middle panels) models with the count rate measured by the particle monitor on the BeppoSAX satellite (bottom panels) with the energy threshold of 20\,MeV for altitude of 474\,km (left), 548\,km (middle) and 597\,km (right).}
\end{figure}

Figure~\ref{fig:beppo_sax} shows a comparison of proton fluxes in SAA by AP8 MIN and AP9 Mean models with the count rate measured by the particle monitor (PM) instrument onboard BeppoSAX satellite at mean altitudes of 474\,km, 548\,km and 597\,km. The energy threshold for proton detections with the PM was 20\,MeV \cite{Campana2014}. The actually measured flux is somewhat in-between the predictions of both models.

\section{CONCLUSIONS}
\label{sec:conclude}

The AE8/AP8 and AE9/AP9 models of near-Earth trapped radiation environment were investigated. We considered 36 different circular low-Earth orbits at three different altitudes (500, 550 and 600\,km) and twelve inclinations ($0^\circ$, $5^\circ$, $10^\circ$, $15^\circ$, $20^\circ$, $30^\circ$, $40^\circ$, $50^\circ$, $60^\circ$, $70^\circ$, $80^\circ$ and $90^\circ$). The main results can be summarized as follows:

\begin{itemize}

\item
AP8 MIN and AP9 models give dramatically different trapped particle fluxes. For low inclinations ($I \leq 20^\circ$ for e$^-$, $I \leq 30^\circ$ for p$^+$) and low energies ($E \lesssim 1$\,MeV for e$^-$, $E \lesssim 10$\,MeV for p$^+$) the AE9 and AP9 50\,\% CL models give up to $\sim 1\,000\times$ (for e$^-$) and up to $\sim 40\,000\times$ (for p$^+$) higher fluxes than the AE8 MAX and AP8 MIN models, respectively. For $I \gtrsim 40^\circ$ different models give comparable results (i.e. within the same order of magnitude).
\vspace{-0.5em}

\item
Higher altitude gives higher flux. A 50\,km difference in orbital altitude gives a factor of roughly $2\times$ difference in orbit-averaged integral flux for the same orbital inclination.

\item
Deepest SAA passings are for $I = ~40-80^\circ$ (AP8 MIN model) or for $I = ~30-70^\circ$ (AP9 50\,\% CL model).
Orbits with $I \gtrsim 40^\circ$ (for e$^-$) or $\gtrsim 30^\circ$ (for p$^+$) face highest average fluxes of particles (up to 6 orders of magnitude higher than near equator).
\vspace{-0.5em}

\item
For e$^-$ by AE8 MAX model, typical duty cycle is $60-90$\,\%, maximal at $I \lesssim 10^\circ$, minimal at $I = \sim 60-80^\circ$.
For p$^+$ by AP8 MIN model, typical duty cycle is $80-100$\,\%, maximal at $I \lesssim 10^\circ$, minimal at $I = \sim 40-70^\circ$.
Lowering altitude by 50\,km increases the duty cycle by about $2-3$\,\%.
AE8 and AE9 give different duty cycle for low $I$ and low flux thresholds since AE9 has excess of low-energy, low-flux e$^-$ near equator compared to AE8.
\vspace{-0.5em}

\item
The measurements from the BeppoSAX satellite indicate that AP9 likely severely overestimate the actual flux values for altitudes below 600\,km and inclinations below $5^\circ$, while AP8 is a likely underestimate.

\end{itemize}

\acknowledgments
 
This work has been carried out in the framework of the HERMES-TP and HERMES-SP collaboration. We acknowledge support from the European Union Horizon 2020 Research and Innovation Framework Programme under grant agreement HERMES-Scientific Pathfinder n. 821896 and from ASI-INAF Accordo Attuativo HERMES Technologic Pathfinder n. 2018-10-HH.0.
The research has been supported by the European Union, co-financed by the European Social Fund (Research and development activities at the E\"{o}tv\"{o}s Lor\'{a}nd University's Campus in Szombathely, EFOP-3.6.1-16-2016-00023).

\bibliography{references} 
\bibliographystyle{spiebib} 

\end{document}